\documentclass[reqno,11pt]{amsart}

\usepackage{amsmath,amssymb,amsthm}
\usepackage{subcaption}
\usepackage[export]{adjustbox}
\usepackage[noend]{algpseudocode}
\usepackage{algorithm}
\usepackage{multicol}
\usepackage{standalone}
\usepackage{url}
\usepackage{hyperref}

\def\endofFact{\hfill\scalebox{.6}{$\Box$}}

\usepackage{eqparbox}
\algnewcommand\LeftComment[1]{%
$\triangleright$ \eqparbox{COMMENT}{#1} \hfill %
}

\usepackage{datetime}
\usepackage{lineno}
\newcommand*\patchAmsMathEnvironmentForLineno[1]{%
\expandafter\let\csname old#1\expandafter\endcsname\csname #1\endcsname
\expandafter\let\csname oldend#1\expandafter\endcsname\csname end#1\endcsname
\renewenvironment{#1}%
{\linenomath\csname old#1\endcsname}%
{\csname oldend#1\endcsname\endlinenomath}}%
\newcommand*\patchBothAmsMathEnvironmentsForLineno[1]{%
\patchAmsMathEnvironmentForLineno{#1}%
\patchAmsMathEnvironmentForLineno{#1*}}%
\AtBeginDocument{%
\patchBothAmsMathEnvironmentsForLineno{equation}%
\patchBothAmsMathEnvironmentsForLineno{align}%
\patchBothAmsMathEnvironmentsForLineno{flalign}%
\patchBothAmsMathEnvironmentsForLineno{alignat}%
\patchBothAmsMathEnvironmentsForLineno{gather}%
\patchBothAmsMathEnvironmentsForLineno{multline}%
}

\newtheorem{theorem}             {Theorem}[section]
\newtheorem{lemma}     	[theorem] {Lemma}

\newtheorem{proposition}[theorem] {Proposition}   
\newtheorem{corollary}	[theorem] {Corollary}
     
\newtheorem{claim}	[theorem] {Claim}  

\newcommand{\oldqed}{}
\def\endofFact{\hfill\scalebox{.6}{$\Box$}}
\newenvironment{claimproof}[1][Proof]{
  \renewcommand{\oldqed}{\qedsymbol}
  \renewcommand{\qedsymbol}{\endofFact}
  \begin{proof}[#1]
}{
  \end{proof}
  \renewcommand{\qedsymbol}{\oldqed}
} 

\usepackage{geometry}
\usepackage{pgf}
\usepackage{tikz}
\usepackage{tkz-graph}
\usetikzlibrary{arrows,calc,fit,positioning,decorations.pathmorphing,decorations.pathreplacing,decorations.markings}
\usepackage{./figs/figstyle}

\usepackage{pgfplots}
\usepackage{enumerate}   

\usepackage[T1]{fontenc}
\usepackage[english]{babel}

\newcommand{\NP}{\mbox{NP}}

\newcommand{\Dag}{dag}
\newcommand{\opt}{\mathrm{opt}}
\newcommand{\maxleaves}{{\sc Maximum Leaf Spanning Arborescence}}
\newcommand{\maxsnp}{MaxSNP}
\newcommand{\calA}{\mathcal{A}}

\newcommand{\calS}{\mathcal{S}}

\newcommand{\Candidates}{\mathit{Candidates}}
\newcommand{\um}{\vspace{1mm}}

\newcommand{\SquareImp}{\textsc{SquareImp}}

\begin{document}

\title{How heavy independent sets help to find \\
arborescences with many leaves in DAGs}
\thanks{
C.~G. Fernandes was partially supported by CNPq (Proc.~308116/2016-0, ~423833/2018-9, 
and~310979/2020-0).
C. N. Lintzmayer was partially supported by CNPq (Proc.~428385/2018-4).}

\author{Cristina G. Fernandes and Carla N. Lintzmayer}

\shortdate
\yyyymmdddate
\settimeformat{ampmtime}
\date{\today, \currenttime}

\address{Institute of Mathematics and Statistics. University of S{\~a}o Paulo. S{\~a}o Paulo, Brazil}
\email{\href{mailto:cris@ime.usp.br}{\texttt{cris@ime.usp.br}}}

\address{Center for Mathematics. Computing and Cognition. Federal University of ABC. Santo Andr{\'e}, S{\~a}o Paulo, Brazil}
\email{\href{mailto:carla.negri@ufabc.edu.br}{\texttt{carla.negri@ufabc.edu.br}}}

\begin{abstract}
    Trees with many leaves have applications on broadcasting, which is a method
    in networks for transferring a message to all recipients simultaneously.  
    Internal nodes of a broadcasting tree require more expensive technology,
    because they have to forward the messages received.
    We address a problem that captures the main goal, 
    which is to find spanning trees with few internal nodes in a given network.
    The \maxleaves\ problem consists of, given a directed graph~$D$, 
    finding a spanning arborescence of~$D$, if one exists, 
    with the maximum number of leaves.  
    This problem is known to be \NP-hard in general and \maxsnp-hard on the
    class of rooted directed acyclic graphs. 
    In this paper, we explore a relation between \maxleaves\ in rooted directed 
    acyclic graphs and maximum weight set packing. 
    The latter problem is related to independent sets on particular classes 
    of intersection graphs.
    Exploiting this relation, we derive a $7/5$-approximation for \maxleaves\ 
    on rooted directed acyclic graphs, improving on the previous $3/2$-approximation. 
    The approach used might lead to improvements on the best approximation
    ratios for the weighted $k$-set packing problem.
\end{abstract}

\maketitle

\section{Introduction}

Broadcasting is a term used to describe the process of sending a message on a
network from a root node to all other nodes of the network.  
A network is modeled as a directed graph in which we broadcast messages
through a minimal subset of the arcs of the network. 
These subsets consist of what we call a spanning arborescence in the network.
The internal nodes of an arborescence receive a message through an arc, and
must duplicate the message and distribute its copies through the outgoing arcs
of the arborescence.
Thus, the internal nodes must be equipped with routers and switches, 
while leaves of the arborescence need only to work as message receptors. 
This situation motivates the search for broadcasting arborescences in networks
with fewer internal nodes and more leaves~\cite{JuttnerM2005,PopeS2015}.
In what follows, we formalize this problem. 

Let~$D$ be a directed graph (digraph for short).  
A node~$r$ in~$D$ is a \emph{root} if there is a directed path in~$D$ from~$r$
to every node in~$D$. 
If~$r$ is a root in~$D$, then we say~$D$ is \emph{$r$-rooted}, or simply
\emph{rooted}.
We say~$D$ is \emph{acyclic} if there is no directed cycle in~$D$.
For short, a directed acyclic graph is called a \emph{dag}.
Any rooted dag has only one root.
An \emph{arborescence} is an $r$-rooted dag~$T$ for which there is a unique
directed path from~$r$ to every node in~$T$. 
The \emph{out-degree} of a node in a digraph is the number of arcs that
start in that node, while 
the \emph{in-degree} of a node is the number of arcs that end in that node.
A node of out-degree~$0$ in an arborescence is called a \emph{leaf}.
Note that the underlying graph of an arborescence is a tree.

In the \maxleaves\ problem, one is given a rooted digraph~$D$, and the
goal is to find a spanning arborescence of~$D$ with the maximum number of
leaves.  
This problem is known to be \NP-hard.
Indeed, its undirected version is listed as the \NP-hard problem ND2 in the
renowned book by Garey and Johnson~\cite{GareyJ1979}, and one can easily reduce
ND2 to \maxleaves.

\maxleaves\ was considered from the viewpoint of fixed parameter
tractability~\cite{AlonFGKS2009,BinkeleRaibleFFLSV2012,CyganFKLMPPS2016,DaligaultT2009}.
Regarding approximation algorithms, which are our interest in this paper, there
exists a 92-approximation for general rooted digraphs.
It came from Daligault and Thomassé's analysis on the parameterized complexity
of the problem~\cite{DaligaultT2009}.
Their work includes the proof that their algorithm has ratio~24 if the digraph
has no directed cycles of length~2.

The case in which the given digraph is a rooted dag was considered in the
literature~\cite{AlonFGKS2009,FernandesL2021,MakurMP2020,SchwartgesSW2012}.  
Specifically, this case was shown to be \maxsnp-hard~\cite{SchwartgesSW2012}
and the best known result for it is a~$3/2$-approximation whose analysis is
tight~\cite{FernandesL2021}.
Thus, improvements require a different algorithm.
In~\cite{FernandesL2021}, an alternative algorithm was proposed for \maxleaves\
on rooted dags.
It uses as a subroutine an approximation for maximum weight 3D-matching, whose
currently best known factor is unfortunately not good enough to achieve an
improvement for \maxleaves\ on rooted dags. 

The best approximations for maximum weight 3D-matchings and for maximum weight 
$k$-set packing in general come from the best approximations for the maximum 
weight independent set on 4-claw free and $(k+1)$-claw free graphs 
respectively~\cite{Berman2000,ChandraH2001,Neuwohner2021}.
In this paper we explore the idea of using the latter approximations 
to obtain better algorithms for \maxleaves\ on rooted dags. 
Concretely, we introduce a particular class of 4-claw free graphs that we call
$\{2,3\}$-intersection graphs, and we design a $7/5$-approximation for the
maximum weight independent set on the class of $\{2,3\}$-intersection graphs
with vertex weights defined in a particular way.  
The new approximation is a tuned version of Berman's approximation for the
maximum weight independent set on 4-claw free graphs~\cite{Berman2000},
inspired also on ideas from the algorithm for \maxleaves\ in~\cite{FernandesL2021}.
From this, we derive a~$7/5$-approximation for \maxleaves\ on rooted dags. 

Berman's algorithm for maximum weight independent set consists of a local
improvement algorithm, as many of the algorithms for maximum (weight) independent set. 
Our tuned algorithm relies on restricted weights, since we are interested 
in a particular class of weight functions, and first modifies
the criterion to decide on whether or not to apply a local improvement.
Then it optimizes some 
particular types of improvements using a maximum matching algorithm.  The way we
modified the decision on when to apply an improvement step might lead to a new 
algorithm for maximum weight set packing, if not for general weights, maybe for 
particular classes of weight functions that, as the ones we used, might be of 
interest for other problems. 

Section~\ref{sec:MIS} presents the maximum weight independent set problem
(wMIS) and summarizes the results known for wMIS on $d$-claw free graphs. 
In particular, we revise the algorithm of Berman known as \SquareImp\ for wMIS.
Section~\ref{sec:maxleavesMIS} discusses the relation between \maxleaves\ on
rooted dags and wMIS on $d$-claw free graphs and, as an intermediate step,
presents a new algorithm for \maxleaves\ on rooted dags that uses
\SquareImp\ as a subroutine.
Section~\ref{sec:23intersectiongraphs} analyzes \SquareImp, showing that 
it is a $3/2$-approximation on a subclass of weighted 4-claw free graphs 
that contains the weighted graphs used by the new algorithm. 
Section~\ref{sec:backtomaxleaves} concludes that the new algorithm is 
a~$3/2$-approximation for \maxleaves\ on rooted dags.
In Section~\ref{sec:better_approx}, we present our main result: a
$7/5$-approximation for \maxleaves\ on rooted dags that uses ingredients from
the $3/2$-approximation in~\cite{FernandesL2021} to tune the
$3/2$-approximation from Section~\ref{sec:maxleavesMIS}.
The new approximation relies on a new $7/5$-approximation on the subclass 
of weighted 4-claw free graphs considered in 
Section~\ref{sec:23intersectiongraphs}.
Further directions are discussed in Section~\ref{sec:remarks}.


\section{Independent sets in $d$-claw free graphs}
\label{sec:MIS}

Let~$G$ be an undirected graph.  
A set~$I$ of vertices of~$G$ is \emph{independent} if the vertices in~$I$ are
pairwise non-adjacent in~$G$. 
A \emph{weighted graph} is a pair $(G,w)$ where~$G$ is an undirected graph
and~$w$ is a function that assigns to each vertex~$v$ of~$G$ a positive
weight~$w_v$. 

The \textsc{Maximum Weight Independent Set} (wMIS) problem consists of, given a
weighted graph~$(G,w)$, finding an independent set~$S$ in~$G$ that
maximizes~$w(S)$, which is the sum of~$w_v$ for all~$v$ in $S$.
In general, wMIS is quite hard to approximate, being
Poly-APX-complete~\cite{BazganEP2005}.  
But we will consider wMIS on a well-known manageable class of graphs. 

An induced subgraph~$C$ of~$G$ is a \emph{$d$-claw} if it consists of an
independent set~$T_C$ of~$d$ vertices, and a \emph{center} vertex that is
adjacent to all~$d$ vertices in~$T_C$.  
For convenience, a singleton is said to be a $1$-claw~$C$ with its unique
vertex in~$T_C$ and no center.
Several problems were studied for the class of $d$-claw free graphs, which are
those in which no induced subgraph is a $d$-claw. 
The reason is that, in many applications, $d$-claw free graphs arise naturally. 
Indeed, we will see in the next section how this class plays a role on
\maxleaves\ in rooted dags.
For convenience, we will use the term claw to refer to a $d$-claw for an
arbitrary~$d$. 

Berman~\cite{Berman2000} presented an algorithm for wMIS on weighted
$d$-claw free graphs with an approximation ratio of $d/2$, enhancing 
on the previous work of Chandra and Halldórsson~\cite{ChandraH2001}.  
Recently Neuwohner~\cite{Neuwohner2021} presented a variation
of Berman's algorithm, showing that its ratio is slightly less than~$d/2$ on
weighted $d$-claw free graphs.
This is the currently best approximation for wMIS on weighted $d$-claw free graphs.
Our algorithm is also based on Berman's.
In what follows, we introduce concepts that are required for the description 
of his algorithm and our results.

For a graph $G=(V,E)$ and a set $S \subseteq V$, we denote by $N(S)$ the
neighborhood of~$S$ in~$G$, that is, $N(S) := \{u \in V : \mbox{$vu \in E$ for
some $v \in S$}\}$.
We use $N(u)$ to denote $N(\{u\})$.
If~$A$ is an independent set and~$C$ is a claw in~$G$, then $(A \cup T_C)
\setminus N(T_C)$ is also an independent set. 
So claws can be used in a greedy way to obtain a larger or heavier independent
set. 
For a weighted graph~$(G,w)$ and a set~$S$ of vertices of~$G$, we denote
by~$w^2(S)$ the sum of the squares $w^2_v$ for all~$v$ in~$S$. 
For an independent set~$A$ in~$G$, we say that a claw \textit{$C$
improves~$w^2(A)$} if~$w^2(A) < w^2((A \cup T_C) \setminus N(T_C))$.
If~$T_C=\{u\}$, we say simply that~$u$ improves $w^2(A)$. 

Algorithm~\ref{alg:11}, called \SquareImp, was proposed by Berman for wMIS 
on weighted $d$-claw free graphs.
This algorithm however might not run in polynomial time, but Berman used the
strategy of Chandra and Halldórsson~\cite{ChandraH2001} to obtain a polynomial
version with a slightly increase in the approximation ratio.  
In particular, on weighted 4-claw free graphs, this leads to a ratio slightly more than~2.  

\begin{algorithm}
  \begin{algorithmic}[1]
    \Require{weighted graph $(G,w)$}
    \Ensure{an independent set in $G$}
    
    \State $A \gets \emptyset$
    \While{there is a claw $C$ in $G$ such that $T_C$ improves $w^2(A)$} 
        \State $A \gets (A \cup T_C) \setminus N(T_C)$
    \EndWhile
    \State \Return $A$
  \end{algorithmic}
  \caption{\SquareImp($G$, $w$)}
  \label{alg:11}
\end{algorithm}

Algorithm \SquareImp\ has a better approximation ratio than its predecessor 
that uses improvements based on the sum of the weights~\cite{ChandraH2001}, 
instead of the sum of the squared weights.

We observe that, since the weights are positive and a singleton is a 1-claw by
definition, the output of \SquareImp\ is always a maximal independent set.
Recall also that we use claw to refer to a $d$-claw for an arbitrary~$d$.

\section{Relation between wMIS and Maximum Leaf Spanning Arborescence}
\label{sec:maxleavesMIS}

We have described in~\cite{FernandesL2021} an algorithm for \maxleaves\ on
rooted dags called \Call{Maxleaves-W3DM}{}.
It uses as a black box an approximation for maximum weight 3D-matching (hence
the W3DM acronym). 
Specifically, it builds, in one of its steps, an instance of the maximum weight
3D-matching, applies an approximation for maximum weight 3D-matching to this
instance, and uses its output to extend the solution being built for
\maxleaves.

The best approximation for maximum weight 3D-matching comes from a reduction to
wMIS on weighted~4-claw free graphs. 
Indeed, given an instance of maximum weight 3D-matching, which consists of a
collection~$\calS$ of 3-sets, each with a positive weight, one can build an
instance of wMIS using the intersection graph for the collection~$\calS$. 
Recall that the intersection graph of a collection of sets consists of the
graph with one vertex for each set in the collection, and two vertices are
adjacent if the corresponding sets intersect.  
For a collection~$\calS$ of 3-sets, the intersection graph is 4-claw free,
because any set in~$\calS$ can intersect at most three disjoint sets.  
Also, in this graph, an independent set corresponds to a 3D-matching
in~$\calS$. 

The idea we will explore is to build directly an instance of wMIS and to apply
an approximation for wMIS on this instance.
An adapted version of a result from~\cite{FernandesL2021} allows us to deduce
that the modified version of \Call{Maxleaves-W3DM}{} that uses an approximation
for wMIS instead of an approximation for maximum weight 3D-matching preserves
the ratio from the approximation for wMIS, up to~$4/3$.

We start by translating the construction in \Call{Maxleaves-W3DM}{} to wMIS
instances.
We will prove that, on these instances, \SquareImp\ runs in polynomial time and
achieves a ratio of~$3/2$.  
Then we will deduce that the modified version of \Call{Maxleaves-W3DM}{} that
uses \SquareImp, instead of an approximation for maximum weight 3D-matching,
achieves a ratio of~$3/2$ for \maxleaves\ on rooted dags. 
This ratio matches the best known for rooted dags, given by another algorithm
presented in~\cite{FernandesL2021}. 

The instances of wMIS in the translated construction use only weights~1 and~2,
hence we call the resulting algorithm \Call{Maxleaves-12MIS}{}.
To describe the algorithm in details, we reproduce some definitions and figures
from~\cite{FernandesL2021}.
 
\begin{figure}[htb]
    \centering
    \resizebox{0.48\textwidth}{!}{\begin{tikzpicture}
    [vertex/.style={circle,draw,fill,inner sep=0pt,minimum size=5pt}]

    \clip (0,-0.3) rectangle (9.5,4.3);

    \node[vertex,label={[xshift=7pt,yshift=-7pt]$a$}] (a) at (1.4,4) {};
    \node[vertex,label={[xshift=7pt,yshift=-7pt]$b$}] (b) at (4.9,4) {};
    \node[vertex,label={[xshift=-7pt,yshift=-7pt]$c$}] (c) at (8.4,4) {};
    \node[vertex,label={[xshift=-7pt,yshift=-9pt]$d$}] (d) at (0.7,3) {};
    \node[vertex,label={[xshift=-7pt,yshift=-9pt]$e$}] (e) at (1.4,3) {};
    \node[vertex,label={[xshift=7pt,yshift=-13pt]$f$}] (f) at (2.1,3) {};
    \node[vertex,label={[xshift=-7pt,yshift=-11pt]$g$}] (g) at (4.2,3) {};
    \node[vertex,label={[xshift=-7pt,yshift=-9pt]$a'$}] (ap) at (4.9,3) {};
    \node[vertex,label={[xshift=-7pt,yshift=-9pt]$b'$}] (bp) at (5.6,3) {};
    \node[vertex,label={[xshift=-7pt,yshift=-5pt]$h$}] (h) at (7.7,3) {};
    \node[vertex,label={[xshift=-7pt,yshift=-9pt]$i$}] (i) at (8.4,3) {};
    \node[vertex,label={[xshift=-7pt,yshift=-9pt]$j$}] (j) at (9.1,3) {};
    \node[vertex,label={[xshift=7pt,yshift=-10pt]$c'$}] (cp) at (0.1,2) {};
    \node[vertex,label={[xshift=-7pt,yshift=-9pt]$k$}] (k) at (1.4,2) {};
    \node[vertex,label={[xshift=7pt,yshift=-14pt]$l$}] (l) at (3.5,2) {};
    \node[vertex,label={[xshift=-7pt,yshift=-9pt]$m$}] (m) at (4.9,2) {};
    \node[vertex,label={[xshift=7pt,yshift=-10pt]$n$}] (n) at (7.7,2) {};
    \node[vertex,label={[xshift=-7pt,yshift=-9pt]$o$}] (o) at (0.7,1) {};
    \node[vertex,label={[xshift=-7pt,yshift=-11pt]$p$}] (p) at (1.4,1) {};
    \node[vertex,label={[xshift=-7pt,yshift=-11pt]$q$}] (q) at (2.1,1) {};
    \node[vertex,label={[xshift=-7pt,yshift=-9pt]$r$}] (r) at (4.2,1) {};
    \node[vertex,label={[xshift=7pt,yshift=-9pt]$s$}] (s) at (4.9,1) {};
    \node[vertex,label={[xshift=7pt,yshift=-9pt]$t$}] (t) at (5.6,1) {};
    \node[vertex,label={[xshift=7pt,yshift=-9pt]$u$}] (u) at (7,1) {};
    \node[vertex,label={[xshift=7pt,yshift=-9pt]$v$}] (v) at (7.7,1) {};
    \node[vertex,label={[xshift=7pt,yshift=-9pt]$w$}] (w) at (8.4,1) {};
    \node[vertex,label={[xshift=-7pt,yshift=-9pt]$d'$}] (dp) at (1,0) {};
    \node[vertex,label={[xshift=7pt,yshift=-9pt]$x$}] (x) at (2,0) {};
    \node[vertex,label={[xshift=-7pt,yshift=-10pt]$e'$}] (ep) at (3.5,0) {};
    \node[vertex,label={[xshift=7pt,yshift=-10pt]$y$}] (y) at (4.5,0) {};
    \node[vertex,label={[xshift=-7pt,yshift=-11pt]$f'$}] (fp) at (6,0) {};
    \node[vertex,label={[xshift=-7pt,yshift=-9pt]$z$}] (z) at (7.2,0) {};
    \node[vertex,label={[xshift=7pt,yshift=-13pt]$g'$}] (gp) at (8,0) {};

    \draw[-latex,gray] (a) -- (d);
    \draw[-latex,gray] (a) -- (e);
    \draw[-latex,gray] (a) -- (f);
    \draw[-latex,line width=2pt] (b) -- (g);
    \draw[-latex,line width=2pt] (b) -- (ap);
    \draw[-latex,line width=2pt] (b) -- (bp);
    \draw[-latex,gray] (b) -- (k);
    \draw[-latex,gray] (c) -- (bp);
    \draw[-latex,line width=2pt] (c) -- (h);
    \draw[-latex,line width=2pt] (c) -- (i);
    \draw[-latex,line width=2pt] (c) -- (j);
    \draw[-latex,line width=2pt] (d) -- (cp);
    \draw[-latex,line width=2pt] (d) -- (k);
    \draw[-latex,gray] (e) -- (cp);
    \draw[-latex,gray] (e) -- (ep);
    \draw[-latex,line width=2pt] (f) -- (b);
    \draw[-latex,line width=2pt] (f) -- (e);
    \draw[-latex,gray] (g) -- (q);
    \draw[-latex,line width=2pt] (g) -- (l);
    \draw[-latex,line width=2pt] (g) -- (m);
    \draw[-latex,gray] (ap) -- (z);
    \draw[-latex,gray] (h) -- (m);
    \draw[-latex,gray] (h) -- (l);
    \draw[-latex,gray] (h) -- (n);
    \draw[-latex,gray] (cp) -- (o);
    \draw[-latex,line width=2pt] (k) -- (o);
    \draw[-latex,line width=2pt] (k) -- (p);
    \draw[-latex,line width=2pt] (k) -- (q);
    \draw[-latex,gray] (l) -- (ep);
    \draw[-latex,gray] (m) -- (r);
    \draw[-latex,line width=2pt] (m) -- (s);
    \draw[-latex,line width=2pt] (m) -- (t);
    \draw[-latex,line width=2pt] (n) -- (u);
    \draw[-latex,line width=2pt] (n) -- (v);
    \draw[-latex,line width=2pt] (n) -- (w);
    \draw[-latex,gray] (p) -- (x);
    \draw[-latex,line width=2pt] (s) -- (y);
    \draw[-latex,line width=2pt] (s) -- (r);
    \draw[-latex,gray] (u) -- (fp);
    \draw[-latex,gray] (w) -- (j);
    \draw[-latex,gray] (dp) -- (o);
    \draw[-latex,gray] (x) -- (dp);
    \draw[-latex,gray] (x) -- (l);
    \draw[-latex,gray] (y) -- (ep);
    \draw[-latex,gray] (fp) -- (t);
    \draw[-latex,gray] (z) -- (v);
    \draw[-latex,gray] (z) -- (gp);
    \draw[-latex,gray] (gp) .. controls +(east:15mm) and +(east:15mm) .. (c);

\end{tikzpicture}}
    \hfill
    \resizebox{0.48\textwidth}{!}{\begin{tikzpicture}
    [vertex/.style={circle,draw,fill,inner sep=0pt,minimum size=5pt}]

    \clip (0,-0.3) rectangle (9.5,4.3);

    \node[vertex,label={[xshift=7pt,yshift=-7pt]$a$}] (a) at (1.4,4) {};
    \node[vertex,label={[xshift=7pt,yshift=-7pt]$b$}] (b) at (4.9,4) {};
    \node[vertex,label={[xshift=-7pt,yshift=-7pt]$c$}] (c) at (8.4,4) {};
    \node[vertex,label={[xshift=-7pt,yshift=-9pt]$d$}] (d) at (0.7,3) {};
    \node[vertex,label={[xshift=-7pt,yshift=-9pt]$e$}] (e) at (1.4,3) {};
    \node[vertex,label={[xshift=7pt,yshift=-13pt]$f$}] (f) at (2.1,3) {};
    \node[vertex,label={[xshift=-7pt,yshift=-11pt]$g$}] (g) at (4.2,3) {};
    \node[vertex,label={[xshift=-7pt,yshift=-9pt]$a'$}] (ap) at (4.9,3) {};
    \node[vertex,label={[xshift=-7pt,yshift=-9pt]$b'$}] (bp) at (5.6,3) {};
    \node[vertex,label={[xshift=-7pt,yshift=-5pt]$h$}] (h) at (7.7,3) {};
    \node[vertex,label={[xshift=-7pt,yshift=-9pt]$i$}] (i) at (8.4,3) {};
    \node[vertex,label={[xshift=-7pt,yshift=-9pt]$j$}] (j) at (9.1,3) {};
    \node[vertex,label={[xshift=7pt,yshift=-10pt]$c'$}] (cp) at (0.1,2) {};
    \node[vertex,label={[xshift=-7pt,yshift=-9pt]$k$}] (k) at (1.4,2) {};
    \node[vertex,label={[xshift=7pt,yshift=-14pt]$l$}] (l) at (3.5,2) {};
    \node[vertex,label={[xshift=-7pt,yshift=-9pt]$m$}] (m) at (4.9,2) {};
    \node[vertex,label={[xshift=7pt,yshift=-10pt]$n$}] (n) at (7.7,2) {};
    \node[vertex,label={[xshift=-7pt,yshift=-9pt]$o$}] (o) at (0.7,1) {};
    \node[vertex,label={[xshift=-7pt,yshift=-11pt]$p$}] (p) at (1.4,1) {};
    \node[vertex,label={[xshift=-7pt,yshift=-11pt]$q$}] (q) at (2.1,1) {};
    \node[vertex,label={[xshift=-7pt,yshift=-9pt]$r$}] (r) at (4.2,1) {};
    \node[vertex,label={[xshift=7pt,yshift=-9pt]$s$}] (s) at (4.9,1) {};
    \node[vertex,label={[xshift=7pt,yshift=-9pt]$t$}] (t) at (5.6,1) {};
    \node[vertex,label={[xshift=7pt,yshift=-9pt]$u$}] (u) at (7,1) {};
    \node[vertex,label={[xshift=7pt,yshift=-9pt]$v$}] (v) at (7.7,1) {};
    \node[vertex,label={[xshift=7pt,yshift=-9pt]$w$}] (w) at (8.4,1) {};
    \node[vertex,label={[xshift=-7pt,yshift=-9pt]$d'$}] (dp) at (1,0) {};
    \node[vertex,label={[xshift=7pt,yshift=-9pt]$x$}] (x) at (2,0) {};
    \node[vertex,label={[xshift=-7pt,yshift=-10pt]$e'$}] (ep) at (3.5,0) {};
    \node[vertex,label={[xshift=7pt,yshift=-10pt]$y$}] (y) at (4.5,0) {};
    \node[vertex,label={[xshift=-7pt,yshift=-11pt]$f'$}] (fp) at (6,0) {};
    \node[vertex,label={[xshift=-7pt,yshift=-9pt]$z$}] (z) at (7.2,0) {};
    \node[vertex,label={[xshift=7pt,yshift=-13pt]$g'$}] (gp) at (8,0) {};

    \draw[-latex,line width=2pt] (a) -- (d);
    \draw[-latex,line width=2pt] (a) -- (e);
    \draw[-latex,line width=2pt] (a) -- (f);
    \draw[-latex,line width=2pt] (b) -- (g);
    \draw[-latex,line width=2pt] (b) -- (ap);
    \draw[-latex,line width=2pt] (b) -- (bp);
    \draw[-latex,line width=2pt] (b) -- (k);
    \draw[-latex,gray] (c) -- (bp);
    \draw[-latex,line width=2pt] (c) -- (h);
    \draw[-latex,line width=2pt] (c) -- (i);
    \draw[-latex,line width=2pt] (c) -- (j);
    \draw[-latex,gray] (d) -- (cp);
    \draw[-latex,gray] (d) -- (k);
    \draw[-latex,gray] (e) -- (cp);
    \draw[-latex,gray] (e) -- (ep);
    \draw[-latex,gray] (f) -- (b);
    \draw[-latex,gray] (f) -- (e);
    \draw[-latex,gray] (g) -- (q);
    \draw[-latex,gray] (g) -- (l);
    \draw[-latex,gray] (g) -- (m);
    \draw[-latex,gray] (ap) -- (z);
    \draw[-latex,line width=2pt] (h) -- (m);
    \draw[-latex,line width=2pt] (h) -- (l);
    \draw[-latex,line width=2pt] (h) -- (n);
    \draw[-latex,gray] (cp) -- (o);
    \draw[-latex,line width=2pt] (k) -- (o);
    \draw[-latex,line width=2pt] (k) -- (p);
    \draw[-latex,line width=2pt] (k) -- (q);
    \draw[-latex,gray] (l) -- (ep);
    \draw[-latex,line width=2pt] (m) -- (r);
    \draw[-latex,line width=2pt] (m) -- (s);
    \draw[-latex,line width=2pt] (m) -- (t);
    \draw[-latex,line width=2pt] (n) -- (u);
    \draw[-latex,line width=2pt] (n) -- (v);
    \draw[-latex,line width=2pt] (n) -- (w);
    \draw[-latex,gray] (p) -- (x);
    \draw[-latex,gray] (s) -- (y);
    \draw[-latex,gray] (s) -- (r);
    \draw[-latex,gray] (u) -- (fp);
    \draw[-latex,gray] (w) -- (j);
    \draw[-latex,gray] (dp) -- (o);
    \draw[-latex,gray] (x) -- (dp);
    \draw[-latex,gray] (x) -- (l);
    \draw[-latex,gray] (y) -- (ep);
    \draw[-latex,gray] (fp) -- (t);
    \draw[-latex,gray] (z) -- (v);
    \draw[-latex,gray] (z) -- (gp);
    \draw[-latex,gray] (gp) .. controls +(east:15mm) and +(east:15mm) .. (c);

\end{tikzpicture}}
    \caption{The bold arcs show a 2-branching and a 3-branching in a same rooted dag.}
    \label{fig:branching}
\end{figure}
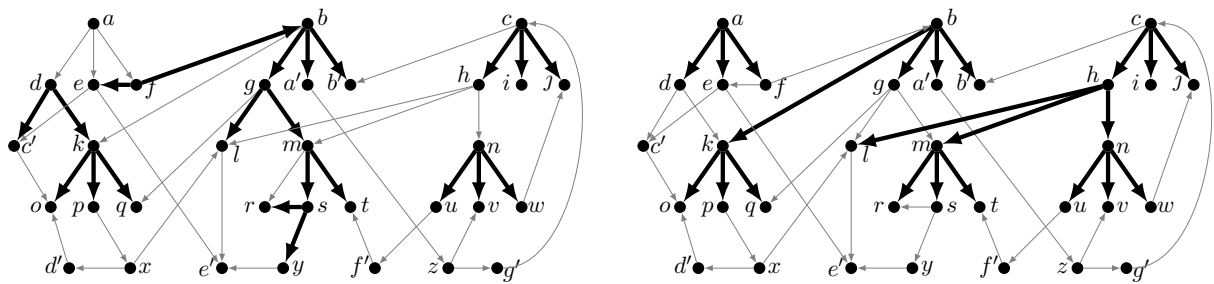

A \emph{branching} in a directed graph is a collection of disjoint
arborescences.
Note that the underlying graph of a branching is a forest.
A node that is not a leaf in a branching is called \emph{internal}. 
For a positive integer~$t$, a~\emph{$t$-branching} is a branching for which
every internal node has out-degree at least~$t$.
See Figure~\ref{fig:branching}.

Algorithm~\ref{alg:greedyexpand} describes a greedy procedure, which we call
\Call{GreedyExpand}{}, that is used in our new algorithm for \maxleaves. 
Each set~$A_v$ in Line~\ref{line:Uv} represents what we call an
\emph{expansion} that can be applied to the arborescence that is being built. 
Formally, an expansion is a set of vertices with in-degree zero in the current
arborescence and that are out-neighbors of a vertex $v$ that has out-degree
zero in the current arborescence.
If $|A_v|=t$, then we call it a~\emph{$t$-expansion}. 
We refer to a~$1$-expansion as a trivial expansion. 

\begin{algorithm}
\begin{algorithmic}[1]
\Require{rooted dag $D$, a positive integer $t$, and a spanning $(t{+}1)$-branching~$F$ of~$D$}
\Ensure{a maximal spanning $t$-branching of $D$ containing~$F$}

\State $F' \gets F$
\For{each $v \in V(D)$ such that $d_{F'}^+(v)=0$}
    \State $A_v \gets \{vu \in A(D) : d_{F'}^-(u)=0\}$ \label{line:Uv}
    \If{$|A_v| \geq t$} 
        \State $F' \gets F' + A_v$
    \EndIf
\EndFor
\State \Return $F'$
\end{algorithmic}
\caption{\textsc{GreedyExpand}($D$, $t$, $F$)}
\label{alg:greedyexpand}
\end{algorithm}

Then algorithm \Call{Maxleaves-12MIS}{} is presented in
Algorithm~\ref{alg:maxleaves-12mis}, and it uses a procedure
\Call{IntersectionGraph}{$V$, $U$} that receives a set~$V$ and a set~$U_v$ for
each $v$ in~$V$, and returns the intersection graph for the collection~$U$ of
sets.
That is, the graph~$G$ whose vertex set is~$V$ and two vertices~$x$ and~$y$ are
adjacent if and only if $U_x \cap U_y \neq \emptyset$.
For the analysis, we will consider~$G$ as the multigraph having $|U_x \cap
U_y|$ parallel edges between~$x$ and~$y$.

\begin{algorithm}
\begin{algorithmic}[1]
\Require{rooted acyclic directed graph $D$}
\Ensure{spanning arborescence of $D$}

\State let $F_0$ be the spanning branching with no arcs
\State $F_1 \gets$ \Call{GreedyExpand}{$D$, $4$, $F_0$}\label{line:F1}
\For{each $v \in V(D)$ such that $d_{F_1}^+(v)=0$} 
    \State $U_v \gets \{u \in V(D): d_{F_1}^-(u)=0 \mbox{ and } vu \in A(D)\}$
\EndFor
\State $\Candidates \gets \{v \in V(D) : d_{F_1}^+(v)=0 \mbox{ and } 2 \leq |U_v| \leq 3\}$\label{line:cand1}
\For{each $v \in \Candidates$ such that $|U_v| = 3$}
   \For{each $u \in U_v$} 
   \State add to $\Candidates$ an element $v_u$ 
   \State $U_{v_u} \gets U_v \setminus \{u\}$
   \EndFor
\EndFor
\Statex \LeftComment{$\Candidates$ correspond to all non-trivial expansions that can be applied to $F_1$.}
\State $G \gets$ \Call{IntersectionGraph}{$\Candidates$, $U$} \label{line9}
\For{each $v \in \Candidates$} 
    \State $w_v \gets |U_v| - 1$ \label{line11}
\EndFor
\State $I \gets$ \Call{SquareImp}{$G$, $w$} \label{alg:call_squareimp}
\State $F_2 \gets F_1$
\For{each $v \in I$} \label{line:forF2} 
    \State $F_2 \gets F_2 + \{vu : u \in U_v\}$
\EndFor\label{line:EndForF2}
\State $T \gets$ \Call{GreedyExpand}{$D$, $1$, $F_2$}\label{line:T}
\State \Return $T$
\end{algorithmic}
\caption{\textsc{MaxLeaves-12MIS}($D$)}
\label{alg:maxleaves-12mis}
\end{algorithm}

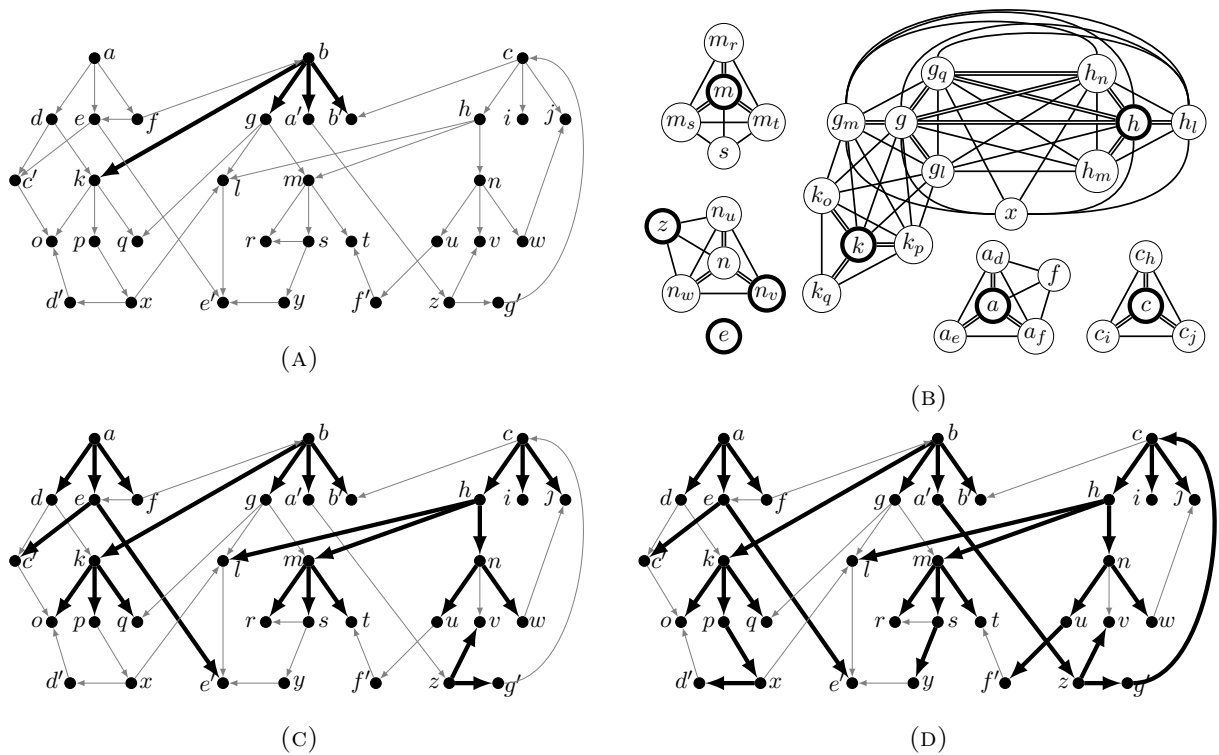
\begin{figure}[htb]
    \centering
    \begin{subfigure}{0.48\textwidth}
        \resizebox{\textwidth}{!}{\begin{tikzpicture}
    [vertex/.style={circle,draw,fill,inner sep=0pt,minimum size=5pt}]

    \clip (0,-0.3) rectangle (9.5,4.3);

    \node[vertex,label={[xshift=7pt,yshift=-7pt]$a$}] (a) at (1.4,4) {};
    \node[vertex,label={[xshift=7pt,yshift=-7pt]$b$}] (b) at (4.9,4) {};
    \node[vertex,label={[xshift=-7pt,yshift=-7pt]$c$}] (c) at (8.4,4) {};
    \node[vertex,label={[xshift=-7pt,yshift=-9pt]$d$}] (d) at (0.7,3) {};
    \node[vertex,label={[xshift=-7pt,yshift=-9pt]$e$}] (e) at (1.4,3) {};
    \node[vertex,label={[xshift=7pt,yshift=-13pt]$f$}] (f) at (2.1,3) {};
    \node[vertex,label={[xshift=-7pt,yshift=-11pt]$g$}] (g) at (4.2,3) {};
    \node[vertex,label={[xshift=-7pt,yshift=-9pt]$a'$}] (ap) at (4.9,3) {};
    \node[vertex,label={[xshift=-7pt,yshift=-9pt]$b'$}] (bp) at (5.6,3) {};
    \node[vertex,label={[xshift=-7pt,yshift=-5pt]$h$}] (h) at (7.7,3) {};
    \node[vertex,label={[xshift=-7pt,yshift=-9pt]$i$}] (i) at (8.4,3) {};
    \node[vertex,label={[xshift=-7pt,yshift=-9pt]$j$}] (j) at (9.1,3) {};
    \node[vertex,label={[xshift=7pt,yshift=-10pt]$c'$}] (cp) at (0.1,2) {};
    \node[vertex,label={[xshift=-7pt,yshift=-9pt]$k$}] (k) at (1.4,2) {};
    \node[vertex,label={[xshift=7pt,yshift=-14pt]$l$}] (l) at (3.5,2) {};
    \node[vertex,label={[xshift=-7pt,yshift=-9pt]$m$}] (m) at (4.9,2) {};
    \node[vertex,label={[xshift=7pt,yshift=-10pt]$n$}] (n) at (7.7,2) {};
    \node[vertex,label={[xshift=-7pt,yshift=-9pt]$o$}] (o) at (0.7,1) {};
    \node[vertex,label={[xshift=-7pt,yshift=-11pt]$p$}] (p) at (1.4,1) {};
    \node[vertex,label={[xshift=-7pt,yshift=-11pt]$q$}] (q) at (2.1,1) {};
    \node[vertex,label={[xshift=-7pt,yshift=-9pt]$r$}] (r) at (4.2,1) {};
    \node[vertex,label={[xshift=7pt,yshift=-9pt]$s$}] (s) at (4.9,1) {};
    \node[vertex,label={[xshift=7pt,yshift=-9pt]$t$}] (t) at (5.6,1) {};
    \node[vertex,label={[xshift=7pt,yshift=-9pt]$u$}] (u) at (7,1) {};
    \node[vertex,label={[xshift=7pt,yshift=-9pt]$v$}] (v) at (7.7,1) {};
    \node[vertex,label={[xshift=7pt,yshift=-9pt]$w$}] (w) at (8.4,1) {};
    \node[vertex,label={[xshift=-7pt,yshift=-9pt]$d'$}] (dp) at (1,0) {};
    \node[vertex,label={[xshift=7pt,yshift=-9pt]$x$}] (x) at (2,0) {};
    \node[vertex,label={[xshift=-7pt,yshift=-10pt]$e'$}] (ep) at (3.5,0) {};
    \node[vertex,label={[xshift=7pt,yshift=-10pt]$y$}] (y) at (4.5,0) {};
    \node[vertex,label={[xshift=-7pt,yshift=-11pt]$f'$}] (fp) at (6,0) {};
    \node[vertex,label={[xshift=-7pt,yshift=-9pt]$z$}] (z) at (7.2,0) {};
    \node[vertex,label={[xshift=7pt,yshift=-13pt]$g'$}] (gp) at (8,0) {};

    \draw[-latex,gray] (a) -- (d);
    \draw[-latex,gray] (a) -- (e);
    \draw[-latex,gray] (a) -- (f);
    \draw[-latex,line width=2pt] (b) -- (g);
    \draw[-latex,line width=2pt] (b) -- (ap);
    \draw[-latex,line width=2pt] (b) -- (bp);
    \draw[-latex,line width=2pt] (b) -- (k);
    \draw[-latex,gray] (c) -- (bp);
    \draw[-latex,gray] (c) -- (h);
    \draw[-latex,gray] (c) -- (i);
    \draw[-latex,gray] (c) -- (j);
    \draw[-latex,gray] (d) -- (cp);
    \draw[-latex,gray] (d) -- (k);
    \draw[-latex,gray] (e) -- (cp);
    \draw[-latex,gray] (e) -- (ep);
    \draw[-latex,gray] (f) -- (b);
    \draw[-latex,gray] (f) -- (e);
    \draw[-latex,gray] (g) -- (q);
    \draw[-latex,gray] (g) -- (l);
    \draw[-latex,gray] (g) -- (m);
    \draw[-latex,gray] (ap) -- (z);
    \draw[-latex,gray] (h) -- (m);
    \draw[-latex,gray] (h) -- (l);
    \draw[-latex,gray] (h) -- (n);
    \draw[-latex,gray] (cp) -- (o);
    \draw[-latex,gray] (k) -- (o);
    \draw[-latex,gray] (k) -- (p);
    \draw[-latex,gray] (k) -- (q);
    \draw[-latex,gray] (l) -- (ep);
    \draw[-latex,gray] (m) -- (r);
    \draw[-latex,gray] (m) -- (s);
    \draw[-latex,gray] (m) -- (t);
    \draw[-latex,gray] (n) -- (u);
    \draw[-latex,gray] (n) -- (v);
    \draw[-latex,gray] (n) -- (w);
    \draw[-latex,gray] (p) -- (x);
    \draw[-latex,gray] (s) -- (y);
    \draw[-latex,gray] (s) -- (r);
    \draw[-latex,gray] (u) -- (fp);
    \draw[-latex,gray] (w) -- (j);
    \draw[-latex,gray] (dp) -- (o);
    \draw[-latex,gray] (x) -- (dp);
    \draw[-latex,gray] (x) -- (l);
    \draw[-latex,gray] (y) -- (ep);
    \draw[-latex,gray] (fp) -- (t);
    \draw[-latex,gray] (z) -- (v);
    \draw[-latex,gray] (z) -- (gp);
    \draw[-latex,gray] (gp) .. controls +(east:15mm) and +(east:15mm) .. (c);

\end{tikzpicture}}
        \caption{}
        \label{figa:f1}
    \end{subfigure}
    \hfill
    \begin{subfigure}{0.48\textwidth}
        \resizebox{\textwidth}{!}{\begin{tikzpicture}
    [vertex/.style={circle,draw,inner sep=1pt,minimum size=5pt,minimum width=15pt}]

    \clip (0.7,-2.4) rectangle (10.2,3.5);

    \node[vertex,line width=2pt] (e) at (2.1,-2) { $e$};

    \node[vertex,line width=2pt] (a) at (6.5,-1.5) {$a$};
    \node[vertex] (a1) at ($(a)+(0,+0.8)$) {$a_d$};
    \node[vertex] (a2) at ($(a)+(-0.7,-0.5)$) {$a_e$};
    \node[vertex] (a3) at ($(a)+(+0.7,-0.5)$) {$a_f$};
    \draw[double,thick] (a) -- (a1);
    \draw[double,thick] (a) -- (a2);
    \draw[double,thick] (a) -- (a3);
    \draw[thick] (a1) -- (a2);
    \draw[thick] (a1) -- (a3);
    \draw[thick] (a2) -- (a3);

    \node[vertex] (f) at (7.5,-1) {$f$};

    \draw[thick] (f) -- (a);
    \draw[thick] (f) -- (a1);
    \draw[thick] (f) -- (a3);

    \node[vertex,line width=2pt] (c) at (9,-1.5) {$c$};
    \node[vertex] (c1) at ($(c)+(0,+0.8)$) {$c_h$};
    \node[vertex] (c2) at ($(c)+(-0.7,-0.5)$) {$c_i$};
    \node[vertex] (c3) at ($(c)+(+0.7,-0.5)$) {$c_j$};
    \draw[double,thick] (c) -- (c1);
    \draw[double,thick] (c) -- (c2);
    \draw[double,thick] (c) -- (c3);
    \draw[thick] (c1) -- (c2);
    \draw[thick] (c1) -- (c3);
    \draw[thick] (c2) -- (c3);

    \node[vertex,line width=2pt] (m) at (2.1,2) {$m$};
    \node[vertex] (m1) at ($(m)+(0,+0.8)$) {$m_r$};
    \node[vertex] (m2) at ($(m)+(-0.7,-0.5)$) {$m_s$};
    \node[vertex] (m3) at ($(m)+(+0.7,-0.5)$) {$m_t$};
    \draw[double,thick] (m) -- (m1);
    \draw[double,thick] (m) -- (m2);
    \draw[double,thick] (m) -- (m3);
    \draw[thick] (m1) -- (m2);
    \draw[thick] (m1) -- (m3);
    \draw[thick] (m2) -- (m3);

    \node[vertex] (s) at (2.1,1) {$s$};

    \draw[thick] (s) -- (m);
    \draw[thick] (s) -- (m2);
    \draw[thick] (s) -- (m3);

    \node[vertex] (n) at (2.1,-0.8) {$n$};
    \node[vertex] (n1) at ($(n)+(0,+0.8)$) {$n_u$};
    \node[vertex] (n2) at ($(n)+(-0.7,-0.5)$) {$n_w$};
    \node[vertex,line width=2pt] (n3) at ($(n)+(+0.7,-0.5)$) {$n_v$};
    \draw[double,thick] (n) -- (n1);
    \draw[double,thick] (n) -- (n2);
    \draw[double,thick] (n) -- (n3);
    \draw[thick] (n1) -- (n2);
    \draw[thick] (n1) -- (n3);
    \draw[thick] (n2) -- (n3);

    \node[vertex,line width=2pt] (z) at (1.1,-0.2) {$z$};

    \draw[thick] (z) -- (n);
    \draw[thick] (z) -- (n1);
    \draw[thick] (z) -- (n2);

    \node[vertex] (g) at (5.0,1.5) {$g$};
    \node[vertex] (g1) at ($(g)+(+0.6,+0.8)$) {$g_q$};
    \node[vertex] (g2) at ($(g)+(-0.9,0.0)$) {$g_m$};
    \node[vertex] (g3) at ($(g)+(+0.6,-0.8)$) {$g_l$};
    \draw[double,thick] (g) -- (g1);
    \draw[double,thick] (g) -- (g2);
    \draw[double,thick] (g) -- (g3);
    \draw[thick] (g1) -- (g2);
    \draw[thick] (g1) -- (g3);
    \draw[thick] (g2) -- (g3);

    \node[vertex] (x) at (6.8,0) {$x$};

    \draw[thick] (x) .. controls +(west:10mm) and +(south:13mm) .. (g);
    \draw[thick] (x) -- (g1);
    \draw[thick] (x) .. controls +(west:10mm) and +(south:13mm) .. (g2);

    \node[vertex,line width=2pt] (h) at (8.8,1.5) {$h$};
    \node[vertex] (h1) at ($(h)+(-0.6,+0.8)$) {$h_n$};
    \node[vertex] (h2) at ($(h)+(+0.9,0)$) {$h_l$};
    \node[vertex] (h3) at ($(h)+(-0.6,-0.8)$) {$h_m$};
    \draw[double,thick] (h) -- (h1);
    \draw[double,thick] (h) -- (h2);
    \draw[double,thick] (h) -- (h3);
    \draw[thick] (h1) -- (h2);
    \draw[thick] (h1) -- (h3);
    \draw[thick] (h2) -- (h3);

    \draw[thick] (x) .. controls +(east:10mm) and +(south:15mm) .. (h);
    \draw[thick] (x) -- (h1);
    \draw[thick] (x) .. controls +(east:10mm) and +(south:15mm) .. (h2);

    \draw[double,thick] (g1) -- (h1);
    \draw[double,thick] (g1) -- (h);
    \draw[thick] (g1) -- (h3);
    \draw[double,thick] (g) -- (h1);
    \draw[double,thick] (g) -- (h);
    \draw[thick] (g) -- (h3);
    \draw[thick] (g3) -- (h1);
    \draw[thick] (g3) -- (h);
    \draw[thick] (g3) -- (h3);
    \draw[thick] (g2) .. controls +(north:21mm) and +(north:10mm) .. (h1);
    \draw[thick] (g2) .. controls +(north:21mm) and +(north:25mm) .. (h);
    \draw[thick] (g2) .. controls +(north:25mm) and +(north:22mm) .. (h2);
    \draw[thick] (g) .. controls +(north:23mm) and +(north:18mm) .. (h2);
    \draw[thick] (g1) .. controls +(north:8mm) and +(north:18mm) .. (h2);

    \node[vertex,line width=2pt] (k) at (4.3,-0.5) {$k$};
    \node[vertex] (k1) at ($(k)+(-0.6,-0.8)$) {$k_q$};
    \node[vertex] (k2) at ($(k)+(-0.6,+0.8)$) {$k_o$};
    \node[vertex] (k3) at ($(k)+(+0.9,0)$) {$k_p$};
    \draw[double,thick] (k) -- (k1);
    \draw[double,thick] (k) -- (k2);
    \draw[double,thick] (k) -- (k3);
    \draw[thick] (k1) -- (k2);
    \draw[thick] (k1) -- (k3);
    \draw[thick] (k2) -- (k3);

    \draw[thick] (g2) -- (k2);
    \draw[thick] (g2) -- (k);
    \draw[thick] (g2) -- (k3);
    \draw[thick] (g) -- (k2);
    \draw[thick] (g) -- (k);
    \draw[thick] (g) -- (k3);
    \draw[thick] (g3) -- (k2);
    \draw[thick] (g3) -- (k);
    \draw[thick] (g3) -- (k3);


\end{tikzpicture}}
        \caption{} 
        \label{figb:intersection}
    \end{subfigure}

    \begin{subfigure}{0.48\textwidth}
        \resizebox{\textwidth}{!}{\begin{tikzpicture}
    [vertex/.style={circle,draw,fill,inner sep=0pt,minimum size=5pt}]

    \clip (0,-0.3) rectangle (9.5,4.3);

    \node[vertex,label={[xshift=7pt,yshift=-7pt]$a$}] (a) at (1.4,4) {};
    \node[vertex,label={[xshift=7pt,yshift=-7pt]$b$}] (b) at (4.9,4) {};
    \node[vertex,label={[xshift=-7pt,yshift=-7pt]$c$}] (c) at (8.4,4) {};
    \node[vertex,label={[xshift=-7pt,yshift=-9pt]$d$}] (d) at (0.7,3) {};
    \node[vertex,label={[xshift=-7pt,yshift=-9pt]$e$}] (e) at (1.4,3) {};
    \node[vertex,label={[xshift=7pt,yshift=-13pt]$f$}] (f) at (2.1,3) {};
    \node[vertex,label={[xshift=-7pt,yshift=-11pt]$g$}] (g) at (4.2,3) {};
    \node[vertex,label={[xshift=-7pt,yshift=-9pt]$a'$}] (ap) at (4.9,3) {};
    \node[vertex,label={[xshift=-7pt,yshift=-9pt]$b'$}] (bp) at (5.6,3) {};
    \node[vertex,label={[xshift=-7pt,yshift=-5pt]$h$}] (h) at (7.7,3) {};
    \node[vertex,label={[xshift=-7pt,yshift=-9pt]$i$}] (i) at (8.4,3) {};
    \node[vertex,label={[xshift=-7pt,yshift=-9pt]$j$}] (j) at (9.1,3) {};
    \node[vertex,label={[xshift=7pt,yshift=-10pt]$c'$}] (cp) at (0.1,2) {};
    \node[vertex,label={[xshift=-7pt,yshift=-9pt]$k$}] (k) at (1.4,2) {};
    \node[vertex,label={[xshift=7pt,yshift=-14pt]$l$}] (l) at (3.5,2) {};
    \node[vertex,label={[xshift=-7pt,yshift=-9pt]$m$}] (m) at (4.9,2) {};
    \node[vertex,label={[xshift=7pt,yshift=-10pt]$n$}] (n) at (7.7,2) {};
    \node[vertex,label={[xshift=-7pt,yshift=-9pt]$o$}] (o) at (0.7,1) {};
    \node[vertex,label={[xshift=-7pt,yshift=-11pt]$p$}] (p) at (1.4,1) {};
    \node[vertex,label={[xshift=-7pt,yshift=-11pt]$q$}] (q) at (2.1,1) {};
    \node[vertex,label={[xshift=-7pt,yshift=-9pt]$r$}] (r) at (4.2,1) {};
    \node[vertex,label={[xshift=7pt,yshift=-9pt]$s$}] (s) at (4.9,1) {};
    \node[vertex,label={[xshift=7pt,yshift=-9pt]$t$}] (t) at (5.6,1) {};
    \node[vertex,label={[xshift=7pt,yshift=-9pt]$u$}] (u) at (7,1) {};
    \node[vertex,label={[xshift=7pt,yshift=-9pt]$v$}] (v) at (7.7,1) {};
    \node[vertex,label={[xshift=7pt,yshift=-9pt]$w$}] (w) at (8.4,1) {};
    \node[vertex,label={[xshift=-7pt,yshift=-9pt]$d'$}] (dp) at (1,0) {};
    \node[vertex,label={[xshift=7pt,yshift=-9pt]$x$}] (x) at (2,0) {};
    \node[vertex,label={[xshift=-7pt,yshift=-10pt]$e'$}] (ep) at (3.5,0) {};
    \node[vertex,label={[xshift=7pt,yshift=-10pt]$y$}] (y) at (4.5,0) {};
    \node[vertex,label={[xshift=-7pt,yshift=-11pt]$f'$}] (fp) at (6,0) {};
    \node[vertex,label={[xshift=-7pt,yshift=-9pt]$z$}] (z) at (7.2,0) {};
    \node[vertex,label={[xshift=7pt,yshift=-13pt]$g'$}] (gp) at (8,0) {};

    \draw[-latex,line width=2pt] (a) -- (d);
    \draw[-latex,line width=2pt] (a) -- (e);
    \draw[-latex,line width=2pt] (a) -- (f);
    \draw[-latex,line width=2pt] (b) -- (g);
    \draw[-latex,line width=2pt] (b) -- (ap);
    \draw[-latex,line width=2pt] (b) -- (bp);
    \draw[-latex,line width=2pt] (b) -- (k);
    \draw[-latex,gray] (c) -- (bp);
    \draw[-latex,line width=2pt] (c) -- (h);
    \draw[-latex,line width=2pt] (c) -- (i);
    \draw[-latex,line width=2pt] (c) -- (j);
    \draw[-latex,gray] (d) -- (cp);
    \draw[-latex,gray] (d) -- (k);
    \draw[-latex,line width=2pt] (e) -- (cp);
    \draw[-latex,line width=2pt] (e) -- (ep);
    \draw[-latex,gray] (f) -- (b);
    \draw[-latex,gray] (f) -- (e);
    \draw[-latex,gray] (g) -- (q);
    \draw[-latex,gray] (g) -- (l);
    \draw[-latex,gray] (g) -- (m);
    \draw[-latex,gray] (ap) -- (z);
    \draw[-latex,line width=2pt] (h) -- (m);
    \draw[-latex,line width=2pt] (h) -- (l);
    \draw[-latex,line width=2pt] (h) -- (n);
    \draw[-latex,gray] (cp) -- (o);
    \draw[-latex,line width=2pt] (k) -- (o);
    \draw[-latex,line width=2pt] (k) -- (p);
    \draw[-latex,line width=2pt] (k) -- (q);
    \draw[-latex,gray] (l) -- (ep);
    \draw[-latex,line width=2pt] (m) -- (r);
    \draw[-latex,line width=2pt] (m) -- (s);
    \draw[-latex,line width=2pt] (m) -- (t);
    \draw[-latex,line width=2pt] (n) -- (u);
    \draw[-latex,gray] (n) -- (v);
    \draw[-latex,line width=2pt] (n) -- (w);
    \draw[-latex,gray] (p) -- (x);
    \draw[-latex,gray] (s) -- (y);
    \draw[-latex,gray] (s) -- (r);
    \draw[-latex,gray] (u) -- (fp);
    \draw[-latex,gray] (w) -- (j);
    \draw[-latex,gray] (dp) -- (o);
    \draw[-latex,gray] (x) -- (dp);
    \draw[-latex,gray] (x) -- (l);
    \draw[-latex,gray] (y) -- (ep);
    \draw[-latex,gray] (fp) -- (t);
    \draw[-latex,line width=2pt] (z) -- (v);
    \draw[-latex,line width=2pt] (z) -- (gp);
    \draw[-latex,gray] (gp) .. controls +(east:15mm) and +(east:15mm) .. (c);

\end{tikzpicture}}
        \caption{}
        \label{figc:f2}
    \end{subfigure}
    \hfill
    \begin{subfigure}{0.48\textwidth}
        \resizebox{\textwidth}{!}{\begin{tikzpicture}
    [vertex/.style={circle,draw,fill,inner sep=0pt,minimum size=5pt}]

    \clip (0,-0.3) rectangle (9.5,4.3);

    \node[vertex,label={[xshift=7pt,yshift=-7pt]$a$}] (a) at (1.4,4) {};
    \node[vertex,label={[xshift=7pt,yshift=-7pt]$b$}] (b) at (4.9,4) {};
    \node[vertex,label={[xshift=-7pt,yshift=-7pt]$c$}] (c) at (8.4,4) {};
    \node[vertex,label={[xshift=-7pt,yshift=-9pt]$d$}] (d) at (0.7,3) {};
    \node[vertex,label={[xshift=-7pt,yshift=-9pt]$e$}] (e) at (1.4,3) {};
    \node[vertex,label={[xshift=7pt,yshift=-13pt]$f$}] (f) at (2.1,3) {};
    \node[vertex,label={[xshift=-7pt,yshift=-11pt]$g$}] (g) at (4.2,3) {};
    \node[vertex,label={[xshift=-7pt,yshift=-9pt]$a'$}] (ap) at (4.9,3) {};
    \node[vertex,label={[xshift=-7pt,yshift=-9pt]$b'$}] (bp) at (5.6,3) {};
    \node[vertex,label={[xshift=-7pt,yshift=-5pt]$h$}] (h) at (7.7,3) {};
    \node[vertex,label={[xshift=-7pt,yshift=-9pt]$i$}] (i) at (8.4,3) {};
    \node[vertex,label={[xshift=-7pt,yshift=-9pt]$j$}] (j) at (9.1,3) {};
    \node[vertex,label={[xshift=7pt,yshift=-10pt]$c'$}] (cp) at (0.1,2) {};
    \node[vertex,label={[xshift=-7pt,yshift=-9pt]$k$}] (k) at (1.4,2) {};
    \node[vertex,label={[xshift=7pt,yshift=-14pt]$l$}] (l) at (3.5,2) {};
    \node[vertex,label={[xshift=-7pt,yshift=-9pt]$m$}] (m) at (4.9,2) {};
    \node[vertex,label={[xshift=7pt,yshift=-10pt]$n$}] (n) at (7.7,2) {};
    \node[vertex,label={[xshift=-7pt,yshift=-9pt]$o$}] (o) at (0.7,1) {};
    \node[vertex,label={[xshift=-7pt,yshift=-11pt]$p$}] (p) at (1.4,1) {};
    \node[vertex,label={[xshift=-7pt,yshift=-11pt]$q$}] (q) at (2.1,1) {};
    \node[vertex,label={[xshift=-7pt,yshift=-9pt]$r$}] (r) at (4.2,1) {};
    \node[vertex,label={[xshift=7pt,yshift=-9pt]$s$}] (s) at (4.9,1) {};
    \node[vertex,label={[xshift=7pt,yshift=-9pt]$t$}] (t) at (5.6,1) {};
    \node[vertex,label={[xshift=7pt,yshift=-9pt]$u$}] (u) at (7,1) {};
    \node[vertex,label={[xshift=7pt,yshift=-9pt]$v$}] (v) at (7.7,1) {};
    \node[vertex,label={[xshift=7pt,yshift=-9pt]$w$}] (w) at (8.4,1) {};
    \node[vertex,label={[xshift=-7pt,yshift=-9pt]$d'$}] (dp) at (1,0) {};
    \node[vertex,label={[xshift=7pt,yshift=-9pt]$x$}] (x) at (2,0) {};
    \node[vertex,label={[xshift=-7pt,yshift=-10pt]$e'$}] (ep) at (3.5,0) {};
    \node[vertex,label={[xshift=7pt,yshift=-10pt]$y$}] (y) at (4.5,0) {};
    \node[vertex,label={[xshift=-7pt,yshift=-11pt]$f'$}] (fp) at (6,0) {};
    \node[vertex,label={[xshift=-7pt,yshift=-9pt]$z$}] (z) at (7.2,0) {};
    \node[vertex,label={[xshift=7pt,yshift=-13pt]$g'$}] (gp) at (8,0) {};

    \draw[-latex,line width=2pt] (a) -- (d);
    \draw[-latex,line width=2pt] (a) -- (e);
    \draw[-latex,line width=2pt] (a) -- (f);
    \draw[-latex,line width=2pt] (b) -- (g);
    \draw[-latex,line width=2pt] (b) -- (ap);
    \draw[-latex,line width=2pt] (b) -- (bp);
    \draw[-latex,line width=2pt] (b) -- (k);
    \draw[-latex,gray] (c) -- (bp);
    \draw[-latex,line width=2pt] (c) -- (h);
    \draw[-latex,line width=2pt] (c) -- (i);
    \draw[-latex,line width=2pt] (c) -- (j);
    \draw[-latex,gray] (d) -- (cp);
    \draw[-latex,gray] (d) -- (k);
    \draw[-latex,line width=2pt] (e) -- (cp);
    \draw[-latex,line width=2pt] (e) -- (ep);
    \draw[-latex,gray] (f) -- (b);
    \draw[-latex,gray] (f) -- (e);
    \draw[-latex,gray] (g) -- (q);
    \draw[-latex,gray] (g) -- (l);
    \draw[-latex,gray] (g) -- (m);
    \draw[-latex,line width=2pt] (ap) -- (z);
    \draw[-latex,line width=2pt] (h) -- (m);
    \draw[-latex,line width=2pt] (h) -- (l);
    \draw[-latex,line width=2pt] (h) -- (n);
    \draw[-latex,gray] (cp) -- (o);
    \draw[-latex,line width=2pt] (k) -- (o);
    \draw[-latex,line width=2pt] (k) -- (p);
    \draw[-latex,line width=2pt] (k) -- (q);
    \draw[-latex,gray] (l) -- (ep);
    \draw[-latex,line width=2pt] (m) -- (r);
    \draw[-latex,line width=2pt] (m) -- (s);
    \draw[-latex,line width=2pt] (m) -- (t);
    \draw[-latex,line width=2pt] (n) -- (u);
    \draw[-latex,gray] (n) -- (v);
    \draw[-latex,line width=2pt] (n) -- (w);
    \draw[-latex,line width=2pt] (p) -- (x);
    \draw[-latex,line width=2pt] (s) -- (y);
    \draw[-latex,gray] (s) -- (r);
    \draw[-latex,line width=2pt] (u) -- (fp);
    \draw[-latex,gray] (w) -- (j);
    \draw[-latex,gray] (dp) -- (o);
    \draw[-latex,line width=2pt] (x) -- (dp);
    \draw[-latex,gray] (x) -- (l);
    \draw[-latex,gray] (y) -- (ep);
    \draw[-latex,gray] (fp) -- (t);
    \draw[-latex,line width=2pt] (z) -- (v);
    \draw[-latex,line width=2pt] (z) -- (gp);
    \draw[-latex,line width=2pt] (gp) .. controls +(east:15mm) and +(east:15mm) .. (c);

\end{tikzpicture}}
        \caption{}
        \label{figd:finalarb}
    \end{subfigure}
    \caption{Illustration of a possible execution of algorithm \textsc{Maxleaves-12MIS}.
      (A) A 4-branching $F_1$ in bold, obtained in Line~\ref{line:F1}, and the
      corresponding set $\Candidates = \{a,c,e,f,g,h,k,m,n,s,x,z\}$.      
      (B)~The corresponding intersection multigraph, obtained from the sets
      $U_g = \{l,m,q\}$, $U_{g_q} = \{l,m\}$, $U_{g_l} = \{m,q\}$, $U_{g_m} =
      \{l,q\}$, $U_h = \{l,m,n\}$, $U_{h_n} = \{l,m\}$, $U_{h_l} = \{m,n\}$,
      $U_{h_m} = \{l,n\}$, etc. 
      (C) The 2-branching~$F_2$ in bold, obtained after Line~\ref{line:EndForF2} 
      from the independent set $I = \{a,c,e,h,k,m,n_v,z\}$, in bold. 
      (D) Final arborescence~$T$ in bold, obtained from~$F_2$ in Line~\ref{line:T}.}
    \label{fig:maxleaves-12mis}
\end{figure}

\Call{Maxleaves-12MIS}{} starts by calling \Call{GreedyExpand}{\mbox{$D$, $4$,
$F_0$}} with the empty spanning branching~$F_0$, which outputs a maximal
$4$-branching~$F_1$.
Then, it considers the collection of non-trivial expansions that can be applied
to~$F_1$. 
Because~$F_1$ is a maximal $4$-branching, this collection contains only
2-expansions and 3-expansions.
Using \Call{IntersectionGraph}{}, algorithm \Call{Maxleaves-12MIS}{} builds the
graph~$G$ whose vertex set is this collection of expansions, with two
expansions adjacent if they are not compatible (that is, if both were applied
to~$F_1$, the result would not be a branching). 
It defines a weight function~$w$ by assigning weight~1 to~$2$-expansions and
weight~2 to~$3$-expansions. 
See Figure~\ref{fig:maxleaves-12mis}.
The algorithm then applies \SquareImp\ to the wMIS instance~$(G,w)$ obtaining
an independent set in~$G$.
This independent set corresponds to a collection of compatible $2$-expansions
and $3$-expansions  that can be applied to~$F_1$, resulting in
a~$2$-branching~$F_2$.
\Call{Maxleaves-12MIS}{} finishes by calling \Call{GreedyExpand}{$D$, $1$,
$F_2$} to obtain a spanning arborescence in~$D$.

In our previous result for \maxleaves~\cite{FernandesL2021}, we used the
collection $\calS = \{U_v \colon v \in \Candidates\}$, with the same weight
function~$w$, as an instance of the maximum weight 3D-matching problem.

The intersection graph~$G$ has no 4-claws because each set~$U_v$ is either a
3-set or a 2-set. 
So \Call{SquareImp}{$G$, $w$} achieves a ratio slightly greater than~2 for
wMIS.
However, our graph~$G$ is not only 4-claw free: it uses only weights~1 and~2,
and has other particularities that we will explore in the next section. 


\section{Weighted $\{2,3\}$-intersection graphs}
\label{sec:23intersectiongraphs}

A pair $(V,U)$ is a \emph{hereditary $\{2,3\}$-collection} if~$V$ is a finite
set and the set $U = \{U_v: v \in V\}$ is a collection of 2-sets and 3-sets such that,
for each $v$ in~$V$ with $U_v = \{a,b,c\}$, there are elements~$v_a$, $v_b$,
and~$v_c$ in~$V$ with~$U_{v_a}=\{b,c\}$, $U_{v_b}=\{a,c\}$,
and~$U_{v_c}=\{a,b\}$. 
 
A $\{2,3\}$-\emph{intersection graph} is the intersection multigraph associated
to a hereditary $\{2,3\}$-collection $(V,U)$, that is, the multigraph whose
vertex set is~$V$ and there are $|U_x \cap U_y|$ parallel edges between any two
vertices~$x$ and~$y$ in~$V$. 
A \emph{weighted} $\{2,3\}$-intersection graph is a $\{2,3\}$-intersection
graph whose weight for a vertex~$v$ is exactly~$|U_v|-1$. 

In a $\{2,3\}$-intersection graph, if~$x$ and~$y$ are neighbors and 
${|U_x \cap U_y| = 1}$, we say they are \emph{single} neighbors.
In a weighted $\{2,3\}$-intersection graph, every weight-2 vertex~$v$ forms
a~$K_4$ with the three weight-1 vertices~$v_a$, $v_b$, and~$v_c$ where $U_v =
\{a,b,c\}$. 
Moreover, there are two parallel edges between~$v$ and each of~$v_a$, $v_b$,
and~$v_c$, and~$v_a$, $v_b$, and~$v_c$ form a triangle.
We use~$K_4^v$ to refer to this~$K_4$.

\begin{claim}
  The weighted graph~$(G,w)$ built in Lines~\ref{line9}-\ref{line11} of
  Algorithm~\ref{alg:maxleaves-12mis} is a weighted $\{2,3\}$-intersection
  graph.
\end{claim}
\begin{claimproof}
  It is enough to argue that $(\Candidates,U)$ is a hereditary
  $\{2,3\}$-collection.
  Indeed $\Candidates$ is a finite set, and every element $v$ in $\Candidates$
  is associated to a set $U_v$ that is a 2-set or a 3-set.  
  Moreover, for every $v$ in $\Candidates$ such that $U_v=\{a,b,c\}$, there are
  three elements $v_a$, $v_b$, and $v_c$ in $\Candidates$ such that $U_{v_a} =
  \{b,c\}$, $U_{v_b} = \{a,c\}$, and $U_{v_c} = \{a,b\}$. 
  Hence $(\Candidates,U)$ is indeed a hereditary $\{2,3\}$-collection, and
  therefore, by the definition of~$w$ in Line~\ref{line11}, $(G,w)$ is a
  weighted $\{2,3\}$-intersection graph. 
\end{claimproof}

There is a straightforward reduction from 3D-matching to maximum independent
set in 4-claw free graphs which implies that wMIS on weighted 4-claw free
graphs is NP-hard~\cite{ChandraH2001,GareyJ1979}. 
We adapt this reduction to prove the following hardness result for wMIS. 

\begin{theorem}
  wMIS is NP-hard on weighted $\{2,3\}$-intersection graphs.
\end{theorem}
\begin{proof}
  We modify the reduction from 3D-matching to the maximum independent set
  problem so that the instance built is an instance of wMIS, specifically, is a
  weighted $\{2,3\}$-intersection graph.

  An instance of 3D-matching consists of the following.
  Let $X$, $Y$, and $Z$ be disjoint sets such that~$|X|=|Y|=|Z|=q$, and
  let~$\calS$ be a subset of $X \times Y \times Z$, that is, each set
  in~$\calS$ is a triple of elements, one in~$X$, one in~$Y$, and one in~$Z$. 
  The goal of the 3D-matching problem is to decide whether there is a
  subcollection of~$\calS$ with exactly~$q$ disjoint sets.

  We build from~$\calS$ an enlarged collection~$\calS'$ that contains~$\calS$
  and all three sets of size~2 contained in a set of~$\calS$.  
  Let~$V$ be a set of size~$|\calS'|$ and associate to each element in~$V$ one
  of the sets~in $\calS'$.
  Note that the pair $(V, \calS')$ is a hereditary $\{2,3\}$-collection. 
  Let $(G,w)$ be the weighted $\{2,3\}$-intersection graph associated to $(V,
  \calS')$.
  Recall that the weight~$w_v$ of a vertex~$v$ in~$G$ whose associated set
  is~$S$ in~$\calS'$ is $|S|-1$.  
  We applied in Algorithm~\ref{alg:maxleaves-12mis} a similar construction on
  the set $\Candidates$. 
  
  Let us prove that there is a solution for the 3D-matching instance if and
  only if there is an independent set in~$G$ of weight at least~$2q$. 
  If there is a solution $M \subseteq \calS$ for the 3D-matching problem, 
  then let~$I$ be the set of vertices of~$G$ corresponding to the sets in~$M$. 
  As~$M$ is a collection of disjoint sets, $I$ is an independent set in~$G$. 
  Each set in~$M$ is a 3-set, therefore its corresponding vertex in~$G$ has
  weight~2. 
  As $|M| = q$, the weight of~$I$ is~$2q$.

  For the other direction, let~$I$ be an independent set in~$G$ of weight at
  least~$2q$. 
  Let~$q_1$ be the number of weight-1 vertices in~$I$ and~$q_2$ be the number
  of weight-2 vertices in~$I$. 
  Let us prove that~$q_1 = 0$ and $q_2 = q$. 
  This would imply that the collection of sets in~$\calS'$ corresponding to~$I$
  is a subset of~$\calS$ with exactly~$q$ disjoint sets, being therefore a
  solution for the 3D-matching instance. 

  The sets corresponding to vertices in~$I$ are pairwise disjoint, and their
  union has size $2q_1 + 3q_2$ and is contained in $X \cup Y \cup Z$, whose
  size is~$3q$.
  Therefore 
  \begin{equation}\label{size}
    2q_1 + 3q_2 \ \leq \ 3q.
  \end{equation}
  This implies that $q_2 \leq q$.
  Moreover, the weight of~$I$ is 
  \begin{equation}\label{eq:2q}
   q_1 + 2q_2 \ \leq \ 3\,\frac{q-q_2}2 + 2q_2  
                \ = \ \frac32 q + \frac12 q_2 \ \leq \ 2q
  \end{equation}
  by~\eqref{size} and because $q_2 \leq q$. 
  As the weight of~$I$ is at least~$2q$, the inequalities in~\eqref{eq:2q} must
  be equalities, which means that $q_2 = q$ and $q_1 = 0$.
  This completes the proof. 
\end{proof}

In what follows, we will present an analysis of \SquareImp\ specific for
weighted $\{2,3\}$-intersection graphs that shows that \SquareImp\ achieves a
ratio significantly better on such graphs than on general weighted 4-claw free
graphs. 
In Section~\ref{sec:better_approx}, we will adapt \textsc{SquareImp} and use
ideas from~\cite{FernandesL2020} to obtain an even better approximation for
wMIS on weighted $\{2,3\}$-intersection graphs. 

The following proposition states properties of weighted $\{2,3\}$-intersection
graphs that will be useful in our analysis.

\begin{proposition}
\label{prop:intersectiongraph}
  Every weighted $\{2,3\}$-intersection graph~$(G,w)$ has the following
  properties: 
  \begin{enumerate}[{\rm (i)}]
    \item there is no 4-claw in~$G$ and every 3-claw in~$G$ has a center of
    weight~2;
    \item every weight-2 vertex~$v$ dominates the neighborhood of~$K_4^v$; 
    \item every single neighbor~$u$ of a weight-2 vertex~$v$ has exactly two
    weight-1 neighbors in~$K_4^v$.
  \end{enumerate}
\end{proposition}
\begin{proof}
  Let $(V,U)$ be the hereditary $\{2,3\}$-collection underlying $(G,w)$. 
  To prove (i), note first that if~$C$ is a $d$-claw with center in~$z$ for
  some~$z$ in~$V$, then, because~$T_C$ is an independent set, each of the sets
  corresponding to a vertex in~$T_C$ intersects~$U_z$ in at least one distinct
  element.
  Since~$|U_z| \leq 3$, we have that $d \leq 3$.  
  Also, if $|U_z| = 2$, that is, $z$ has weight~1, then $d \leq 2$.

  For (ii), suppose that~$x$ is a neighbor of a vertex~$y$ in~$K_4^v$, which
  means $U_x \cap U_y \neq \emptyset$.
  Observe that~$U_y \subseteq U_v$ for every~$y$ in $K_4^v$.
  Therefore $U_x \cap U_v \supseteq U_x \cap U_y \neq \emptyset$.

  For (iii), suppose $U_v=\{a,b,c\}$, and let~$u$ be a single neighbor of~$v$. 
  Without loss of generality, we may assume that $U_u \cap U_v=\{a\}$.  
  Recall that the vertices of~$K_4^v$ are~$v$, $v_a$, $v_b$, and~$v_c$. 
  Then~$u$ is a neighbor of~$v_b$ and~$v_c$, but not of~$v_a$.  
\end{proof}


We tailored a better and tighter analysis of \SquareImp\ for weighted
$\{2,3\}$-intersection graphs. 

\begin{theorem}
\label{thm:squareimp32}
  \textsc{SquareImp}{} is a~$\frac32$-approximation for wMIS on weighted
  $\{2,3\}$-intersection graphs.
\end{theorem}
\begin{proof}
  Let $(G,w)$ be a weighted $\{2,3\}$-intersection graph. 
  Let $A^*$ be an independent set in~$G$ that maximizes~$w(A^*)$ and let~$A$ be
  the independent set produced by \Call{SquareImp}{$G$,$w$}.
  We shall prove that $w(A^*) \leq \frac32\,w(A)$. 
  We will do this using the strategy of Berman~\cite{Berman2000}: each vertex
  in~$A^*$ will distribute its weight among its neighbors in~$A$ so that no
  vertex in~$A$ gets more than~$3/2$ its own weight. 

  For the sake of the argument in this proof, we consider~$G$ as the
  corresponding intersection multigraph.
  Vertices in $A^* \cap A$ keep their weights.  
  Because~$A$ is maximal, every vertex in $A^* \setminus A$ has at least one
  neighbor in~$A$.
  Also, by Proposition~\ref{prop:intersectiongraph}(i), every vertex in~$A^*$
  of weight~$z$ has at most~$z+1$ edges going to its neighbors in~$A$.

  In Figure~\ref{fig:weights32}, we show how each vertex in $A^* \setminus A$
  distributes its weight to its neighbors in~$A$.  
  We represent vertices in~$A^*$ by red squares and vertices in~$A$ by blue
  circles.  
  The number on top of each vertex in~$A^*$ is its weight.  
  The number below a vertex in~$A$ denotes its weight when that weight matters
  for the distribution.
  The number on each edge is the amount of weight distributed from the vertex
  in~$A^*$ to the vertex in~$A$.
  Recall that, in this argument, $G$ is a multigraph, so some of the blue round
  vertices connected to a red square vertex might be the same, receiving some
  weight from the same red square vertex through two or three edges. 
  An example of such distribution can be seen in Figure~\ref{fig:example_dist}.

  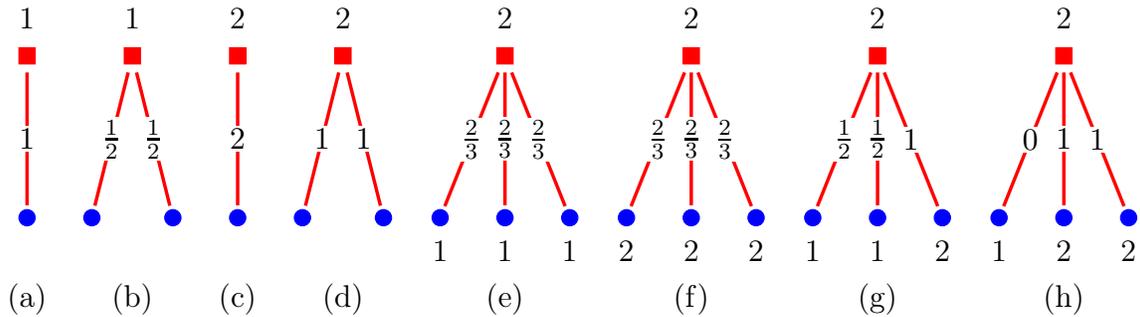
\begin{figure}[htb]
    \centering
    \resizebox{0.95\textwidth}{!}{\begin{tikzpicture}
  [LabelStyle/.style={inner sep=1pt}]

  \node[label=above:1](r0) [red vertex] at (0,4) {};
  \node(f10) [blue vertex] at ($(r0)+(0,-2)$) {};
  \Edge[color=red,label={ $1$},lw=1.2pt,style={pos=0.5}](r0)(f10);
  \node at ($(r0)+(0,-3)$) {(a)};

  \node[label=above:1](r1) [red vertex] at ($(r0)+(1.3,0)$) {};
  \node(f11) [blue vertex] at ($(r1)+(-0.5,-2)$) {};
  \node(f31) [blue vertex] at ($(r1)+(+0.5,-2)$) {};
  \Edge[color=red,label={ $\frac12$},lw=1.2pt,style={pos=0.5}](r1)(f11);
  \Edge[color=red,label={ $\frac12$},lw=1.2pt,style={pos=0.5}](r1)(f31);
  \node at ($(r1)+(0,-3)$) {(b)};

  \node[label=above:2](r2) [red vertex] at ($(r1)+(1.3,0)$) {};
  \node(f12) [blue vertex] at ($(r2)+(0,-2)$) {};
  \Edge[color=red,label={ $2$},lw=1.2pt,style={pos=0.5}](r2)(f12);
  \node at ($(r2)+(0,-3)$) {(c)};

  \node[label=above:2](r3) [red vertex] at ($(r2)+(1.3,0)$) {};
  \node(f13) [blue vertex] at ($(r3)+(-0.5,-2)$) {};
  \node(f33) [blue vertex] at ($(r3)+(+0.5,-2)$) {};
  \Edge[color=red,label={ $1$},lw=1.2pt,style={pos=0.5}](r3)(f13);
  \Edge[color=red,label={ $1$},lw=1.2pt,style={pos=0.5}](r3)(f33);
  \node at ($(r3)+(0,-3)$) {(d)};

  \node[label=above:2](r4) [red vertex] at ($(r3)+(2,0)$) {};
  \node[label=below:1](f14) [blue vertex] at ($(r4)+(-0.8,-2)$) {};
  \node[label=below:1](f24) [blue vertex] at ($(r4)+(0,-2)$) {};
  \node[label=below:1](f34) [blue vertex] at ($(r4)+(+0.8,-2)$) {};
  \Edge[color=red,label={ $\frac23$},lw=1.2pt,style={pos=0.5}](r4)(f14);
  \Edge[color=red,label={ $\frac23$},lw=1.2pt,style={pos=0.5}](r4)(f24);
  \Edge[color=red,label={ $\frac23$},lw=1.2pt,style={pos=0.5}](r4)(f34);
  \node at ($(r4)+(0,-3)$) {(e)};

  \node[label=above:2](r5) [red vertex] at ($(r4)+(2.3,0)$) {};
  \node[label=below:2](f15) [blue vertex] at ($(r5)+(-0.8,-2)$) {};
  \node[label=below:2](f25) [blue vertex] at ($(r5)+(0,-2)$) {};
  \node[label=below:2](f35) [blue vertex] at ($(r5)+(0.8,-2)$) {};
  \Edge[color=red,label={ $\frac23$},lw=1.2pt,style={pos=0.5}](r5)(f15);
  \Edge[color=red,label={ $\frac23$},lw=1.2pt,style={pos=0.5}](r5)(f25);
  \Edge[color=red,label={ $\frac23$},lw=1.2pt,style={pos=0.5}](r5)(f35);
  \node at ($(r5)+(0,-3)$) {(f)};

  \node[label=above:2](r6) [red vertex] at ($(r5)+(2.3,0)$) {};
  \node[label=below:1](f16) [blue vertex] at ($(r6)+(-0.8,-2)$) {};
  \node[label=below:1](f26) [blue vertex] at ($(r6)+(0,-2)$) {};
  \node[label=below:2](f36) [blue vertex] at ($(r6)+(+0.8,-2)$) {};
  \Edge[color=red,label={ $\frac12$},lw=1.2pt,style={pos=0.5}](r6)(f16);
  \Edge[color=red,label={ $\frac12$},lw=1.2pt,style={pos=0.5}](r6)(f26);
  \Edge[color=red,label={ $1$},lw=1.2pt,style={pos=0.5}](r6)(f36);
  \node at ($(r6)+(0,-3)$) {(g)};

  \node[label=above:2](r7) [red vertex] at ($(r6)+(2.3,0)$) {};
  \node[label=below:1](f17) [blue vertex] at ($(r7)+(-0.8,-2)$) {};
  \node[label=below:2](f27) [blue vertex] at ($(r7)+(0,-2)$) {};
  \node[label=below:2](f37) [blue vertex] at ($(r7)+(+0.8,-2)$) {};
  \Edge[color=red,label={ $0$},lw=1.2pt,style={pos=0.5}](r7)(f17);
  \Edge[color=red,label={ $1$},lw=1.2pt,style={pos=0.5}](r7)(f27);
  \Edge[color=red,label={ $1$},lw=1.2pt,style={pos=0.5}](r7)(f37);
  \node at ($(r7)+(0,-3)$) {(h)};

\end{tikzpicture}}
    \caption{Weight distribution for Theorem~\ref{thm:squareimp32}.}
    \label{fig:weights32}
  \end{figure}

  \begin{figure}[htb]
  \centering
  \begin{subfigure}{\textwidth}
    \centering
    \resizebox{0.8\textwidth}{!}{\begin{tikzpicture}
    [vertex/.style={circle,draw,fill,inner sep=0pt,minimum size=5pt}]


    \def\h{10mm};

    \node (P) at (0,0) {};

    \node[vertex,label={[xshift=-7pt,yshift=-9pt]$c'$}]  (cp)  at ($(P)+(0*\h,0)$) {};
    \node[vertex,label={[xshift=-7pt,yshift=-9pt]$b'$}]  (bp)  at ($(P)+(1*\h,0)$) {};
    \node[vertex,label={[xshift=-7pt,yshift=-10pt]$c$}] (c) at ($(P)+(2*\h,0)$) {};
    \node[vertex,label={[xshift=-7pt,yshift=-10pt]$b$}]  (b)  at ($(P)+(3*\h,0)$) {};
    \node[vertex,label={[xshift=-7pt,yshift=-10pt]$a$}]  (a)  at ($(P)+(4*\h,0)$) {};
    \node[vertex,label={[xshift=-7pt,yshift=-11pt]$a'$}]  (ap)  at ($(P)+(5*\h,0)$) {};
    \node[vertex,label={[xshift=-7pt,yshift=-11pt]$z'$}] (zp) at ($(P)+(6*\h,0)$) {};
    \node[vertex,label={[xshift=-7pt,yshift=-9pt]$z$}]  (z)  at ($(P)+(7*\h,0)$) {};
    \node[vertex,label={[xshift=-7pt,yshift=-9pt]$i$}]  (i)  at ($(P)+(8*\h,0)$) {};
    \node[vertex,label={[xshift=-7pt,yshift=-9pt]$u$}] (u) at ($(P)+(9*\h,0)$) {};
    \node[vertex,label={[xshift=-7pt,yshift=-9pt]$h$}]  (h)  at ($(P)+(10*\h,0)$) {};
    \node[vertex,label={[xshift=-7pt,yshift=-11pt]$g$}] (g) at ($(P)+(11*\h,0)$) {};
    \node[vertex,label={[xshift=-7pt,yshift=-9pt]$d$}]  (d)  at ($(P)+(12*\h,0)$) {};
    \node[vertex,label={[xshift=-7pt,yshift=-9pt]$e$}]  (e)  at ($(P)+(13*\h,0)$) {};
    \node[vertex,label={[xshift=-7pt,yshift=-11pt]$f$}]  (f)  at ($(P)+(14*\h,0)$) {};

    \node[vertex,label={[xshift=-7pt,yshift=-9pt]$o$}] (o) at ($(cp)+(\h/2,1.2)$) {};
    \node[vertex,label={[xshift=-7pt,yshift=-10pt]$p$}] (p) at ($(bp)+(\h/2,1.2)$) {};
    \node[vertex,label={[xshift=-8pt,yshift=-10pt]$q$}] (q) at ($(b)+(0,1.2)$) {};
    \node[vertex,label={[xshift=7pt,yshift=-10pt]$r$}] (r) at ($(a)+(\h/2,1.2)$) {};
    \node[vertex,label={[xshift=7pt,yshift=-9pt]$s$}] (s) at ($(ap)+(\h/2,1.2)$) {};
    \node[vertex,label={[xshift=7pt,yshift=-9pt]$t$}]  (t) at ($(zp)+(\h/2,1.2)$) {};
    \node[vertex,label={[xshift=7pt,yshift=-9pt]$v$}]  (v) at ($(i)+(0,1.2)$) {};
    \node[vertex,label={[xshift=7pt,yshift=-9pt]$y$}]  (y) at ($(h)+(0,1.2)$) {};
    \node[vertex,label={[xshift=7pt,yshift=-9pt]$x$}]  (x) at ($(g)+(\h/2,1.2)$) {};
    \node[vertex,label={[xshift=7pt,yshift=-9pt]$w$}]  (w) at ($(d)+(\h/2,1.2)$) {};
    \node[vertex,label={[xshift=7pt,yshift=-9pt]$l$}]  (l) at ($(e)+(\h/2,1.2)$) {};

    \draw[-latex,blue] (o) -- (cp);
    \draw[-latex,blue] (o) -- (c);
    \draw[-latex,blue] (p) -- (bp);
    \draw[-latex,blue] (p) -- (b);
    \draw[-latex,red] (q) -- (c);
    \draw[-latex,red] (q) -- (b);
    \draw[-latex,red] (q) -- (a);
    \draw[-latex,blue] (r) -- (ap);
    \draw[-latex,blue] (r) -- (a);
    \draw[-latex,red] (s) -- (ap);
    \draw[-latex,red] (s) -- (zp);
    \draw[-latex,blue] (t) -- (zp);
    \draw[-latex,blue] (t) -- (z);
    \draw[-latex,red] (v) -- (z);
    \draw[-latex,red] (v) -- (i);
    \draw[-latex,red] (v) -- (u);
    \draw[-latex,blue] (y) -- (u);
    \draw[-latex,red] (y) -- (h);
    \draw[-latex,red] (y) -- (g);
    \draw[-latex,dashed,blue] (y) -- (h);
    \draw[-latex,dashed,blue](y) -- (g);
    \draw[-latex,gray] (x) -- (g);
    \draw[-latex,gray] (x) -- (d);
    \draw[-latex,red] (w) -- (d);
    \draw[-latex,red] (w) -- (e);
    \draw[-latex,dashed,blue] (w) -- (d);
    \draw[-latex,dashed,blue] (w) -- (e);
    \draw[-latex,gray] (l) -- (e);
    \draw[-latex,gray] (l) -- (f);

    \node[vertex,label={[xshift=-7pt,yshift=-7pt]$j$}]  (j) at ($(q)+(\h/2,1.2)$) {};
    \node[vertex,label={[xshift=7pt,yshift=-5pt]$k$}]  (k) at ($(v)+(0,2.4)$) {};

    \draw[-latex,line width=2pt] (j) -- (o);
    \draw[-latex,line width=2pt] (j) -- (p);
    \draw[-latex,line width=2pt] (j) -- (q);
    \draw[-latex,line width=2pt] (j) -- (r);
    \draw[-latex,line width=2pt] (j) -- (s);
    \draw[-latex,line width=2pt] (j) -- (t);
    \draw[-latex,line width=2pt] (k) -- (j);
    \draw[-latex,line width=2pt] (k) -- (v);
    \draw[-latex,line width=2pt] (k) -- (y);
    \draw[-latex,line width=2pt] (k) -- (x);
    \draw[-latex,line width=2pt] (k) -- (w);
    \draw[-latex,line width=2pt] (k) -- (l);

\end{tikzpicture}}
    \caption{Digraph $D$ with a 4-branching in bold.}
    \label{figa:dist_D}
  \end{subfigure}

  \begin{subfigure}{\textwidth}
    \centering
    \resizebox{0.8\textwidth}{!}{\begin{tikzpicture}
    [vertex/.style={circle,draw,inner sep=0pt,minimum size=5pt,minimum width=15pt}]

    \node[vertex,line width=2pt,dashed,red] (v) at (0,0) {$v$};
    \node[vertex] (v1) at ($(v)+(0,0.8)$) {$v_i$};
    \node[vertex] (v2) at ($(v)+(-0.7,-0.5)$) {$v_u$};
    \node[vertex] (v3) at ($(v)+(+0.7,-0.5)$) {$v_z$};
    \draw[double,thick] (v) -- (v1);
    \draw[double,thick] (v) -- (v2);
    \draw[double,thick] (v) -- (v3);
    \draw[thick] (v1) -- (v2);
    \draw[thick] (v1) -- (v3);
    \draw[thick] (v2) -- (v3);

    \node[vertex,line width=2pt,blue] (y) at ($(v)+(1.8,0.8)$) {$y$};
    \node[vertex] (y1) at ($(y)+(-0.7,0.5)$) {$y_g$};
    \node[vertex] (y2) at ($(y)+(0,-0.8)$) {$y_h$};
    \node[vertex,line width=2pt,dashed,red] (y3) at ($(y)+(+0.7,0.5)$) {$y_u$};
    \draw[double,thick] (y) -- (y1);
    \draw[double,thick] (y) -- (y2);
    \draw[double,thick] (y) -- (y3);
    \draw[thick] (y1) -- (y2);
    \draw[thick] (y1) -- (y3);
    \draw[thick] (y2) -- (y3);

    \draw[thick] (v1) -- (y1);
    \draw[thick] (v1) -- (y);
    \draw[thick] (v1) -- (y2);
    \draw[thick] (v3) -- (y1);
    \draw[thick] (v3) -- (y);
    \draw[thick] (v3) -- (y2);
    \draw[thick] (v) -- (y1);
    \draw[thick] (v) -- (y);
    \draw[thick] (v) -- (y2);

    \node[vertex] (x) at ($(y)+(1.2,-0.5)$) {$x$};
    \draw[thick] (x) -- (y);
    \draw[thick] (x) -- (y2);
    \draw[thick] (x) -- (y3);
    \node[vertex,line width=1.5pt,blue] (w) at ($(x)+(1,0)$) {$w$};
    \node[vertex,line width=1.5pt,dashed,red] () at (w) {};
    \draw[thick] (x) -- (w);
    \node[vertex] (l) at ($(w)+(0,1)$) {$l$};
    \draw[thick] (l) -- (w);

    \node[vertex,line width=2pt,blue] (t) at ($(v)+(-1.2,0.5)$) {$t$};
    \draw[thick] (t) -- (v);
    \draw[thick] (t) -- (v1);
    \draw[thick] (t) -- (v2);
    \node[vertex,line width=2pt,dashed,red] (s) at ($(t)+(-0.6,-1)$) {$s$};
    \draw[thick] (s) -- (t);
    \node[vertex,line width=2pt,blue] (r) at ($(s)+(-1,0)$) {$r$};
    \draw[thick] (s) -- (r);

    \node[vertex,line width=2pt,dashed,red] (q) at ($(r)+(-1.2,0.5)$) {$q$};
    \node[vertex] (q1) at ($(q)+(-0.7,0.5)$) {$q_a$};
    \node[vertex] (q2) at ($(q)+(0,-0.8)$) {$q_b$};
    \node[vertex] (q3) at ($(q)+(+0.7,0.5)$) {$q_c$};
    \draw[double,thick] (q) -- (q1);
    \draw[double,thick] (q) -- (q2);
    \draw[double,thick] (q) -- (q3);
    \draw[thick] (q1) -- (q2);
    \draw[thick] (q1) -- (q3);
    \draw[thick] (q2) -- (q3);

    \draw[thick] (r) -- (q);
    \draw[thick] (r) -- (q2);
    \draw[thick] (r) -- (q3);

    \node[vertex,line width=2pt,blue] (p) at ($(q)+(0,1.2)$) {$p$};
    \node[vertex,line width=2pt,blue] (o) at ($(q)+(-1.2,-0.5)$) {$o$};
    \draw[thick] (p) -- (q);
    \draw[thick] (p) -- (q1);
    \draw[thick] (p) -- (q3);
    \draw[thick] (o) -- (q);
    \draw[thick] (o) -- (q1);
    \draw[thick] (o) -- (q2);
\end{tikzpicture}}
    \caption{Corresponding intersection graph $G$ with bold blue vertices
    in~$A$ and red dashed ones in~$A^*$.}
    \label{figb:dist_G}
  \end{subfigure}

  \caption{Example of weight distribution for Theorem~\ref{thm:squareimp32}:
  vertex $v$ distributes weight~$1$ to each $y$ and $t$; $s$ distributes $\frac12$
  to each $t$ and $r$; $q$ distributes $\frac23$ to $r$, $p$, and $o$; and $y_u$
  distributes $\frac12+\frac12=1$ to $y$.}
  \label{fig:example_dist}
  \end{figure}
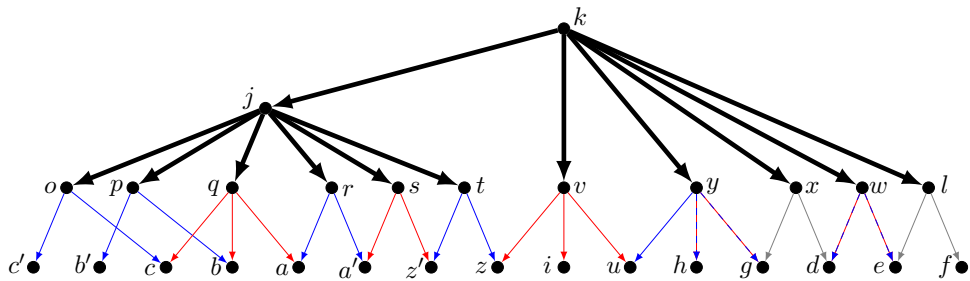
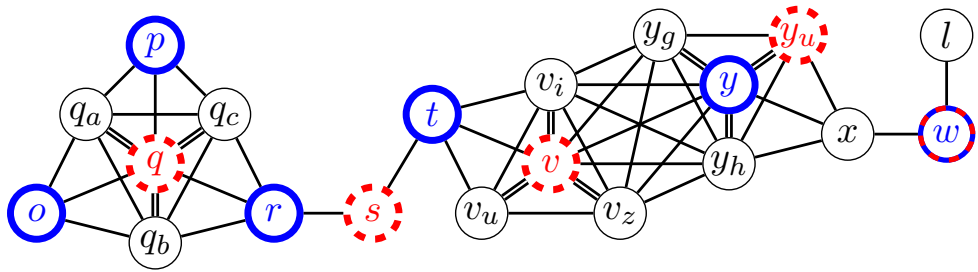

  Let us argue that Configurations~(c) and~(e) in Figure~\ref{fig:weights32}
  cannot happen.  
  The red square vertex in Configuration~(e) would itself improve $w^2(A)$, so
  this configuration does not occur. 
  Now, suppose, for a contradiction, that there is a weight-2 vertex $v \in A^*
  \setminus A$ whose only neighbor in~$A$ is a single neighbor~$u$ of~$v$, as
  in Configuration~(c).  
  By Proposition~\ref{prop:intersectiongraph}(iii), vertex~$u$ must be adjacent
  to exactly two weight-1 vertices in~$K_4^v$.
  But then the third weight-1 vertex in~$K_4^v$ would have no neighbor in~$A$
  by Proposition~\ref{prop:intersectiongraph}(ii), and it would thus
  improve~$w^2(A)$, a contradiction.  
  Hence, Configuration~(c) of Figure~\ref{fig:weights32} also does not occur. 

  Now, let us prove that no vertex in~$A$ gets more than~$3/2$ of its weight. 
  First consider a weight-2 vertex~$u$ in~$A$.
  Such a vertex~$u$ receives weight from~$A^*$ through at most three edges, 
  by Proposition~\ref{prop:intersectiongraph}(i). 
  The value that~$u$ receives through each edge is in $\{0,\frac12,\frac23,1\}$. 
  Thus~$u$ receives at most~3 in total.
  Now consider a weight-1 vertex~$u$ in~$A$.
  Such a vertex~$u$ receives weight from~$A^*$ through at most two edges, by
  Proposition~\ref{prop:intersectiongraph}(i). 
  The value that~$u$ receives through each edge is in $\{0,\frac12,1\}$. 
  The only way to receive in total more than~$3/2$ is by receiving~1 through
  the two edges. 
  These possibilities are summarized in Figure~\ref{fig:weights32receiving},
  and each leads to a claw in~$G$ that improves~$w^2(A)$.
  Indeed, Configuration~(a) in Figure~\ref{fig:weights32receiving} contains
  an improving 1-claw and Configuration~(b) is itself an improving 2-claw. 
  In Configuration~(c), $z$ is a single neighbor of~$v$, so there exists a 
  weight-1 vertex~$x$ in~$K_4^v$ (not depicted) that is not a neighbor of~$z$.
  Vertex~$x$ and the red square weight-1 vertex~$y$ form the set~$T_C$ 
  of an improving 2-claw~$C$ centered at the blue round vertex~$u$. 
  The same argument applied twice shows that there is an improving 2-claw
  if Configuration~(d) occurs.
  Therefore these possibilities cannot occur and every~$u$ in~$A$ receives at
  most~$3/2$ of its weight~$w_u$. 
  
  \begin{figure}[htb]
    \centering
    \resizebox{0.9\textwidth}{!}{\begin{tikzpicture}[
    LabelStyle/.style={inner sep=1pt},
    hl/.style={fill=green!30, circle, minimum size=15pt},
    nn/.style={}
  ]

  \node (h0) [nn] at (0,0) {};
  \node[label=below:1,label=right:$u$](s0) [blue vertex] at ($(h0)$) {};
  \node[label=above:2](r0) [red vertex] at ($(s0)+(0,+2)$) {};
  \Edge[color=red,label={ 1},lw=1.2pt,style={pos=0.5,bend right}](r0)(s0);
  \Edge[color=red,label={ 1},lw=1.2pt,style={pos=0.5,bend left}](r0)(s0);
  \node at ($(h0)+(0,-1)$) {(a)};

  \node (h1) [nn] at ($(h0)+(3.5,0)$) {};
  \node[label=below:1,label=right:$u$](s1) [blue vertex] at ($(h1)$) {};
  \node[label=above:1](r11) [red vertex] at ($(s1)+(-1,+2)$) {};
  \node[label=above:1](r12) [red vertex] at ($(s1)+(+1,+2)$) {};
  \Edge[color=red,label={ 1},lw=1.2pt,style={pos=0.5}](s1)(r11);
  \Edge[color=red,label={ 1},lw=1.2pt,style={pos=0.5}](s1)(r12);
  \node at ($(h1)+(0,-1)$) {(b)};

  \node (h2) [nn] at ($(h1)+(4.5,0)$) {};
  \node[label=below:1,label=right:$u$](s2) [blue vertex] at ($(h2)$) {};
  \node[label=above:2,label=left:$v$](r21) [red vertex] at ($(s2)+(-1,+2)$) {};
  \node[label=above:1,label=right:$y$](r22) [red vertex] at ($(s2)+(+1,+2)$) {};
  \Edge[color=red,label={ 1},lw=1.2pt,style={pos=0.5}](s2)(r21);
  \Edge[color=red,label={ 1},lw=1.2pt,style={pos=0.5}](s2)(r22);

  \node (h21) [nn] at ($(h2)+(-2,0)$) {};
  \node[label=left:$z$] (s22) [blue vertex] at ($(h21)$) {};
  \Edge[color=red,label={ 1},lw=1.2pt,style={pos=0.5}](r21)(s22);
  \node at ($(h2)+(0,-1)$) {(c)};

  \node (h3) [nn] at ($(h2)+(4.5,0)$) {};
  \node[label=below:1,label=right:$u$](s3) [blue vertex] at ($(h3)$) {};
  \node[label=above:2](r31) [red vertex] at ($(s3)+(-1,+2)$) {};
  \node[label=above:2](r32) [red vertex] at ($(s3)+(+1,+2)$) {};
  \Edge[color=red,label={ 1},lw=1.2pt,style={pos=0.5}](s3)(r31);
  \Edge[color=red,label={ 1},lw=1.2pt,style={pos=0.5}](s3)(r32);

  \node (h31) [nn] at ($(h3)+(-2,0)$) {};
  \node (s32) [blue vertex] at ($(h31)$) {};
  \Edge[color=red,label={ 1},lw=1.2pt,style={pos=0.5}](r31)(s32);

  \node (h32) [nn] at ($(h3)+(+2,0)$) {};
  \node (s33) [blue vertex] at ($(h32)$) {};
  \Edge[color=red,label={ 1},lw=1.2pt,style={pos=0.5}](r32)(s33);
  \node at ($(h3)+(0,-1)$) {(d)};
\end{tikzpicture}}
    \caption{Configurations that imply on an improving claw centered at $u$.}
    \label{fig:weights32receiving}
  \end{figure}
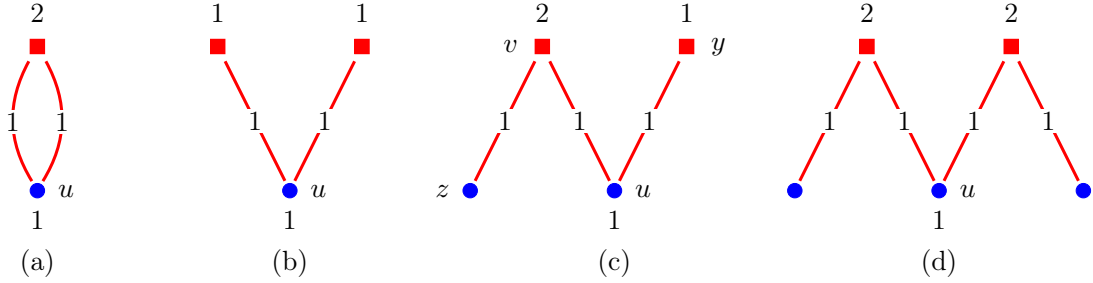

  We point out that this analysis is tight, since a vertex in~$A$ might receive
  exactly~$3/2$ of its weight, as shown in Figure~\ref{fig:tight32}, which
  contains no improving claw.
  Indeed, for these examples, \SquareImp\ might produce an independent set of
  weight~2 although there is one of weight~3. 
  
  \begin{figure}[htb]
    \centering
    \resizebox{0.7\textwidth}{!}{\begin{tikzpicture}
  [LabelStyle/.style={inner sep=1pt}]

  \node[label=below:2](s1) [blue vertex] at (2,2) {};
  \node[label=above:1](r1) [red vertex] at (0,4) {};
  \Edge[color=red,label={ $1$},lw=1.2pt,style={pos=0.5}](r1)(s1);
  \node[label=above:1](r2) [red vertex] at (2,4) {};
  \Edge[color=red,label={ $1$},lw=1.2pt,style={pos=0.5}](r2)(s1);
  \node[label=above:1](r3) [red vertex] at (4,4) {};
  \Edge[color=red,label={ $1$},lw=1.2pt,style={pos=0.5}](r3)(s1);
  \node at ($(r2)+(0,-3)$) {(a)};


  \node[label=above:1](r1) [red vertex] at (7,4) {};
  \node[label=above:1](r2) [red vertex] at (9,4) {};
  \node[label=above:1](r3) [red vertex] at (11,4) {};

  \node[label=below:1](s1) [blue vertex] at (8,2) {};
  \Edge[color=red,label={ $1$},lw=1.2pt,style={pos=0.5}](s1)(r1);
  \Edge[color=red,label={ $\frac12$},lw=1.2pt,style={pos=0.5}](s1)(r2);

  \node[label=below:1](s3) [blue vertex] at (10,2) {};
  \Edge[color=red,label={ $\frac12$},lw=1.2pt,style={pos=0.5}](s3)(r2);
  \Edge[color=red,label={ $1$},lw=1.2pt,style={pos=0.5}](s3)(r3);

  \node at ($(r2)+(0,-3)$) {(b)};

\end{tikzpicture}}
    \caption{Tight examples for the analysis of Theorem~\ref{thm:squareimp32}.}
    \label{fig:tight32}
  \end{figure}
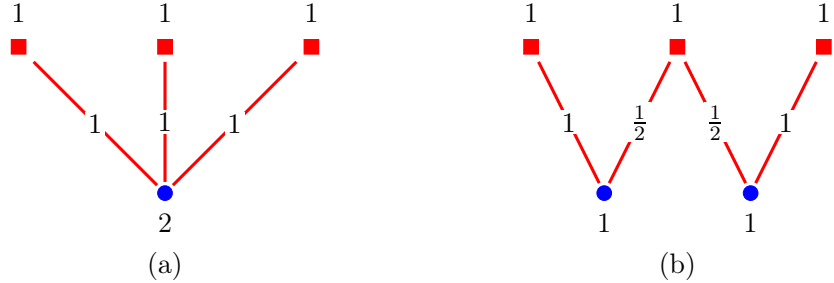

  Finally, let us argue that \Call{SquareImp}{$G$, $w$} runs in polynomial time.
  Let~$n$ be the number of vertices of~$G$. 
  Because~$w^2(A)$ increases in every iteration of \SquareImp\ and all weights
  in~$w$ are~1 or~2, \Call{SquareImp}{$G$, $w$} does at most~$4n$ iterations. 
  In each iteration, one can test in polynomial time all $d$-claws in~$G$, for
  $d \in \{1,2,3\}$, which is enough because~$G$ is 4-claw free.   
\end{proof}

\section{Back to Maximum Leaf Spanning Arborescence}
\label{sec:backtomaxleaves}

In this section we go back to \maxleaves, and derive a new~$3/2$-approximation
for rooted dags from \SquareImp. 

Recall that in \Call{MaxLeaves-12MIS}{} we built the intersection graph~$G$
having as vertex set the set $\Candidates$, which contains vertices of
out-degree~0 with two or three out-neighbors of in-degree~0.
The edges were added to~$G$ according to the sets~$U_v$ defined for each $v \in
\Candidates$, while~$w_v$ was set to~$|U_v|-1$.
Now note that the set $\calS = \{U_v \colon v \in \Candidates\}$, with the same
weight function~$w$, is an instance of the weighted 3D-matching problem.
In the weighted 3D-matching, one wants to find a collection $\calS' \subseteq
\calS$ of pairwise disjoint sets as heavy as possible.
One can see that any optimal solution for the weighted 3D-matching on
$(\calS,w)$ provides an optimal solution for the wMIS on $(G,w)$ and
vice-versa.
In~\cite{FernandesL2020}, we presented a theorem using the weighted 3D-matching
nomenclature whose proof can be adapted to stablish the following.
We present the proof of this theorem in Appendix~\ref{proofsDAM}.
(See~\cite[Theorem~4.3]{FernandesL2021} for details.)

\begin{theorem}
\label{thm:weighted_approx}
  If $\calA$ is an $\alpha$-approximation algorithm for the wMIS on weighted
  $\{2,3\}$-intersection graphs, then algorithm \Call{MaxLeaves-12MIS}{} using
  $\calA$ in Line~\ref{alg:call_squareimp} instead of \SquareImp\ is a
  $\max\{\frac43,\alpha\}$-approximation for \maxleaves\ on rooted directed
  acyclic graphs.
\end{theorem}

Applying Theorem~\ref{thm:weighted_approx} with \SquareImp\ as algorithm
$\mathcal{A}$, and using Theorem~\ref{thm:squareimp32}, we derive the following
result. 

\begin{corollary}\label{cor:32}
  Algorithm \Call{MaxLeaves-12MIS}{} is a $3/2$-approximation for the
  \maxleaves\ on rooted directed acyclic graphs.
\end{corollary}

\newcommand{\SquarePImp}{\textsc{Square$^+$Imp}}
\newcommand{\AuxiliaryGraph}{\textsc{AuxiliaryGraph}}
\newcommand{\AugmentingPath}{\textsc{AugmentingPath}}
\newcommand{\NULL}{\textsc{Null}}

\section{Boosting \SquareImp\ with maximum matchings}
\label{sec:better_approx}

In this section, we modify \SquareImp{} to achieve a ratio for wMIS better
than~3/2 on weighted $\{2,3\}$-intersection graphs. 
The improvement depends on two ideas, each coming from one of the tight
examples in Figure~\ref{fig:tight32}. 

The example on Figure~\ref{fig:tight32}(a) is a claw itself, but it is not
improving with the weights we assigned to the vertices of~$G$.  
However, it ends up being a good exchange for both wMIS and \maxleaves{} problems.  
Hence we decided to adapt \SquareImp, defining~$w_+^2(A)$ to be the sum of
$(w_v+1)^2$ for every~$v$ in~$A$, and applying the exchange whenever~$w_+^2(A)$
increases. 
This has the effect of doing all the previous improvements and also the one in
Figure~\ref{fig:tight32}(a).

The example on Figure~\ref{fig:tight32}(b) can be generalized into a longer
path on weight-1 vertices, alternating between red square and blue round
vertices.
So in fact it is a class of examples, and it does not contain an improving
claw, but it indicates a way to increase the independent set by one: exchange
the blue round vertices by the red square vertices in the path.  
In analogy to matching, we refer to such an exchange as an augmenting path
improvement.

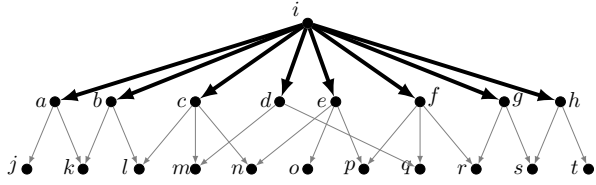
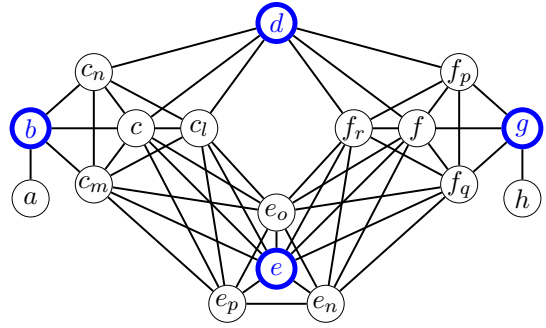
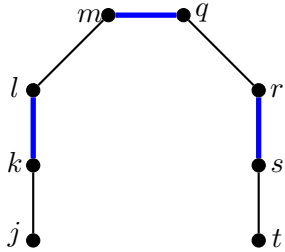
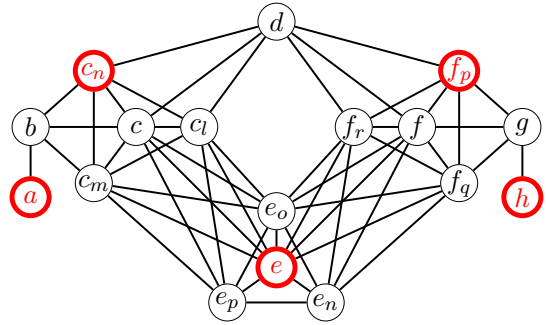
\begin{figure}[htb]
    \centering
    \begin{subfigure}{0.49\textwidth}
        \centering
        \resizebox{\textwidth}{!}{\begin{tikzpicture}
    [vertex/.style={circle,draw,fill,inner sep=0pt,minimum size=5pt}]


    \def\h{10mm};

    \node (P) at (0,0) {};

    \node[vertex,label={[xshift=-7pt,yshift=-9pt]$j$}] (v0) at ($(P)+(0*\h,0)$) {};
    \node[vertex,label={[xshift=-7pt,yshift=-9pt]$k$}] (v1) at ($(P)+(1*\h,0)$) {};
    \node[vertex,label={[xshift=-7pt,yshift=-9pt]$l$}] (v2) at ($(P)+(2*\h,0)$) {};
    \node[vertex,label={[xshift=-7pt,yshift=-9pt]$m$}] (v3) at ($(P)+(3*\h,0)$) {};
    \node[vertex,label={[xshift=-7pt,yshift=-9pt]$n$}]  (x) at ($(P)+(4*\h,0)$) {};
    \node[vertex,label={[xshift=-7pt,yshift=-9pt]$o$}]  (y) at ($(P)+(5*\h,0)$) {};
    \node[vertex,label={[xshift=-7pt,yshift=-9pt]$p$}]  (z) at ($(P)+(6*\h,0)$) {};
    \node[vertex,label={[xshift=-7pt,yshift=-9pt]$q$}] (v4) at ($(P)+(7*\h,0)$) {};
    \node[vertex,label={[xshift=-7pt,yshift=-9pt]$r$}] (v5) at ($(P)+(8*\h,0)$) {};
    \node[vertex,label={[xshift=-7pt,yshift=-9pt]$s$}] (v6) at ($(P)+(9*\h,0)$) {};
    \node[vertex,label={[xshift=-7pt,yshift=-9pt]$t$}] (v7) at ($(P)+(10*\h,0)$) {};

    \node[vertex,label={[xshift=-7pt,yshift=-9pt]$a$}] (a) at ($(v0)+(\h/2,1.2)$) {};
    \node[vertex,label={[xshift=-7pt,yshift=-9pt]$b$}] (b) at ($(v1)+(\h/2,1.2)$) {};
    \node[vertex,label={[xshift=-7pt,yshift=-9pt]$c$}] (c) at ($(v3)+(0,1.2)$) {};
    \node[vertex,label={[xshift=-7pt,yshift=-9pt]$d$}] (d) at ($(x)+(\h/2,1.2)$) {};
    \node[vertex,label={[xshift=-7pt,yshift=-9pt]$e$}] (e) at ($(y)+(\h/2,1.2)$) {};
    \node[vertex,label={[xshift=7pt,yshift=-9pt]$f$}] (f) at ($(v4)+(0,1.2)$) {};
    \node[vertex,label={[xshift=7pt,yshift=-9pt]$g$}] (g) at ($(v5)+(\h/2,1.2)$) {};
    \node[vertex,label={[xshift=7pt,yshift=-9pt]$h$}] (h) at ($(v6)+(\h/2,1.2)$) {};

    \node[vertex,label={[xshift=-6pt,yshift=-3pt]$i$}] (i) at ($(y)+(0,2.6)$) {};

    \draw[-latex,line width=2pt] (i) -- (a);
    \draw[-latex,line width=2pt] (i) -- (b);
    \draw[-latex,line width=2pt] (i) -- (c);
    \draw[-latex,line width=2pt] (i) -- (d);
    \draw[-latex,line width=2pt] (i) -- (e);
    \draw[-latex,line width=2pt] (i) -- (f);
    \draw[-latex,line width=2pt] (i) -- (g);
    \draw[-latex,line width=2pt] (i) -- (h);

    \draw[-latex,gray] (a) -- (v0);
    \draw[-latex,gray] (a) -- (v1);
    \draw[-latex,gray] (b) -- (v1);
    \draw[-latex,gray] (b) -- (v2);
    \draw[-latex,gray] (c) -- (v2);
    \draw[-latex,gray] (c) -- (v3);
    \draw[-latex,gray] (c) -- (x);
    \draw[-latex,gray] (d) -- (v3);
    \draw[-latex,gray] (d) -- (v4);
    \draw[-latex,gray] (e) -- (x);
    \draw[-latex,gray] (e) -- (y);
    \draw[-latex,gray] (e) -- (z);
    \draw[-latex,gray] (f) -- (z);
    \draw[-latex,gray] (f) -- (v4);
    \draw[-latex,gray] (f) -- (v5);
    \draw[-latex,gray] (g) -- (v5);
    \draw[-latex,gray] (g) -- (v6);
    \draw[-latex,gray] (h) -- (v6);
    \draw[-latex,gray] (h) -- (v7);

\end{tikzpicture}}
        \caption{4-branching $F_1$ in bold.}
        \label{figa:f1_bad}
    \end{subfigure}
    \hfill
    \begin{subfigure}{0.49\textwidth}
        \centering
        \resizebox{0.9\textwidth}{!}{\begin{tikzpicture}
    [vertex/.style={circle,draw,inner sep=0pt,minimum size=5pt,minimum width=15pt},
    vertex2/.style={draw=line width=2pt,blue,double=red,circle,inner sep=0pt,minimum size=5pt,minimum width=15pt}]

    \node[vertex] (c) at (-2,2) {$c$};
    \node[vertex] (c1) at ($(c)+(-0.6,+0.8)$) {$c_n$};
    \node[vertex] (c2) at ($(c)+(-0.6,-0.8)$) {$c_m$};
    \node[vertex] (c3) at ($(c)+(+0.9,0)$) {$c_l$};
    \draw[thick] (c) -- (c1);
    \draw[thick] (c) -- (c2);
    \draw[thick] (c) -- (c3);
    \draw[thick] (c1) -- (c2);
    \draw[thick] (c1) -- (c3);
    \draw[thick] (c2) -- (c3);

    \node[vertex] (f) at (2,2) {$f$};
    \node[vertex] (f1) at ($(f)+(+0.6,+0.8)$) {$f_p$};
    \node[vertex] (f2) at ($(f)+(-0.9,0)$) {$f_r$};
    \node[vertex] (f3) at ($(f)+(+0.6,-0.8)$) {$f_q$};
    \draw[thick] (f) -- (f1);
    \draw[thick] (f) -- (f2);
    \draw[thick] (f) -- (f3);
    \draw[thick] (f1) -- (f2);
    \draw[thick] (f1) -- (f3);
    \draw[thick] (f2) -- (f3);

    \node[vertex,line width=2pt,blue] (e) at (0,0) {$e$};
    \node[vertex] (e1) at ($(e)+(-0.7,-0.5)$) {$e_p$};
    \node[vertex] (e2) at ($(e)+(0,+0.8)$) {$e_o$};
    \node[vertex] (e3) at ($(e)+(+0.7,-0.5)$) {$e_n$};
    \draw[thick] (e) -- (e1);
    \draw[thick] (e) -- (e2);
    \draw[thick] (e) -- (e3);
    \draw[thick] (e1) -- (e2);
    \draw[thick] (e1) -- (e3);
    \draw[thick] (e2) -- (e3);

    \node[vertex,line width=2pt,blue] (b) at ($(c)+(-1.5,0)$) {$b$};
    \node[vertex] (a) at ($(b)+(0,-1)$) {$a$};
    \node[vertex,line width=2pt,blue] (d) at (0,3.5) {$d$};
    \node[vertex,line width=2pt,blue] (g) at ($(f)+(1.5,0)$) {$g$};
    \node[vertex] (h) at ($(g)+(0,-1)$) {$h$};

    \draw[thick] (a) -- (b);
    \draw[thick] (b) -- (c);
    \draw[thick] (b) -- (c1);
    \draw[thick] (b) -- (c2);

    \draw[thick] (h) -- (g);
    \draw[thick] (g) -- (f);
    \draw[thick] (g) -- (f1);
    \draw[thick] (g) -- (f3);

    \draw[thick] (d) -- (c1);
    \draw[thick] (d) -- (c3);
    \draw[thick] (d) -- (c);
    \draw[thick] (d) -- (f);
    \draw[thick] (d) -- (f2);
    \draw[thick] (d) -- (f1);

    \draw[thick] (c) -- (e);
    \draw[thick] (c2) -- (e);
    \draw[thick] (c3) -- (e);
    \draw[thick] (c) -- (e1);
    \draw[thick] (c2) -- (e1);
    \draw[thick] (c3) -- (e1);
    \draw[thick] (c) -- (e2);
    \draw[thick] (c2) -- (e2);
    \draw[thick] (c3) -- (e2);

    \draw[thick] (f) -- (e);
    \draw[thick] (f2) -- (e);
    \draw[thick] (f3) -- (e);
    \draw[thick] (f) -- (e3);
    \draw[thick] (f2) -- (e3);
    \draw[thick] (f3) -- (e3);
    \draw[thick] (f) -- (e2);
    \draw[thick] (f2) -- (e2);
    \draw[thick] (f3) -- (e2);


\end{tikzpicture}}
        \caption{Intersection graph $G$ and independent set $A$ in bold blue.}
        \label{figb:intersection_bad}
    \end{subfigure}

    \begin{subfigure}{0.49\textwidth}
        \centering
        \resizebox{0.5\textwidth}{!}{\begin{tikzpicture}
    [vertex/.style={circle,draw,fill,inner sep=0pt,minimum size=5pt}]

    \node (X) at (0,0) {};
    \node[vertex,label={[xshift=-7pt,yshift=-9pt]$m$}] (v3) at ($(X)+(-0.5,0)$) {};
    \node[vertex,label={[xshift=7pt,yshift=-9pt]$q$}] (v4) at ($(X)+(+0.5,0)$) {};

    \node[vertex,label={[xshift=-7pt,yshift=-9pt]$l$}] (v2) at ($(X)+(-1.5,-1)$) {};
    \node[vertex,label={[xshift=-7pt,yshift=-9pt]$k$}] (v1) at ($(X)+(-1.5,-2)$) {};
    \node[vertex,label={[xshift=-7pt,yshift=-9pt]$j$}] (v0) at ($(X)+(-1.5,-3)$) {};

    \node[vertex,label={[xshift=7pt,yshift=-9pt]$r$}] (v5) at ($(X)+(1.5,-1)$) {};
    \node[vertex,label={[xshift=7pt,yshift=-9pt]$s$}] (v6) at ($(X)+(1.5,-2)$) {};
    \node[vertex,label={[xshift=7pt,yshift=-9pt]$t$}] (v7) at ($(X)+(1.5,-3)$) {};

    \draw[thick] (v0) -- (v1);
    \draw[line width=2pt,blue] (v1) -- (v2);
    \draw[thick] (v2) -- (v3);
    \draw[line width=2pt,blue] (v3) -- (v4);
    \draw[thick] (v4) -- (v5);
    \draw[line width=2pt,blue] (v5) -- (v6);
    \draw[thick] (v6) -- (v7);


\end{tikzpicture}}
        \caption{Graph~$H$ built from $G$ and $A$, with matching $M$ in bold blue.}
        \label{figc:graphH}
    \end{subfigure}
    \hfill
    \begin{subfigure}{0.49\textwidth}
        \centering
        \resizebox{0.9\textwidth}{!}{\begin{tikzpicture}
    [vertex/.style={circle,draw,inner sep=0pt,minimum size=5pt,minimum width=15pt},
    vertex2/.style={draw=blue,double=line width=2pt,red,circle,inner sep=0pt,minimum size=5pt,minimum width=15pt}]

    \node[vertex] (c) at (-2,2) {$c$};
    \node[vertex,line width=2pt,red] (c1) at ($(c)+(-0.6,+0.8)$) {$c_n$};
    \node[vertex] (c2) at ($(c)+(-0.6,-0.8)$) {$c_m$};
    \node[vertex] (c3) at ($(c)+(+0.9,0)$) {$c_l$};
    \draw[thick] (c) -- (c1);
    \draw[thick] (c) -- (c2);
    \draw[thick] (c) -- (c3);
    \draw[thick] (c1) -- (c2);
    \draw[thick] (c1) -- (c3);
    \draw[thick] (c2) -- (c3);

    \node[vertex] (f) at (2,2) {$f$};
    \node[vertex,line width=2pt,red] (f1) at ($(f)+(+0.6,+0.8)$) {$f_p$};
    \node[vertex] (f2) at ($(f)+(-0.9,0)$) {$f_r$};
    \node[vertex] (f3) at ($(f)+(+0.6,-0.8)$) {$f_q$};
    \draw[thick] (f) -- (f1);
    \draw[thick] (f) -- (f2);
    \draw[thick] (f) -- (f3);
    \draw[thick] (f1) -- (f2);
    \draw[thick] (f1) -- (f3);
    \draw[thick] (f2) -- (f3);

    \node[vertex,line width=2pt,red] (e) at (0,0) {$e$};
    \node[vertex] (e1) at ($(e)+(-0.7,-0.5)$) {$e_p$};
    \node[vertex] (e2) at ($(e)+(0,+0.8)$) {$e_o$};
    \node[vertex] (e3) at ($(e)+(+0.7,-0.5)$) {$e_n$};
    \draw[thick] (e) -- (e1);
    \draw[thick] (e) -- (e2);
    \draw[thick] (e) -- (e3);
    \draw[thick] (e1) -- (e2);
    \draw[thick] (e1) -- (e3);
    \draw[thick] (e2) -- (e3);

    \node[vertex] (b) at ($(c)+(-1.5,0)$) {$b$};
    \node[vertex,line width=2pt,red] (a) at ($(b)+(0,-1)$) {$a$};
    \node[vertex] (d) at (0,3.5) {$d$};
    \node[vertex] (g) at ($(f)+(1.5,0)$) {$g$};
    \node[vertex,line width=2pt,red] (h) at ($(g)+(0,-1)$) {$h$};

    \draw[thick] (a) -- (b);
    \draw[thick] (b) -- (c);
    \draw[thick] (b) -- (c1);
    \draw[thick] (b) -- (c2);

    \draw[thick] (h) -- (g);
    \draw[thick] (g) -- (f);
    \draw[thick] (g) -- (f1);
    \draw[thick] (g) -- (f3);

    \draw[thick] (d) -- (c1);
    \draw[thick] (d) -- (c3);
    \draw[thick] (d) -- (c);
    \draw[thick] (d) -- (f);
    \draw[thick] (d) -- (f2);
    \draw[thick] (d) -- (f1);

    \draw[thick] (c) -- (e);
    \draw[thick] (c2) -- (e);
    \draw[thick] (c3) -- (e);
    \draw[thick] (c) -- (e1);
    \draw[thick] (c2) -- (e1);
    \draw[thick] (c3) -- (e1);
    \draw[thick] (c) -- (e2);
    \draw[thick] (c2) -- (e2);
    \draw[thick] (c3) -- (e2);

    \draw[thick] (f) -- (e);
    \draw[thick] (f2) -- (e);
    \draw[thick] (f3) -- (e);
    \draw[thick] (f) -- (e3);
    \draw[thick] (f2) -- (e3);
    \draw[thick] (f3) -- (e3);
    \draw[thick] (f) -- (e2);
    \draw[thick] (f2) -- (e2);
    \draw[thick] (f3) -- (e2);


\end{tikzpicture}}
        \caption{Set $A' = A \oplus P$ in bold red.}
        \label{figd:aug_path_imp}
    \end{subfigure}
  \caption{The independent set $A=\{b,d,e,g\}$ has no improving claw. 
  Path $\langle j,k,l,m,q,r,s,t \rangle$ is an augmenting path in~$H$.  
  The independent set ${A' = \{a,c_n,f_p,h,e\}}$, obtained from the augmenting 
  path improvement, has an improving claw where $T_C = \{c,f\}$.}
  \label{fig:badpath}
\end{figure}

The idea to search for such improvements is inspired on algorithm
\Call{MaxExpand}{} from~\cite{FernandesL2020}.
Let~$(G,w)$ be a weighted $\{2,3\}$-intersection graph where~$G$ is the
intersection graph for the pair~$(V,U)$.
Let~$A$ be an independent set in~$G$. 
Consider an auxiliary graph~$H$ as follows.
Let~$X$ be the union of~$U_v$ for all weight-2 vertices in~$A$. 
We think of the elements of~$X$ as forbidden.
The vertex set of~$H$ is the union of~$U_v$ for all weight-1 vertices of~$G$
such that~$U_v$ has no forbidden element, that is, $U_v$ does not intersect
with~$X$.
There is an edge between two vertices~$x$ and~$y$ of~$H$ if there is a
vertex~$v$ in~$G$ with $U_v=\{x,y\}$. 
So edges of~$H$ are associated to weight-1 vertices from~$G$.
Let~$M$ be the set of edges in~$H$ corresponding to weight-1 vertices in~$A$. 
Note that~$M$ is a matching in~$H$, and that~$H$ and~$M$ can be obtained in
polynomial time from~$G$ and~$A$.
See Figure~\ref{fig:badpath} for an example.

We recall the matching nomenclature.
A vertex in~$H$ is \emph{$M$-covered} if it is the end of an edge in~$M$, and is
\emph{$M$-uncovered} otherwise.
An \emph{$M$-alternating path} in~$H$ is one which alternates edges not in~$M$
with edges from~$M$.
An \emph{$M$-augmenting path} in~$H$ is an $M$-alternating path that starts and
ends at $M$-uncovered vertices. 

Given an $M$-augmenting path~$P$ in~$H$, we denote by $A \oplus P$ the
independent set obtained from~$A$ by removing all vertices of~$A$ associated to
edges in ${E(P) \cap M}$ and including all vertices from~$G$ associated to edges
in ${E(P) \setminus M}$.
Note that $A \oplus P$ is larger than~$A$, and in fact heavier than~$A$, as we
only exchanged weight-1 vertices.
We call such change on~$A$ an \textit{augmenting path improvement}.

We adapt \SquareImp\ in the following way. 
After a claw improvement phase is complete, we check whether there exists an
augmenting path improvement.
If such an improvement exists, we perform it and go back to the claw
improvement phase.  
We repeat this until no augmenting path improvement exists.  
We call \SquarePImp\ the resulting algorithm, which is presented in
Algorithm~\ref{alg:22}.
The routine \AugmentingPath$(H, M)$ returns an $M$-augmenting path in~$H$, if
one exists, and $\NULL$ otherwise.
There are polynomial-time algorithms for this in the
literature~\cite{Edmonds1965}.

\begin{algorithm}
  \begin{algorithmic}[1]
    \Require{{\small weighted $\{2,3\}$-intersection graph $(G,w)$}}
    \Ensure{{\small an independent set in $G$}}

    \State let $V$ and $U$ be such that $G=\Call{IntersectionGraph}{V, U}$
    \State $A \gets \emptyset$

    \Repeat

        \While{there is a claw $C$ in $G$ such that $T_C$ improves $w^2_+(A)$} 
            \State $A \gets (A \cup T_C) \setminus N(T_C)$
        \EndWhile

        \um

        \State $X \gets \bigcup\{U_v : \mbox{$v \in A$ and $w_v = 2$}\}$
        \State $V' \gets \bigcup\{U_v : \mbox{$v \in V$, $w_v = 1$, and $U_v \cap X = \emptyset$}\}$
        \State $E' \gets \{xy : \mbox{there is a vertex $v \in V$ such that $U_v=\{x,y\} \subseteq V'$}\}$
        \State let $H$ be the graph $(V',E')$ \label{alg:graphH}

        \um

        \State $M \gets \{xy \in E(H) : \mbox{there is a vertex $v \in A$ such that $U_v=\{x,y\}$}\}$
        \State $P \gets \AugmentingPath(H, M)$

        \If{$P \neq \NULL$} 
            \State $A \gets A \oplus P$
        \EndIf
    
    \Until{$P = \NULL$}

    \State \Return $A$
  \end{algorithmic}
  \caption{\SquarePImp($G$, $w$)}
  \label{alg:22}
\end{algorithm}

The same argument that we used on \Call{SquareImp}{$G$, $w$} assures that the
\Call{Square$^+$Imp}{$G$, $w$} runs in polynomial time. 
Indeed, let~$n$ be the number of vertices of~$G$. 
Because $w^2_+(A)$ starts from zero and increases by a positive integer value
in every iteration of both loops and $w^2_+(A) \leq 9n$,
algorithm \Call{Square$^+$Imp}{$G$, $w$} does at most a linear number of iterations
in~$n$.

The example in Figure~\ref{fig:badpath} shows that, after an augmenting path
improvement, there might be feasible claw improvements to be done. 

The proof of the next result partially mimics the proof of
Theorem~\ref{thm:squareimp32}, using a different weight distribution rule.
In one of the cases, however, it bounds the \emph{average} of the receiving
weights, instead of the maximum receiving weight per vertex.

\begin{theorem}\label{thm:squareimp75}
  \SquarePImp\ is a~$\frac75$-approximation for wMIS on weighted
  $\{2,3\}$-intersection graphs.
\end{theorem}
\begin{proof}
  Let $(G,w)$ be a weighted $\{2,3\}$-intersection graph. 
  Let~$A^*$ be an independent set in~$G$ that maximizes~$w(A^*)$ and let~$A$ be
  the independent set produced by \Call{Square$^+$Imp}{$G$,$w$}.
  We shall prove that $w(A^*) \leq \frac75\,w(A)$ by distributing differently
  the weight on vertices in~$A^*$ among their neighbors in~$A$ so that no
  weight-2 vertex in~$A$ gets more than~$14/5$ and weight-1 vertices in~$A$
  get, on average, no more than~$7/5$.

  At the end of \SquarePImp, no claw is improving for~$A$. 
  For the sake of the argument, again we consider~$G$ as the corresponding
  intersection multigraph.
  Vertices in $A^* \cap A$ keep their weights.  
  Because~$A$ is maximal, every vertex in $A^* \setminus A$ has at least one
  neighbor in~$A$.  
  By Proposition~\ref{prop:intersectiongraph}(i), every vertex in~$A^*$ of
  weight~$z$ has at most~$z+1$ edges going to its neighbors in~$A$.

  In Figure~\ref{fig:weights75}, we show how each vertex in $A^* \setminus A$
  distributes its weight to its neighbors in~$A$.  
  As in Figure~\ref{fig:weights32}, we represent vertices in~$A^*$ by red
  squares and vertices in~$A$ by blue circles.
  The number on top of each vertex in~$A^*$ is its weight.  
  The number below a vertex in~$A$ denotes its weight when that weight matters
  for the distribution.
  The number next to each edge is the amount transferred from the vertex
  in~$A^*$ to the vertex in~$A$.

  The same argument used in Theorem~\ref{thm:squareimp32} shows that 
  Configuration~(e) of Figure~\ref{fig:weights75} does not occur.

  \begin{figure}[htb]
    \centering
    \resizebox{0.95\textwidth}{!}{\begin{tikzpicture}
  [LabelStyle/.style={inner sep=1pt}]

  \node[label=above:1](r0) [red vertex] at (0,4) {};
  \node(f10) [blue vertex] at ($(r0)+(0,-2)$) {};
  \Edge[color=red,label={ 1},lw=1.2pt,style={pos=0.5}](r0)(f10);
  \node at ($(r0)+(0,-3)$) {(a)};

  \node[label=above:1](r1) [red vertex] at ($(r0)+(1.2,0)$) {};
  \node[label=below:1](f11) [blue vertex] at ($(r1)+(-0.4,-2)$) {};
  \node[label=below:1](f31) [blue vertex] at ($(r1)+(+0.4,-2)$) {};
  \Edge[color=red,label={ $\frac12$},lw=1.2pt,style={pos=0.5}](r1)(f11);
  \Edge[color=red,label={ $\frac12$},lw=1.2pt,style={pos=0.5}](r1)(f31);
  \node at ($(r1)+(0,-3)$) {(b)};

  \node[label=above:1](r12) [red vertex] at ($(r1)+(1.6,0)$) {};
  \node[label=below:2](f11) [blue vertex] at ($(r12)+(-0.4,-2)$) {};
  \node[label=below:2](f31) [blue vertex] at ($(r12)+(+0.4,-2)$) {};
  \Edge[color=red,label={ $\frac12$},lw=1.2pt,style={pos=0.5}](r12)(f11);
  \Edge[color=red,label={ $\frac12$},lw=1.2pt,style={pos=0.5}](r12)(f31);
  \node at ($(r12)+(0,-3)$) {(c)};

  \node[label=above:1](r15) [red vertex] at ($(r12)+(1.6,0)$) {};
  \node[label=below:1](f11) [blue vertex] at ($(r15)+(-0.4,-2)$) {};
  \node[label=below:2](f31) [blue vertex] at ($(r15)+(+0.4,-2)$) {};
  \Edge[color=red,label={ $\frac25$},lw=1.2pt,style={pos=0.5}](r15)(f11);
  \Edge[color=red,label={ $\frac35$},lw=1.2pt,style={pos=0.5}](r15)(f31);
  \node at ($(r15)+(0,-3)$) {(d)};

  \node[label=above:2](r2) [red vertex] at ($(r15)+(1.2,0)$) {};
  \node(f12) [blue vertex] at ($(r2)+(0,-2)$) {};
  \Edge[color=red,label={ 2},lw=1.2pt,style={pos=0.5}](r2)(f12);
  \node at ($(r2)+(0,-3)$) {(e)};

  \node[label=above:2](r3) [red vertex] at ($(r2)+(1.2,0)$) {};
  \node(f13) [blue vertex] at ($(r3)+(-0.4,-2)$) {};
  \node(f33) [blue vertex] at ($(r3)+(+0.4,-2)$) {};
  \Edge[color=red,label={ 1},lw=1.2pt,style={pos=0.5}](r3)(f13);
  \Edge[color=red,label={ 1},lw=1.2pt,style={pos=0.5}](r3)(f33);
  \node at ($(r3)+(0,-3)$) {(f)};

  \node[label=above:2](r4) [red vertex] at ($(r3)+(1.8,0)$) {};
  \node[label=below:1](f14) [blue vertex] at ($(r4)+(-0.7,-2)$) {};
  \node[label=below:1](f24) [blue vertex] at ($(r4)+(0,-2)$) {};
  \node[label=below:1](f34) [blue vertex] at ($(r4)+(+0.7,-2)$) {};
  \Edge[color=red,label={ $\frac23$},lw=1.2pt,style={pos=0.5}](r4)(f14);
  \Edge[color=red,label={ $\frac23$},lw=1.2pt,style={pos=0.5}](r4)(f24);
  \Edge[color=red,label={ $\frac23$},lw=1.2pt,style={pos=0.5}](r4)(f34);
  \node at ($(r4)+(0,-3)$) {(g)};

  \node[label=above:2](r5) [red vertex] at ($(r4)+(2.1,0)$) {};
  \node[label=below:2](f15) [blue vertex] at ($(r5)+(-0.7,-2)$) {};
  \node[label=below:2](f25) [blue vertex] at ($(r5)+(0,-2)$) {};
  \node[label=below:2](f35) [blue vertex] at ($(r5)+(0.7,-2)$) {};
  \Edge[color=red,label={ $\frac23$},lw=1.2pt,style={pos=0.5}](r5)(f15);
  \Edge[color=red,label={ $\frac23$},lw=1.2pt,style={pos=0.5}](r5)(f25);
  \Edge[color=red,label={ $\frac23$},lw=1.2pt,style={pos=0.5}](r5)(f35);
  \node at ($(r5)+(0,-3)$) {(h)};

  \node[label=above:2](r6) [red vertex] at ($(r5)+(2.1,0)$) {};
  \node[label=below:1](f16) [blue vertex] at ($(r6)+(-0.7,-2)$) {};
  \node[label=below:1](f26) [blue vertex] at ($(r6)+(0,-2)$) {};
  \node[label=below:2](f36) [blue vertex] at ($(r6)+(+0.7,-2)$) {};
  \Edge[color=red,label={ $\frac23$},lw=1.2pt,style={pos=0.5}](r6)(f16);
  \Edge[color=red,label={ $\frac23$},lw=1.2pt,style={pos=0.5}](r6)(f26);
  \Edge[color=red,label={ $\frac23$},lw=1.2pt,style={pos=0.5}](r6)(f36);
  \node at ($(r6)+(0,-3)$) {(i)};

  \node[label=above:2](r7) [red vertex] at ($(r6)+(2.1,0)$) {};
  \node[label=below:1](f17) [blue vertex] at ($(r7)+(-0.7,-2)$) {};
  \node[label=below:2](f27) [blue vertex] at ($(r7)+(0,-2)$) {};
  \node[label=below:2](f37) [blue vertex] at ($(r7)+(+0.7,-2)$) {};
  \Edge[color=red,label={ $\frac25$},lw=1.2pt,style={pos=0.5}](r7)(f17);
  \Edge[color=red,label={ $\frac45$},lw=1.2pt,style={pos=0.5}](r7)(f27);
  \Edge[color=red,label={ $\frac45$},lw=1.2pt,style={pos=0.5}](r7)(f37);
  \node at ($(r7)+(0,-3)$) {(j)};

\end{tikzpicture}}
    \caption{Weight distribution for Theorem~\ref{thm:squareimp75}.}
    \label{fig:weights75}
  \end{figure}
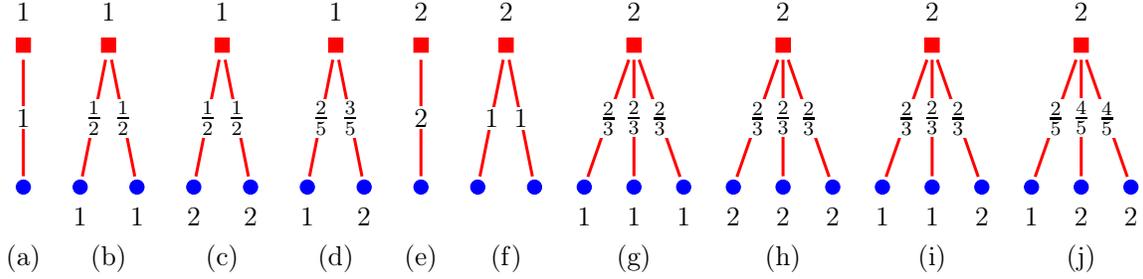

  First, let us prove that no weight-2 vertex~$u$ in~$A$ gets more than~$14/5$.
  By Proposition~\ref{prop:intersectiongraph}(i), such a vertex~$u$ receives
  weight from $A^*$ through at most three edges.
  The value that~$u$ receives through each edge is in
  $\{0,\frac12,\frac35,\frac23,\frac45,1\}$. 
  Thus, the only way to receive in total more than~$14/5$ is by receiving~1
  through the three edges. 
  These possibilities are summarized in Figure~\ref{fig:75w2receiving} and, for
  each, there exists an improving claw, which contradicts the fact that~$A$ is
  the output of \SquarePImp.
  Indeed, Configuration~(a) in Figure~\ref{fig:75w2receiving} is an improving
  claw itself.  
  In Configuration~(b), call~$v$ the weight-2 red square vertex and~$u'$ the
  blue round vertex other than~$u$. 
  Observe that~$u'$ is a single neighbor of~$v$.
  Because~$G$ is a weighted $\{2,3\}$-intersection graph, there is a weight-1
  vertex~$y$ in~$K_4^v$ such that $N(y) \subseteq N(v) \setminus \{u'\}$.
  The two weight-1 red square vertices and~$y$ form a 3-claw with~$u$ that
  improves $w_+^2(A)$. 
  The same argument can be used to derive an improving 3-claw from
  Configurations~(c) and~(d) in Figure~\ref{fig:75w2receiving}.
  So these possibilities cannot occur and every weight-2 vertex~$u$ in~$A$
  receives at most $1+1+\frac45 = \frac{14}{5}$. 

  \begin{figure}[htb]
    \centering
    \resizebox{0.9\textwidth}{!}{\begin{tikzpicture}
  [LabelStyle/.style={inner sep=1pt}]

  \node[label=below:2](s1) [blue vertex] at (1,2) {};
  \node[label=above:1](r1) [red vertex] at (0,4) {};
  \Edge[color=red,label={1},lw=1.2pt,style={pos=0.45}](r1)(s1);
  \node[label=above:1](r2) [red vertex] at (1,4) {};
  \Edge[color=red,label={1},lw=1.2pt,style={pos=0.45}](r2)(s1);
  \node[label=above:1](r3) [red vertex] at (2,4) {};
  \Edge[color=red,label={1},lw=1.2pt,style={pos=0.45}](r3)(s1);
  \node at ($(s1)+(0,-1)$) {(a)};

  \node[label=below:2](s1) [blue vertex] at (4,2) {};
  \node at ($(s1)+(0.3,0)$) {$u$};
  \node[label=above:2](r1) [red vertex] at (3,4) {};
  \node at ($(r1)+(-0.3,0)$) {$v$};
  \node(b1) [blue vertex] at (3,2) {};
  \node at ($(b1)+(-0.3,0)$) {$u'$};
  \Edge[color=red,label={1},lw=1.2pt,style={pos=0.45}](r1)(s1);
  \Edge[color=red,label={1},lw=1.2pt,style={pos=0.45}](r1)(b1);
  \node[label=above:1](r2) [red vertex] at (4,4) {};
  \Edge[color=red,label={1},lw=1.2pt,style={pos=0.45}](r2)(s1);
  \node[label=above:1](r3) [red vertex] at (5,4) {};
  \Edge[color=red,label={1},lw=1.2pt,style={pos=0.45}](r3)(s1);
  \node at ($(s1)+(0,-1)$) {(b)};

  \node[label=below:2](s1) [blue vertex] at (7,2) {};
  \node[label=above:2](r1) [red vertex] at (6,4) {};
  \node(b1) [blue vertex] at (6,2) {};
  \Edge[color=red,label={1},lw=1.2pt,style={pos=0.45}](r1)(s1);
  \Edge[color=red,label={1},lw=1.2pt,style={pos=0.45}](r1)(b1);
  \node[label=above:1](r2) [red vertex] at (7,4) {};
  \Edge[color=red,label={1},lw=1.2pt,style={pos=0.45}](r2)(s1);
  \node[label=above:2](r3) [red vertex] at (8,4) {};
  \node(b3) [blue vertex] at (8,2) {};
  \Edge[color=red,label={1},lw=1.2pt,style={pos=0.45}](r3)(s1);
  \Edge[color=red,label={1},lw=1.2pt,style={pos=0.45}](r3)(b3);
  \node at ($(s1)+(0,-1)$) {(c)};

  \node[label=below:2](s1) [blue vertex] at (10,2) {};
  \node[label=above:2](r1) [red vertex] at (9,4) {};
  \node(b1) [blue vertex] at (9,2) {};
  \Edge[color=red,label={1},lw=1.2pt,style={pos=0.45}](r1)(s1);
  \Edge[color=red,label={1},lw=1.2pt,style={pos=0.45}](r1)(b1);
  \node[label=above:2](r2) [red vertex] at (10,4) {};
  \node(b2) [blue vertex] at (11,2) {};
  \Edge[color=red,label={1},lw=1.2pt,style={pos=0.45}](r2)(s1);
  \Edge[color=red,label={1},lw=1.2pt,style={pos=0.24}](r2)(b2);
  \node[label=above:2](r3) [red vertex] at (11,4) {};
  \node(b3) [blue vertex] at (12,2) {};
  \Edge[color=red,label={1},lw=1.2pt,style={pos=0.75}](r3)(s1);
  \Edge[color=red,label={1},lw=1.2pt,style={pos=0.45}](r3)(b3);
  \node at ($(s1)+(0,-1)$) {(d)};

\end{tikzpicture}}
    \caption{Configurations that imply on an improving 3-claw centered in~$u$.}
    \label{fig:75w2receiving}
  \end{figure}
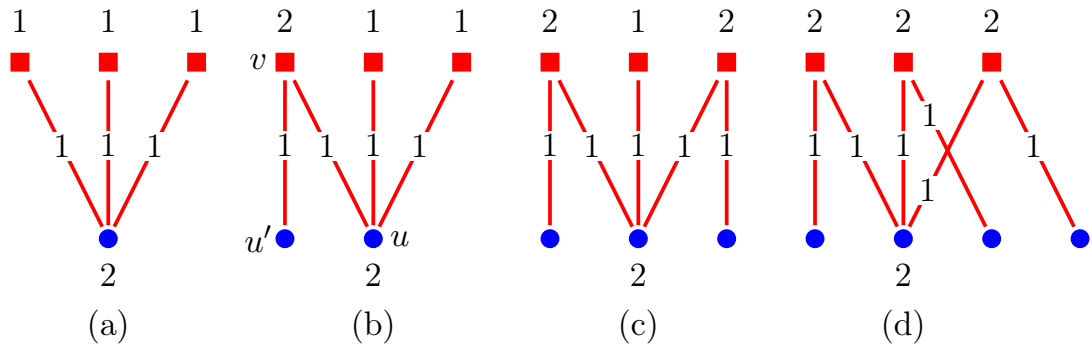


  Now we will analyze how much the weight-1 vertices in~$A$ receive from the
  vertices in~$A^*$.
  We will prove that, on average, each weight-1 vertex in~$A$ receives at
  most~$7/5$. 
  Note that a weight-1 vertex in $A$ receives through each edge a value in
  $\{0,\frac25,\frac12,\frac23,1\}$. 

  Let $Y = \bigcup\{U_v \colon v \in A\}$ and $Z = \bigcup\{U_v \colon v \in
  A^*\}$. 
  We will modify a little bit the interpretation of the weight distribution,
  considering that each vertex~$v$ in~$A^*$ in fact distributes its
  weight~$w_v$ among the elements in~$Z \cap Y$.
  That is, we will consider that the blue round vertices in
  Figure~\ref{fig:weights75} represent elements in the set $Z \cap Y$. 
  Then, a vertex~$v$ in~$A$ receives from~$A^*$ the sum of the weights that its
  elements in $U_v \cap Z$ received.

  Note that the vertex set~$V'$ of the graph~$H$ defined in
  Line~\ref{alg:graphH} is exactly $Z \setminus X$.
  Moreover, each weight-1 vertex~$v$ in~$A$ is associated to an edge~$e_v$
  in~$M$, and receives from~$A^*$ exactly the sum of the weights received by
  the ends of~$e_v$.

  An \textit{$A^*$-edge} is an edge of~$H$ that is contained in the set~$U_v$
  for a $v \in A^*$. 
  Consider the spanning subgraph~$H'$ of~$H$ containing only the edges in~$M$
  and the~$A^*$-edges. 
  See Figure~\ref{fig:distributionH} for an example.

  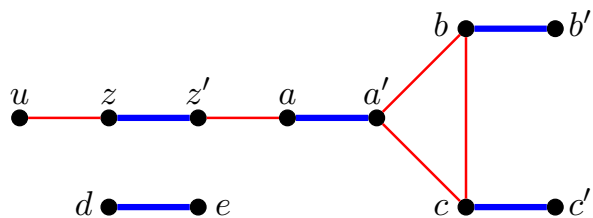
\begin{figure}[htb]
    \centering
    \resizebox{0.5\textwidth}{!}{\begin{tikzpicture}
    [vertex/.style={circle,draw,fill,inner sep=0pt,minimum size=5pt}]

    \node (X) at (0,0) {};
    \node[vertex,label={[xshift=0pt,yshift=-2pt]$u$}] (u) at ($(X)$) {};
    \node[vertex,label={[xshift=0pt,yshift=-2pt]$z$}] (z) at ($(u)+(1,0)$) {};
    \node[vertex,label={[xshift=0pt,yshift=-2pt]$z'$}] (zp) at ($(z)+(1,0)$) {};
    \node[vertex,label={[xshift=0pt,yshift=-2pt]$a$}] (a) at ($(zp)+(1,0)$) {};
    \node[vertex,label={[xshift=0pt,yshift=-2pt]$a'$}] (ap) at ($(a)+(1,0)$) {};

    \node[vertex,label={[xshift=-8pt,yshift=-9pt]$b$}] (b) at ($(ap)+(1,1)$) {};
    \node[vertex,label={[xshift=-8pt,yshift=-9pt]$c$}] (c) at ($(ap)+(1,-1)$) {};
    \node[vertex,label={[xshift=8pt,yshift=-9pt]$b'$}] (yp) at ($(b)+(1,0)$) {};
    \node[vertex,label={[xshift=8pt,yshift=-9pt]$c'$}] (cp) at ($(c)+(1,0)$) {};

    \node[vertex,label={[xshift=-8pt,yshift=-9pt]$d$}] (d) at ($(z)+(0,-1)$) {};
    \node[vertex,label={[xshift=8pt,yshift=-9pt]$e$}] (e) at ($(d)+(1,0)$) {};

    \draw[thick,red] (u) -- (z);
    \draw[line width=2pt,blue] (z) -- (zp);
    \draw[thick,red] (zp) -- (a);
    \draw[line width=2pt,blue] (a) -- (ap);
    \draw[thick,red] (ap) -- (b);
    \draw[thick,red] (ap) -- (c);
    \draw[thick,red] (b) -- (c);
    \draw[line width=2pt,blue] (b) -- (yp);
    \draw[line width=2pt,blue] (c) -- (cp);
    \draw[line width=2pt,blue] (d) -- (e);

\end{tikzpicture}}

    \caption{Graph $H'$ built from the intersection graph in Figure~\ref{figb:dist_G}. 
      Bold blue edges are associated to vertices in~$A$. All the other are $A^*$-edges. 
      Vertex $z$ receives~$1$ from $v$ in Figure~\ref{figb:dist_G}, 
      while $z'$ receives $1/2$ from~$s$, as $zz'$ is associated with vertex~$t$.  
      Vertex $a$ also receives $1/2$ from $s$.  
      Eaxh of vertices $a'$, $b$, and $c$ receives $2/3$ from $q$.}

    \label{fig:distributionH}
  \end{figure}

  \begin{claim}
  \label{clm:um}
    There is at most one $M$-uncovered vertex in each component of~$H'$. 
  \end{claim}
  \begin{claimproof}
    There are two types of $A^*$-edges in~$H'$: the isolated ones, that share
    no end with another $A^*$-edge in~$H'$, and the ones in a triangle.
    Indeed, as an $A^*$-edge~$e$ is contained in~$U_v$ for some~$v \in A^*$,
    either $e = U_v$ or there is a vertex~$x$ in~$U_v$ not in~$e$. 
    If $x \in V'$, then the three pairs of elements in~$U_v$ correspond to
    $A^*$-edges in $H'$, including~$e$, that form a triangle in~$H'$. 
    If $x \not\in V'$ or $e=U_v$, then~$e$ is an isolated $A^*$-edge in~$H'$.
    This implies that any path in~$H'$ corresponds to an alternating path
    in~$H'$: if the path contains two consecutive $A^*$-edges, these two edges
    lie in a triangle in~$H'$, and the third edge in this triangle can be used
    to shortcut the common vertex of the consecutive $A^*$-edges.  
    So, if there were two $M$-uncovered vertices in the same component of~$H'$,
    there would be an augmenting path between them, contradicting the fact
    that~$M$ is a maximum matching in~$H$. 
  \end{claimproof}
    
  Now we will prove that the average weight that~$A^*$ assigns to
  an~$M$-covered vertex in~$H'$ is at most~$7/10$.  
  This implies that each weight-1 vertex in~$A$ receives at most~$7/5$ from
  $A^*$.  
  The analysis considers one connected component~$H''$ of~$H'$ at a time.  

  Every~$M$-covered vertex in $H''$ that receives~1 is like the blue round
  vertex from Configuration~\ref{fig:weights75}(a), or like one of the blue
  round vertices in Configuration~\ref{fig:weights75}(f). 
  Each of these configurations corresponds to an $M$-uncovered vertex in~$H$.  
  Thus, by Claim~\ref{clm:um}, in~$H''$, there are at most two vertices
  receiving~1 from~$A^*$, each one adjacent in~$H''$ to the only $M$-uncovered
  vertex in~$H''$. 

  If no vertex in~$H''$ receives~1 from~$A^*$, then every~$M$-covered vertex
  in~$H''$ would receive at most~$2/3 < 7/10$ and the statement holds for the
  $M$-covered vertices in~$H''$. 
  The rest of the proof follows from the next two claims. 

  \begin{claim}
  \label{claim:two_unmatched}
    Let $H''$ be a connected component of $H'$ that contains an $M$-uncovered
    vertex $u$ incident to exactly two $A^*$-edges.  In average,
    every~$M$-covered vertex in~$H''$ receives at most~$7/10$.
  \end{claim}
  \begin{claimproof}
    Let $a$ and $b$ be the other ends of the two $A^*$-edges incident to~$u$.
    Both~$a$ and~$b$ receive~1 from a vertex~$v$ in~$A^*$, 
    as in Configuration~\ref{fig:weights75}(f).  
    Note that $ab \not\in M$, otherwise there is a trivial augmenting claw. 
    See Figure~\ref{fig:two_unmatched-case1}.

    \begin{figure}[htb]
      \centering
      \resizebox{0.5\textwidth}{!}{\begin{tikzpicture}
    [vertexH/.style={circle,draw,fill,inner sep=0pt,minimum size=5pt},
    vertexG/.style={circle,draw,inner sep=0pt,minimum size=5pt,minimum width=15pt}]

    \node (X) at (0,0) {};

    \node[vertexH,label={below:$u$}] (u) at ($(X)+(-1,-0.8)$) {};
    \node[vertexH,label={[xshift=0pt,yshift=0pt]$a$}] (a) at ($(u)+(-0.6,1)$) {};
    \node[vertexH,label={[xshift=0pt,yshift=0pt]$b$}] (b) at ($(u)+(0.6,1)$) {};

    \draw[thick,red] (u) -- (a);
    \draw[thick,red] (u) -- (b);
    \draw[line width=2pt,blue] (a) -- (b);
    \draw[thick,red] (a) -- (b);

    \node[vertexG,line width=2pt,dashed,red] (v) at ($(X)+(2.5,-0.5)$) {$v$};
    \node[vertexG,line width=2pt,blue] (v1) at ($(v)+(0,0.8)$) {$v_u$};
    \node[vertexG] (v2) at ($(v)+(-0.7,-0.5)$) {$v_b$};
    \node[vertexG] (v3) at ($(v)+(+0.7,-0.5)$) {$v_a$};
    \draw[thick] (v) -- (v1);
    \draw[thick] (v) -- (v2);
    \draw[thick] (v) -- (v3);
    \draw[thick] (v1) -- (v2);
    \draw[thick] (v1) -- (v3);
    \draw[thick] (v2) -- (v3);

\end{tikzpicture}}
      \caption{Here $U_v=\{a,b,u\}$.  
        Because~$u$ is $M$-uncovered, $v_u$ is the only neighbor of $v$ in $A$,
        thus the claw with $T_C = \{v\}$ centered at $v_u$ is improving.}
      \label{fig:two_unmatched-case1}
    \end{figure}
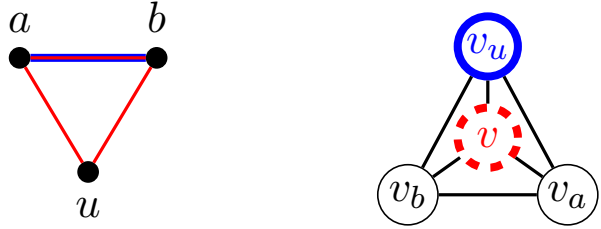
    
    Let~$a'$ and~$b'$ be such that $aa' \in M$ and $bb' \in M$. 
    If each of~$a'$ and~$b'$ receives at most $2/5$ from~$A^*$, then the claim
    holds. 
    Otherwise, we may assume~$a'$ receives at least~$1/2$, like the blue round
    vertex in Configuration~\ref{fig:weights75}(b), or like one of the blue
    round vertices in Configurations~\ref{fig:weights75}(g), or
    Configuration~\ref{fig:weights75}(i), where its sibling of weight 2 is not
    in~$H$.  
    So there is at least one $A^*$-edge incident to~$a'$ in~$H'$. 

    If there are two $A^*$-edges incident to~$a'$, we are in
    Configuration~\ref{fig:weights75}(g).
    Let~$x$ be the weight-2 vertex of~$G$ corresponding to the red square
    vertex in this configuration, and to the two $A^*$-edges incident to~$a'$
    in~$H'$. 
    Thus, there is an augmenting claw~$C$ with $T_C = \{v,x\}$, as there are at
    most four weight-1 vertices in~$A \cap N(T_C)$, and~$9 + 9 = 18 > 16 = 4
    \cdot 4$. 
    See Figure~\ref{fig:two_unmatched-case2}.

    \begin{figure}[htb]
      \centering
      \begin{subfigure}{0.47\textwidth}
        \centering
        \resizebox{0.88\textwidth}{!}{\begin{tikzpicture}
    [vertexH/.style={circle,draw,fill,inner sep=0pt,minimum size=5pt},
    vertexG/.style={circle,draw,inner sep=0pt,minimum size=5pt,minimum width=15pt}]

    \node (X) at (0,0) {};

    \node[vertexH,label={[xshift=0pt,yshift=-2pt]$u$}] (u) at ($(X)$) {};
    \node[vertexH,label={[xshift=-6pt,yshift=-9pt]$a$}] (a) at ($(u)+(0.5,-1)$) {};
    \node[vertexH,label={[xshift=0pt,yshift=-2pt]$b$}] (b) at ($(u)+(1,0)$) {};
    \node[vertexH,label={[xshift=-6pt,yshift=-9pt]$a'$}] (ap) at ($(a)+(0,-1)$) {};
    \node[vertexH,label={[xshift=0pt,yshift=-2pt]$b'$}] (bp) at ($(b)+(1,0)$) {};
    \node[vertexH,label={[xshift=-6pt,yshift=-9pt]$c$}] (c) at ($(ap)+(-0.5,-1)$) {};
    \node[vertexH,label={[xshift=6pt,yshift=-9pt]$d$}] (d) at ($(ap)+(0.5,-1)$) {};
    \node[vertexH,label={[xshift=-6pt,yshift=-9pt]$c'$}] (cp) at ($(c)+(0,-1)$) {};
    \node[vertexH,label={[xshift=8pt,yshift=-9pt]$d'$}] (dp) at ($(d)+(0,-1)$) {};

    \draw[thick,red] (u) -- (a);
    \draw[thick,red] (u) -- (b);
    \draw[thick,red] (a) -- (b);
    \draw[thick,red] (ap) -- (c);
    \draw[thick,red] (ap) -- (d);
    \draw[thick,red] (c) -- (d);
    \draw[line width=2pt,blue] (a) -- (ap);
    \draw[line width=2pt,blue] (b) -- (bp);
    \draw[line width=2pt,blue] (c) -- (cp);
    \draw[line width=2pt,blue] (d) -- (dp);

    \node[vertexG,line width=2pt,dashed,red] (v) at ($(X)+(3.5,-0.7)$) {$v$};
    \node[vertexG] (v1) at ($(v)+(0,0.8)$) {$v_a$};
    \node[vertexG] (v2) at ($(v)+(-0.7,-0.5)$) {$v_b$};
    \node[vertexG] (v3) at ($(v)+(+0.7,-0.5)$) {$v_u$};
    \draw[thick] (v) -- (v1);
    \draw[thick] (v) -- (v2);
    \draw[thick] (v) -- (v3);
    \draw[thick] (v1) -- (v2);
    \draw[thick] (v1) -- (v3);
    \draw[thick] (v2) -- (v3);

    \node[vertexG,line width=2pt,blue] (q) at ($(v)+(0.9,0.5)$) {$q$};
    \draw[thick] (q) -- (v);
    \draw[thick] (q) -- (v1);
    \draw[thick] (q) -- (v3);

    \node[vertexG,line width=2pt,blue] (r) at ($(v)+(0,-1.2)$) {$r$};
    \draw[thick] (r) -- (v);
    \draw[thick] (r) -- (v2);
    \draw[thick] (r) -- (v3);

    \node[vertexG,line width=2pt,dashed,red] (x) at ($(r)+(0,-1.2)$) {$x$};
    \node[vertexG] (x1) at ($(x)+(-0.7,0.5)$) {$x_d$};
    \node[vertexG] (x2) at ($(x)+(0.7,0.5)$) {$x_c$};
    \node[vertexG] (x3) at ($(x)+(0,-0.8)$) {$x_{a'}$};
    \draw[thick] (x) -- (x1);
    \draw[thick] (x) -- (x2);
    \draw[thick] (x) -- (x3);
    \draw[thick] (x1) -- (x2);
    \draw[thick] (x1) -- (x3);
    \draw[thick] (x2) -- (x3);

    \draw[thick] (r) -- (x);
    \draw[thick] (r) -- (x1);
    \draw[thick] (r) -- (x2);

    \node[vertexG,line width=2pt,blue] (s) at ($(x)+(-0.9,-0.5)$) {$s$};
    \node[vertexG,line width=2pt,blue] (t) at ($(x)+(0.9,-0.5)$) {$t$};

    \draw[thick] (s) -- (x);
    \draw[thick] (s) -- (x1);
    \draw[thick] (s) -- (x3);
    \draw[thick] (t) -- (x);
    \draw[thick] (t) -- (x2);
    \draw[thick] (t) -- (x3);

\end{tikzpicture}}
        \caption{The claw with $T_C = \{v,x\}$ is improving.}
        \label{fig:two_unmatched-case2}
      \end{subfigure}
      \hfill
      \begin{subfigure}{0.47\textwidth}
        \centering
        \resizebox{0.97\textwidth}{!}{\begin{tikzpicture}
    [vertexH/.style={circle,draw,fill,inner sep=0pt,minimum size=5pt},
    vertexG/.style={circle,draw,inner sep=0pt,minimum size=5pt,minimum width=15pt}]

    \node (X) at (0,0) {};

    \node[vertexH,label={[xshift=0pt,yshift=-2pt]$u$}] (u) at ($(X)$) {};
    \node[vertexH,label={[xshift=-6pt,yshift=-9pt]$a$}] (a) at ($(u)+(0.5,-1)$) {};
    \node[vertexH,label={[xshift=0pt,yshift=-2pt]$b$}] (b) at ($(u)+(1,0)$) {};
    \node[vertexH,label={[xshift=-6pt,yshift=-9pt]$a'$}] (ap) at ($(a)+(0,-1)$) {};
    \node[vertexH,label={[xshift=0pt,yshift=-2pt]$b'$}] (bp) at ($(b)+(1,0)$) {};
    \node[vertexH,label={[xshift=-6pt,yshift=-9pt]$c$}] (c) at ($(ap)+(0,-1)$) {};
    \node[vertexH,label={[xshift=-6pt,yshift=-9pt]$c'$}] (cp) at ($(c)+(0,-1)$) {};

    \draw[thick,red] (u) -- (a);
    \draw[thick,red] (u) -- (b);
    \draw[thick,red] (a) -- (b);
    \draw[thick,red] (ap) -- (c);
    \draw[line width=2pt,blue] (a) -- (ap);
    \draw[line width=2pt,blue] (b) -- (bp);
    \draw[line width=2pt,blue] (c) -- (cp);

    \node[vertexG,line width=2pt,dashed,red] (v) at ($(X)+(3.3,-0.6)$) {$v$};
    \node[vertexG] (v1) at ($(v)+(0,0.8)$) {$v_a$};
    \node[vertexG] (v2) at ($(v)+(-0.7,-0.5)$) {$v_b$};
    \node[vertexG] (v3) at ($(v)+(+0.7,-0.5)$) {$v_u$};
    \draw[thick] (v) -- (v1);
    \draw[thick] (v) -- (v2);
    \draw[thick] (v) -- (v3);
    \draw[thick] (v1) -- (v2);
    \draw[thick] (v1) -- (v3);
    \draw[thick] (v2) -- (v3);

    \node[vertexG,line width=2pt,blue] (q) at ($(v)+(0.9,0.5)$) {$q$};
    \draw[thick] (q) -- (v);
    \draw[thick] (q) -- (v1);
    \draw[thick] (q) -- (v3);

    \node[vertexG,line width=2pt,blue] (r) at ($(v)+(0,-1.2)$) {$r$};
    \draw[thick] (r) -- (v);
    \draw[thick] (r) -- (v2);
    \draw[thick] (r) -- (v3);

    \node[vertexG,line width=2pt,dashed,red] (x) at ($(r)+(0,-1.2)$) {$x$};
    \node[vertexG,label={[xshift=-10pt,yshift=-8pt]$y$}] (x1) at ($(x)+(-0.7,0.5)$) {$x_d$};
    \node[vertexG] (x2) at ($(x)+(0.7,0.5)$) {$x_c$};
    \node[vertexG] (x3) at ($(x)+(0,-0.8)$) {$x_{a'}$};
    \draw[thick] (x) -- (x1);
    \draw[thick] (x) -- (x2);
    \draw[thick] (x) -- (x3);
    \draw[thick] (x1) -- (x2);
    \draw[thick] (x1) -- (x3);
    \draw[thick] (x2) -- (x3);

    \draw[thick] (r) -- (x);
    \draw[thick] (r) -- (x1);
    \draw[thick] (r) -- (x2);

    \node[vertexG,line width=2pt,blue] (s) at ($(x)+(-0.9,-0.5)$) {$s$};
    \draw[thick] (s) -- (x);
    \draw[thick] (s) -- (x1);
    \draw[thick] (s) -- (x3);

    \node[vertexG,line width=2pt,blue] (t) at ($(x)+(1.6,-0.7)$) {$t$};
    \node[vertexG] (t1) at ($(t)+(0,0.8)$) {$t_e$};
    \node[vertexG] (t2) at ($(t)+(-0.7,-0.5)$) {$t_f$};
    \node[vertexG,line width=2pt,dashed,red] (t3) at ($(t)+(+0.7,-0.5)$) {$t_d$};
    \draw[thick] (t) -- (t1);
    \draw[thick] (t) -- (t2);
    \draw[thick] (t) -- (t3);
    \draw[thick] (t1) -- (t2);
    \draw[thick] (t1) -- (t3);
    \draw[thick] (t2) -- (t3);

    \draw[thick] (t) -- (x);
    \draw[thick] (t) -- (x2);
    \draw[thick] (t) -- (x3);
    \draw[thick] (t1) -- (x);
    \draw[thick] (t1) -- (x2);
    \draw[thick] (t1) -- (x3);
    \draw[thick] (t2) -- (x);
    \draw[thick] (t2) -- (x2);
    \draw[thick] (t2) -- (x3);

\end{tikzpicture}}
        \caption{The claw with $T_C = \{v,x'\}$ is improving.}
        \label{fig:two_unmatched-case3}
      \end{subfigure}

      \caption{In each case, part of $H''$ is to the left while the corresponding 
        part of~$G$ is to the right. In $H''$, bold blue arcs are associated with 
        vertices in $A$, which are bold blue in $G$. Dashed red vertices are in $A^*$.}
      \label{fig:two_unmatched}
    \end{figure}
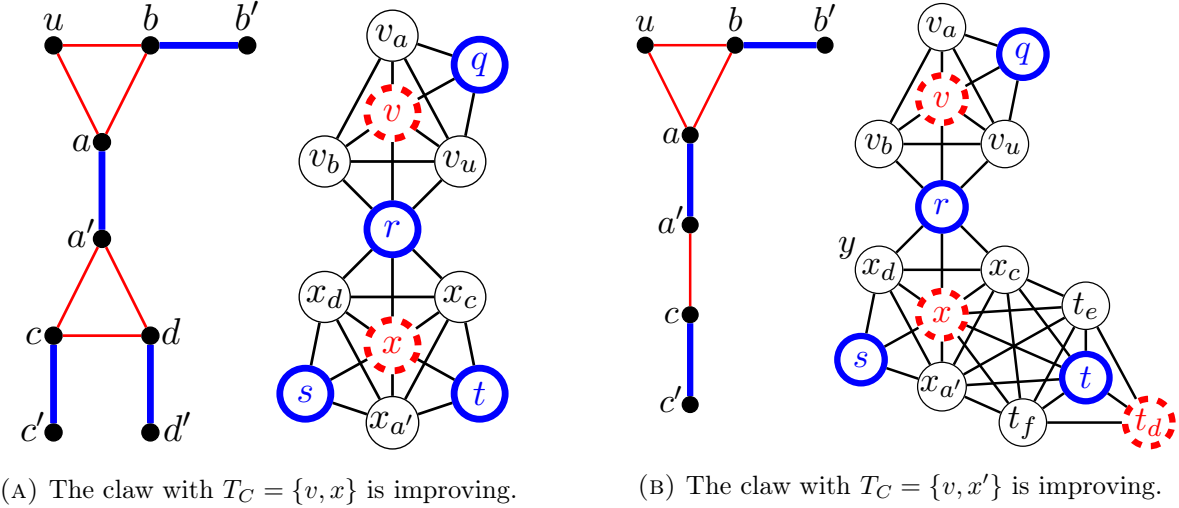

    Otherwise, we are either in Configuration~\ref{fig:weights75}(b) or in
    Configuration~\ref{fig:weights75}(i).
    Let $x$ be the red square vertex in the corresponding configuration. 
    If we are in Configuration~\ref{fig:weights75}(b), let $y = x$, that is,
    $y$ is the weight-1 vertex of $G$ corresponding to the only $A^*$-edge
    incident to $a'$ in $H'$. 
    If we are in Configuration~\ref{fig:weights75}(i), let~$y$ be the weight-1
    vertex in~$K_4^x$ corresponding to the only $A^*$-edge incident to~$a'$. 
    Again, the claw $C$ with $T_C = \{v,y\}$ is augmenting, with at most three
    weight-1 vertices in~$A \cap N(T_C)$, and~$9 + 4 = 13 > 12 = 3 \cdot 4$. 
    See Figure~\ref{fig:two_unmatched-case3}. 
  \end{claimproof}

  \begin{claim}
  \label{claim:one_unmatched}
    Let $H''$ be a connected component of $H'$ that contains an $M$-uncovered
    vertex $u$ incident to only one $A^*$-edge $e$. 
    In average, every~$M$-covered vertex in~$H''$ receives at most~$7/10$.
  \end{claim}
  \begin{claimproof}
    Let $z$ be the other end of~$e$.
    Vertex~$z$ receives~1 from $A^*$, and is either like the blue round vertex
    in Configuration~\ref{fig:weights75}(a), or like one of the blue round
    vertices in Configuration~\ref{fig:weights75}(f) when its sibling has
    weight-2 and thus is not in~$H$.  

    If there is a vertex $x$ in~$H''$ that receives at most~$2/5$ from~$A^*$,
    then the claim holds.
    Indeed, every $M$-covered vertex of~$H''$ other than~$x$ and $z$ receives
    at most~$2/3$ from~$A^*$.
    Thus, because $2/3 < 7/10$ and $(1+2/5)/2 = 7/10$, the claim holds. 
    So we may assume that every vertex in~$H''$ receives at least~$1/2$. 
    If there is one vertex that receives~$1/2$ in~$H''$, then this vertex is
    like one of the blue round vertices in Configuration~\ref{fig:weights75}(b)
    and its sibling is also in~$H''$. 
    Therefore, there would be two vertices in~$H''$ receiving~$1/2$, and all
    the others except for $z$ receive at most $2/3$. 
    Hence the average in~$H''$ would be at most $(1+1/2+1/2)/3 = 2/3 < 7/10$.  
    Thus we may assume that every vertex in~$H''$ except for $z$ receives~$2/3$. 

    The number of $M$-covered vertices in~$H''$ is always even, and at least
    two because $z$ is $M$-covered.  
    If there are at least ten $M$-covered vertices in~$H''$, the claim holds
    because $(1 + 9 \cdot 2/3)/10 = 7/10$.
    So we may assume there are at most eight $M$-covered vertices in~$H''$, and
    an odd number of them receive~$2/3$. 

    From the configurations in Figure~\ref{fig:weights75}, a vertex in~$A^*$
    cannot send~$2/3$ to only one vertex in~$H''$. 
    It either sends~$2/3$ to three vertices in~$H''$
    (Configuration~\ref{fig:weights75}(g)), or to two vertices in~$H''$
    (Configuration~\ref{fig:weights75}(i)).     
    As the number of $M$-covered vertices in~$H''$ is even, but~$z$ receives~1,
    the number of vertices that receive~$2/3$ in $H''$ is odd. 
    Thus there must be exactly one group of three vertices $x,y,y'$ in~$H''$
    receiving~$2/3$ from the same vertex~$t$ in~$A^*$
    (Configuration~\ref{fig:weights75}(g)).
    If one of~$xy$, $xy'$ or $yy' \in M$, then~$t$ is an improving claw.  
    Therefore, no two vertices in~$x,y,y'$ are matched to each other in~$M$.

    Let~$v$ be the vertex in~$A^*$ corresponding to~$e$.  
    Let~$z'$ be the vertex such that $zz' \in M$.
    Vertex~$z'$ is also in~$H''$. 
    Note that $z' \not\in \{x,y,y'\}$, otherwise there is an improving claw~$C$
    with $T_C=\{v,t\}$.

    Thus~$H''$ must contain exactly eight $M$-covered vertices. 
    See Figure~\ref{fig:one_unmatched}.
    The vertices adjacent to~$x$, $y$, and~$y'$ in~$M$ must be distinct and two
    of them, say~$w$ and~$w'$, receive~$2/3$ from the same weight-2 vertex~$s$
    in $A^*$ (Configuration~\ref{fig:weights75}(i)). 
    Let~$s'$ be the weight-1 vertex in~$K_4^s$ corresponding to the only
    $A^*$-edge $ww'$ in~$H''$ coming from~$s$. 
    Then the claw~$C$ with $T_C = \{t,s'\}$ is improving (because $9+4 = 13 >
    12 = 3 \cdot 4$).  
  \end{claimproof}

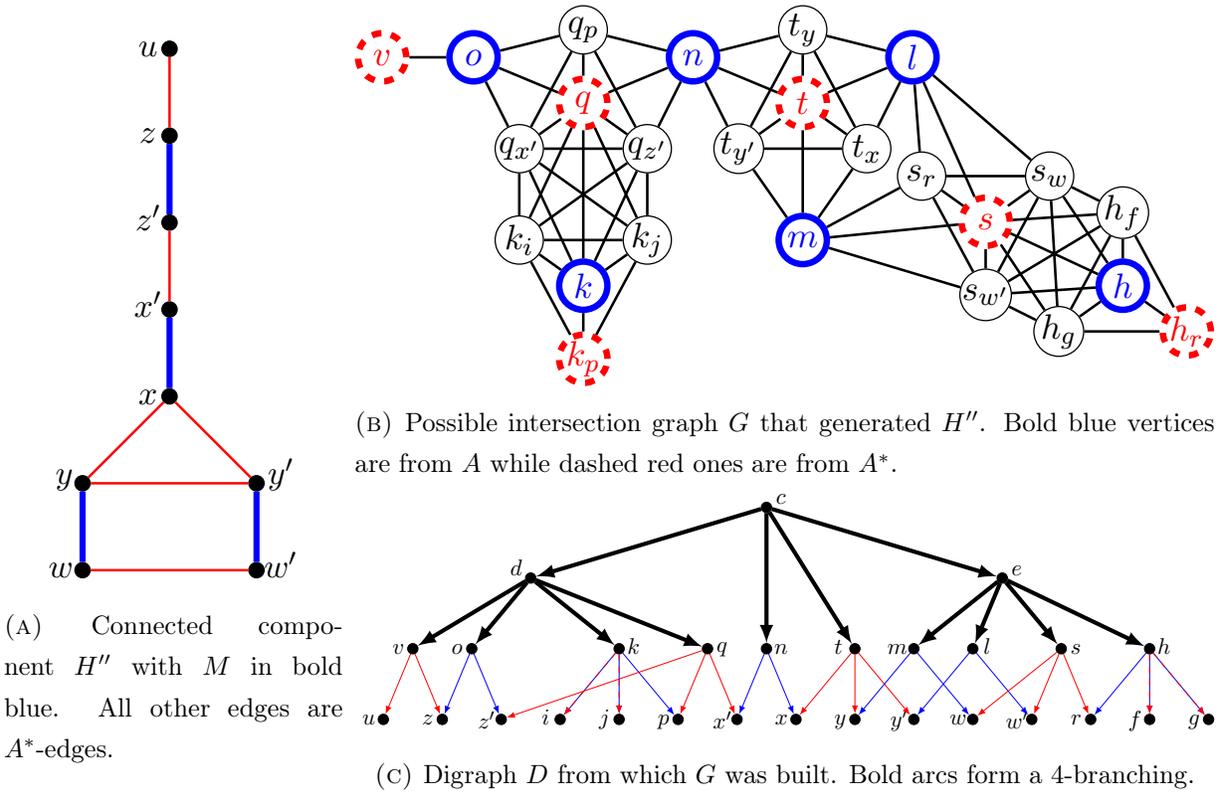
\begin{figure}[htb]
  \centering
  \begin{minipage}{0.28\textwidth}
    \begin{subfigure}{\textwidth}
      \centering
      \resizebox{0.8\textwidth}{!}{\begin{tikzpicture}
    [vertex/.style={circle,draw,fill,inner sep=0pt,minimum size=5pt}]

    \node (X) at (0,0) {};
    \node[vertex,label={[xshift=-7pt,yshift=-9pt]$u$}] (u) at ($(X)$) {};
    \node[vertex,label={[xshift=-7pt,yshift=-9pt]$z$}] (z) at ($(u)+(0,-1)$) {};
    \node[vertex,label={[xshift=-7pt,yshift=-9pt]$z'$}] (zp) at ($(z)+(0,-1)$) {};
    \node[vertex,label={[xshift=-7pt,yshift=-9pt]$x'$}] (xp) at ($(zp)+(0,-1)$) {};
    \node[vertex,label={[xshift=-7pt,yshift=-9pt]$x$}] (x) at ($(xp)+(0,-1)$) {};

    \node[vertex,label={[xshift=-6pt,yshift=-9pt]$y$}] (y) at ($(x)+(-1,-1)$) {};
    \node[vertex,label={[xshift=8pt,yshift=-9pt]$y'$}] (yp) at ($(x)+(1,-1)$) {};
    \node[vertex,label={[xshift=-7pt,yshift=-9pt]$w$}] (w) at ($(y)+(0,-1)$) {};
    \node[vertex,label={[xshift=8pt,yshift=-9pt]$w'$}] (wp) at ($(yp)+(0,-1)$) {};

    \draw[thick,red] (u) -- (z);
    \draw[line width=2pt,blue] (z) -- (zp);
    \draw[thick,red] (zp) -- (xp);
    \draw[line width=2pt,blue] (xp) -- (x);
    \draw[thick,red] (x) -- (y);
    \draw[thick,red] (x) -- (yp);
    \draw[thick,red] (y) -- (yp);
    \draw[line width=2pt,blue] (y) -- (w);
    \draw[line width=2pt,blue] (yp) -- (wp);
    \draw[thick,red] (w) -- (wp);

\end{tikzpicture}}
      \caption{Connected component~$H''$ with $M$ in bold blue. 
        All other edges are $A^*$-edges.}
      \label{figa:one_unmatched-C}
    \end{subfigure}
  \end{minipage}
  \hfill
  \begin{minipage}{0.71\textwidth}
    \begin{subfigure}{\textwidth}
      \centering
      \resizebox{\textwidth}{!}{\begin{tikzpicture}
    [vertex/.style={circle,draw,inner sep=0pt,minimum size=5pt,minimum width=15pt},
    vertex2/.style={draw=line width=2pt,blue,double=line width=2pt,dashed,red,circle,inner sep=0pt,minimum size=5pt,minimum width=15pt}]

    \node[vertex,line width=2pt,dashed,red] (q) at (0,0) {$q$};
    \node[vertex] (q1) at ($(q)+(0,0.8)$) {$q_p$};
    \node[vertex] (q2) at ($(q)+(-0.7,-0.5)$) {$q_{x'}$};
    \node[vertex] (q3) at ($(q)+(+0.7,-0.5)$) {$q_{z'}$};
    \draw[thick] (q) -- (q1);
    \draw[thick] (q) -- (q2);
    \draw[thick] (q) -- (q3);
    \draw[thick] (q1) -- (q2);
    \draw[thick] (q1) -- (q3);
    \draw[thick] (q2) -- (q3);

    \node[vertex,line width=2pt,blue] (o) at ($(q)+(-1.2,0.5)$) {$o$};
    \node[vertex,line width=2pt,dashed,red] (v) at ($(o)+(-1,0)$) {$v$};
    \draw[thick] (o) -- (v);
    \draw[thick] (o) -- (q);
    \draw[thick] (o) -- (q1);
    \draw[thick] (o) -- (q2);

    \node[vertex,line width=2pt,blue] (k) at ($(q)+(0,-2)$) {$k$};
    \node[vertex] (k1) at ($(k)+(-0.7,0.5)$) {$k_i$};
    \node[vertex] (k2) at ($(k)+(+0.7,0.5)$) {$k_j$};
    \node[vertex,line width=2pt,dashed,red] (k3) at ($(k)+(0,-0.8)$) {$k_p$};
    \draw[thick] (k) -- (k1);
    \draw[thick] (k) -- (k2);
    \draw[thick] (k) -- (k3);
    \draw[thick] (k1) -- (k2);
    \draw[thick] (k1) -- (k3);
    \draw[thick] (k2) -- (k3);

    \draw[thick] (q2) -- (k1);
    \draw[thick] (q2) -- (k);
    \draw[thick] (q2) -- (k2);
    \draw[thick] (q3) -- (k1);
    \draw[thick] (q3) -- (k);
    \draw[thick] (q3) -- (k2);
    \draw[thick] (q) -- (k1);
    \draw[thick] (q) -- (k);
    \draw[thick] (q) -- (k2);

    \node[vertex,line width=2pt,blue] (n) at ($(q)+(1.2,0.5)$) {$n$};
    \draw[thick] (n) -- (q);
    \draw[thick] (n) -- (q1);
    \draw[thick] (n) -- (q3);

    \node[vertex,line width=2pt,dashed,red] (t) at ($(n)+(1.2,-0.5)$) {$t$};
    \node[vertex] (t1) at ($(t)+(0,0.8)$) {$t_y$};
    \node[vertex] (t2) at ($(t)+(-0.7,-0.5)$) {$t_{y'}$};
    \node[vertex] (t3) at ($(t)+(+0.7,-0.5)$) {$t_x$};
    \draw[thick] (t) -- (t1);
    \draw[thick] (t) -- (t2);
    \draw[thick] (t) -- (t3);
    \draw[thick] (t1) -- (t2);
    \draw[thick] (t1) -- (t3);
    \draw[thick] (t2) -- (t3);

    \draw[thick] (n) -- (t);
    \draw[thick] (n) -- (t1);
    \draw[thick] (n) -- (t2);

    \node[vertex,line width=2pt,blue] (m) at ($(t)+(0,-1.5)$) {$m$};
    \draw[thick] (m) -- (t);
    \draw[thick] (m) -- (t2);
    \draw[thick] (m) -- (t3);

    \node[vertex,line width=2pt,blue] (l) at ($(t)+(1.2,0.5)$) {$l$};
    \draw[thick] (l) -- (t);
    \draw[thick] (l) -- (t1);
    \draw[thick] (l) -- (t3);

    \node[vertex,line width=2pt,dashed,red] (s) at ($(t)+(2,-1.3)$) {$s$};
    \node[vertex] (s1) at ($(s)+(-0.7,0.5)$) {$s_r$};
    \node[vertex] (s2) at ($(s)+(0,-0.8)$) {$s_{w'}$};
    \node[vertex] (s3) at ($(s)+(0.7,0.5)$) {$s_w$};
    \draw[thick] (s) -- (s1);
    \draw[thick] (s) -- (s2);
    \draw[thick] (s) -- (s3);
    \draw[thick] (s1) -- (s2);
    \draw[thick] (s1) -- (s3);
    \draw[thick] (s2) -- (s3);

    \draw[thick] (m) -- (s);
    \draw[thick] (m) -- (s1);
    \draw[thick] (m) -- (s2);
    \draw[thick] (l) -- (s);
    \draw[thick] (l) -- (s1);
    \draw[thick] (l) -- (s3);

    \node[vertex,line width=2pt,blue] (h) at ($(s)+(1.5,-0.7)$) {$h$};
    \node[vertex] (h1) at ($(h)+(0,0.8)$) {$h_f$};
    \node[vertex] (h2) at ($(h)+(-0.7,-0.5)$) {$h_g$};
    \node[vertex,line width=2pt,dashed,red] (h3) at ($(h)+(+0.7,-0.5)$) {$h_r$};
    \draw[thick] (h) -- (h1);
    \draw[thick] (h) -- (h2);
    \draw[thick] (h) -- (h3);
    \draw[thick] (h1) -- (h2);
    \draw[thick] (h1) -- (h3);
    \draw[thick] (h2) -- (h3);

    \draw[thick] (h1) -- (s);
    \draw[thick] (h1) -- (s2);
    \draw[thick] (h1) -- (s3);
    \draw[thick] (h) -- (s);
    \draw[thick] (h) -- (s2);
    \draw[thick] (h) -- (s3);
    \draw[thick] (h2) -- (s);
    \draw[thick] (h2) -- (s2);
    \draw[thick] (h2) -- (s3);

\end{tikzpicture}}
      \caption{Possible intersection graph $G$ that generated $H''$. Bold
        blue vertices are from $A$ while dashed red ones are from $A^*$.}
      \label{figa:one_unmatched-G}
    \end{subfigure}
    
    \begin{subfigure}{\textwidth}
      \centering
      \resizebox{\textwidth}{!}{\begin{tikzpicture}
    [vertex/.style={circle,draw,fill,inner sep=0pt,minimum size=5pt}]


    \def\h{10mm};

    \node (P) at (0,0) {};

    \node[vertex,label={[xshift=-7pt,yshift=-9pt]$u$}]  (u)  at ($(P)+(0*\h,0)$) {};
    \node[vertex,label={[xshift=-7pt,yshift=-9pt]$z$}]  (z)  at ($(P)+(1*\h,0)$) {};
    \node[vertex,label={[xshift=-7pt,yshift=-11pt]$z'$}] (zp) at ($(P)+(2*\h,0)$) {};
    \node[vertex,label={[xshift=-7pt,yshift=-9pt]$i$}]  (i)  at ($(P)+(3*\h,0)$) {};
    \node[vertex,label={[xshift=-7pt,yshift=-11pt]$j$}]  (j)  at ($(P)+(4*\h,0)$) {};
    \node[vertex,label={[xshift=-7pt,yshift=-11pt]$p$}]  (p)  at ($(P)+(5*\h,0)$) {};
    \node[vertex,label={[xshift=-7pt,yshift=-11pt]$x'$}] (xp) at ($(P)+(6*\h,0)$) {};
    \node[vertex,label={[xshift=-7pt,yshift=-9pt]$x$}]  (x)  at ($(P)+(7*\h,0)$) {};
    \node[vertex,label={[xshift=-7pt,yshift=-11pt]$y$}]  (y)  at ($(P)+(8*\h,0)$) {};
    \node[vertex,label={[xshift=-7pt,yshift=-12pt]$y'$}] (yp) at ($(P)+(9*\h,0)$) {};
    \node[vertex,label={[xshift=-7pt,yshift=-9pt]$w$}]  (w)  at ($(P)+(10*\h,0)$) {};
    \node[vertex,label={[xshift=-7pt,yshift=-11pt]$w'$}] (wp) at ($(P)+(11*\h,0)$) {};
    \node[vertex,label={[xshift=-7pt,yshift=-9pt]$r$}]  (r)  at ($(P)+(12*\h,0)$) {};
    \node[vertex,label={[xshift=-7pt,yshift=-12pt]$f$}]  (f)  at ($(P)+(13*\h,0)$) {};
    \node[vertex,label={[xshift=-7pt,yshift=-11pt]$g$}]  (g)  at ($(P)+(14*\h,0)$) {};

    \node[vertex,label={[xshift=-7pt,yshift=-9pt]$v$}] (v) at ($(u)+(\h/2,1.2)$) {};
    \node[vertex,label={[xshift=-7pt,yshift=-9pt]$o$}] (o) at ($(z)+(\h/2,1.2)$) {};
    \node[vertex,label={[xshift=7pt,yshift=-9pt]$k$}]  (k) at ($(j)+(0,1.2)$) {};
    \node[vertex,label={[xshift=7pt,yshift=-10pt]$q$}] (q) at ($(p)+(\h/2,1.2)$) {};
    \node[vertex,label={[xshift=7pt,yshift=-9pt]$n$}]  (n) at ($(xp)+(\h/2,1.2)$) {};
    \node[vertex,label={[xshift=-8pt,yshift=-9pt]$t$}]  (t) at ($(y)+(0,1.2)$) {};
    \node[vertex,label={[xshift=-8pt,yshift=-9pt]$m$}]  (m) at ($(yp)+(0,1.2)$) {};
    \node[vertex,label={[xshift=7pt,yshift=-9pt]$l$}]  (l) at ($(w)+(0,1.2)$) {};
    \node[vertex,label={[xshift=7pt,yshift=-9pt]$s$}]  (s) at ($(wp)+(\h/2,1.2)$) {};
    \node[vertex,label={[xshift=7pt,yshift=-9pt]$h$}]  (h) at ($(f)+(0,1.2)$) {};

    \draw[-latex,red] (v) -- (u);
    \draw[-latex,red] (v) -- (z);
    \draw[-latex,blue] (o) -- (z);
    \draw[-latex,blue] (o) -- (zp);
    \draw[-latex,blue] (k) -- (i);
    \draw[-latex,blue] (k) -- (j);
    \draw[-latex,dashed,red] (k) -- (i);
    \draw[-latex,dashed,red] (k) -- (j);
    \draw[-latex,blue] (k) -- (p);
    \draw[-latex,red] (q) -- (zp);
    \draw[-latex,red] (q) -- (p);
    \draw[-latex,red] (q) -- (xp);
    \draw[-latex,blue] (n) -- (xp);
    \draw[-latex,blue] (n) -- (x);
    \draw[-latex,red] (t) -- (x);
    \draw[-latex,red] (t) -- (y);
    \draw[-latex,red] (t) -- (yp);
    \draw[-latex,blue] (m) -- (y);
    \draw[-latex,blue] (m) -- (w);
    \draw[-latex,blue] (l) -- (yp);
    \draw[-latex,blue] (l) -- (wp);
    \draw[-latex,red] (s) -- (w);
    \draw[-latex,red] (s) -- (wp);
    \draw[-latex,red] (s) -- (r);
    \draw[-latex,blue] (h) -- (r);
    \draw[-latex,blue] (h) -- (f);
    \draw[-latex,blue] (h) -- (g);
    \draw[-latex,dashed,red] (h) -- (f);
    \draw[-latex,dashed,red] (h) -- (g);

    \node[vertex,label={[xshift=-7pt,yshift=-5pt]$d$}]  (d) at ($(zp)+(\h/2,2.4)$) {};
    \node[vertex,label={[xshift=7pt,yshift=-5pt]$e$}]  (e) at ($(l)+(\h/2,1.2)$) {};
    \node[vertex,label={[xshift=7pt,yshift=-5pt]$c$}]  (c) at ($(n)+(0,2.4)$) {};

    \draw[-latex,line width=2pt] (d) -- (v);
    \draw[-latex,line width=2pt] (d) -- (o);
    \draw[-latex,line width=2pt] (d) -- (k);
    \draw[-latex,line width=2pt] (d) -- (q);
    \draw[-latex,line width=2pt] (e) -- (m);
    \draw[-latex,line width=2pt] (e) -- (l);
    \draw[-latex,line width=2pt] (e) -- (s);
    \draw[-latex,line width=2pt] (e) -- (h);
    \draw[-latex,line width=2pt] (c) -- (d);
    \draw[-latex,line width=2pt] (c) -- (e);
    \draw[-latex,line width=2pt] (c) -- (n);
    \draw[-latex,line width=2pt] (c) -- (t);

\end{tikzpicture}}
      \caption{Digraph $D$ from which $G$ was built. Bold arcs form a 4-branching.}
      \label{figa:one_unmatched-D}
    \end{subfigure}
  \end{minipage}
  \caption{Illustration of Claim~\ref{claim:one_unmatched}.}
  \label{fig:one_unmatched}
\end{figure}

From Claims~\ref{claim:two_unmatched} and \ref{claim:one_unmatched}, we
conclude the proof of Theorem~\ref{thm:squareimp75}. 
\end{proof}

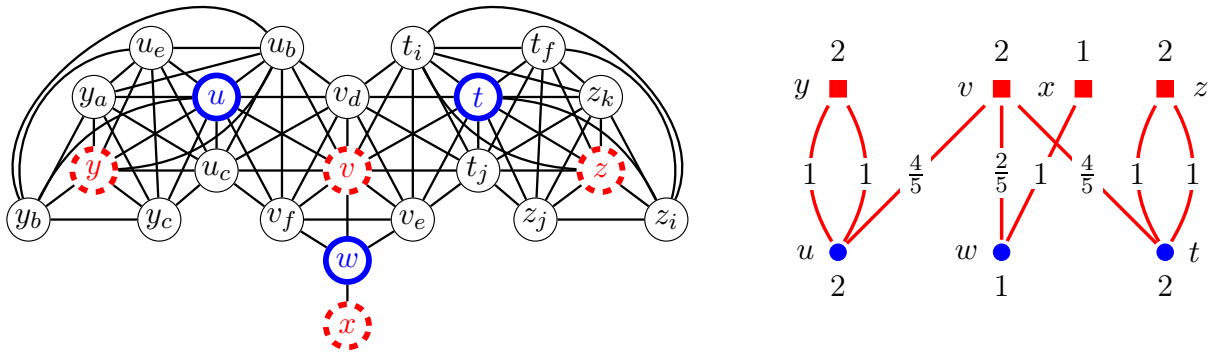
\begin{figure}[h]
  \centering
  \resizebox{\textwidth}{!}{\begin{tikzpicture}
  [vertex/.style={circle,draw,inner sep=0pt,minimum size=5pt,minimum width=15pt},
  LabelStyle/.style={inner sep=2pt}]

    \clip (-4.2,0.6) rectangle (10.6,5.1);

  \node[vertex,line width=2pt,dashed,red] (v) at (0,3) {$v$};
  \node[vertex] (v1) at ($(v)+(0,0.9)$) {$v_d$};
  \node[vertex] (v2) at ($(v)+(-0.8,-0.6)$) {$v_f$};
  \node[vertex] (v3) at ($(v)+(+0.8,-0.6)$) {$v_e$};
  \draw[thick] (v) -- (v1);
  \draw[thick] (v) -- (v2);
  \draw[thick] (v) -- (v3);
  \draw[thick] (v1) -- (v2);
  \draw[thick] (v1) -- (v3);
  \draw[thick] (v2) -- (v3);

  \node[vertex,line width=2pt,blue] (w) at ($(v)+(0,-1.1)$) {$w$};
  \draw[thick] (w) -- (v);
  \draw[thick] (w) -- (v2);
  \draw[thick] (w) -- (v3);
  \node[vertex,line width=2pt,dashed,red] (x) at ($(v)+(0,-1.9)$) {$x$};
  \draw[thick] (w) -- (x);

  \node[vertex,line width=2pt,dashed,red] (y) at ($(v)+(-3.1,0)$) {$y$};
  \node[vertex] (y1) at ($(y)+(0,0.9)$) {$y_a$};
  \node[vertex] (y2) at ($(y)+(-0.8,-0.6)$) {$y_b$};
  \node[vertex] (y3) at ($(y)+(+0.8,-0.6)$) {$y_c$};
  \draw[thick] (y) -- (y1);
  \draw[thick] (y) -- (y2);
  \draw[thick] (y) -- (y3);
  \draw[thick] (y1) -- (y2);
  \draw[thick] (y1) -- (y3);
  \draw[thick] (y2) -- (y3);

  \node[vertex,line width=2pt,dashed,red] (z) at ($(v)+(3.1,0)$) {$z$};
  \node[vertex] (z1) at ($(z)+(0,0.9)$) {$z_k$};
  \node[vertex] (z2) at ($(z)+(-0.8,-0.6)$) {$z_j$};
  \node[vertex] (z3) at ($(z)+(+0.8,-0.6)$) {$z_i$};
  \draw[thick] (z) -- (z1);
  \draw[thick] (z) -- (z2);
  \draw[thick] (z) -- (z3);
  \draw[thick] (z1) -- (z2);
  \draw[thick] (z1) -- (z3);
  \draw[thick] (z2) -- (z3);

  \node[vertex,line width=2pt,blue] (u) at ($(v)+(-1.6,0.9)$) {$u$};
  \node[vertex] (u1) at ($(u)+(0,-0.9)$) {$u_c$};
  \node[vertex] (u2) at ($(u)+(-0.8,0.6)$) {$u_e$};
  \node[vertex] (u3) at ($(u)+(+0.8,0.6)$) {$u_b$};
  \draw[thick] (u) -- (u1);
  \draw[thick] (u) -- (u2);
  \draw[thick] (u) -- (u3);
  \draw[thick] (u1) -- (u2);
  \draw[thick] (u1) -- (u3);
  \draw[thick] (u2) -- (u3);

  \node[vertex,line width=2pt,blue] (t) at ($(v)+(1.6,0.9)$) {$t$};
  \node[vertex] (t1) at ($(t)+(0,-0.9)$) {$t_j$};
  \node[vertex] (t2) at ($(t)+(-0.8,0.6)$) {$t_i$};
  \node[vertex] (t3) at ($(t)+(+0.8,0.6)$) {$t_f$};
  \draw[thick] (t) -- (t1);
  \draw[thick] (t) -- (t2);
  \draw[thick] (t) -- (t3);
  \draw[thick] (t1) -- (t2);
  \draw[thick] (t1) -- (t3);
  \draw[thick] (t2) -- (t3);



  \draw[thick] (v) -- (u);
  \draw[thick] (v) -- (t);
  \draw[thick] (v2) -- (u);
  \draw[thick] (v1) -- (u);
  \draw[thick] (v) -- (u1);
  \draw[thick] (v) -- (u3);
  \draw[thick] (v2) -- (u1);
  \draw[thick] (v2) -- (u3);
  \draw[thick] (v1) -- (u1);
  \draw[thick] (v1) -- (u3);
  \draw[thick] (v3) -- (t1);
  \draw[thick] (v1) -- (t1);
  \draw[thick] (v3) -- (t2);
  \draw[thick] (v1) -- (t2);
  \draw[thick] (v) -- (t1);
  \draw[thick] (v) -- (t2);
  \draw[thick] (v3) -- (t);
  \draw[thick] (v1) -- (t);

  \draw[thick] (y) -- (u);
  \draw[thick] (y3) -- (u);
  \draw[thick] (y1) -- (u);
  \draw[thick] (u1) -- (y);
  \draw[thick] (u2) -- (y);
  \draw[thick] (u1) -- (y1);
  \draw[thick] (u2) -- (y1);
  \draw[thick] (u3) -- (y1);
  \draw[thick] (u2) -- (y3);
  \draw[thick] (u1) -- (y3);
  \Edge[style={bend left}](u3)(y);
  \Edge[style={bend right}](u)(y2);
  \Edge[style={bend right=80}](u3)(y2);
  \Edge[style={bend right=50}](u2)(y2);

  \draw[thick] (z) -- (t);
  \draw[thick] (z1) -- (t);
  \draw[thick] (z2) -- (t);
  \draw[thick] (z) -- (t1);
  \draw[thick] (z) -- (t3);
  \draw[thick] (z1) -- (t1);
  \draw[thick] (z2) -- (t1);
  \draw[thick] (z1) -- (t2);
  \draw[thick] (z2) -- (t3);
  \draw[thick] (z1) -- (t3);
  \Edge[style={bend right}](t2)(z);
  \Edge[style={bend left}](t)(z3);
  \Edge[style={bend left=80}](t2)(z3);
  \Edge[style={bend left=50}](t3)(z3);

  \node[label=above:2,label=left:$y$](r0) [red vertex] at (6,4) {};
  \node[label=below:2,label=left:$u$](f0) [blue vertex] at ($(r0)+(0,-2)$) {};
  \Edge[color=red,label={ 1},lw=1.2pt,style={pos=0.5,bend right}](r0)(f0);
  \Edge[color=red,label={ 1},lw=1.2pt,style={pos=0.5,bend left}](r0)(f0);

  \node[label=above:2,label=right:$z$](r1) [red vertex] at ($(r0)+(4,0)$) {};
  \node[label=below:2,label=right:$t$](f1) [blue vertex] at ($(r1)+(0,-2)$) {};
  \Edge[color=red,label={ 1},lw=1.2pt,style={pos=0.5,bend right}](r1)(f1);
  \Edge[color=red,label={ 1},lw=1.2pt,style={pos=0.5,bend left}](r1)(f1);

  \node[label=above:2,label=left:$v$](r2) [red vertex] at ($(r0)+(2,0)$) {};
  \node[label=above:1,label=left:$x$](r3) [red vertex] at ($(r0)+(3,0)$) {};
  \node[label=below:1,label=left:$w$](f2) [blue vertex] at ($(r2)+(0,-2)$) {};
  \Edge[color=red,label={ $\frac45$},lw=1.2pt,style={pos=0.5}](r2)(f0);
  \Edge[color=red,label={ $\frac45$},lw=1.2pt,style={pos=0.5}](r2)(f1);
  \Edge[color=red,label={ $\frac25$},lw=1.2pt,style={pos=0.5}](r2)(f2);
  \Edge[color=red,label={ 1},lw=1.2pt,style={pos=0.5}](r3)(f2);

\end{tikzpicture}}
  \caption{Tight example for the $7/5$-approximation (parallel edges omitted).}
  \label{fig:tight75}
\end{figure}

This analysis is tight.
For the example in Figure~\ref{fig:tight75}, \SquarePImp\ might produce the
independent set marked in bold blue, of weight~5, while the heaviest
independent set is marked in dashed red, and has weight~7.
Note that there is no improving claw and no augmenting path. 

Applying Theorem~\ref{thm:weighted_approx} with \SquarePImp\ as algorithm
$\mathcal{A}$, and using Theorem~\ref{thm:squareimp75}, we strengthen
Corollary~\ref{cor:32} and derive our main contribution on \maxleaves.

\begin{theorem}
  Algorithm \Call{MaxLeaves-12MIS}{} using \SquarePImp\ instead of \SquareImp\ 
  is a~$7/5$-approximation for the \maxleaves\ on rooted directed acyclic graphs.
\end{theorem}

\section{Final remarks}
\label{sec:remarks}

One might ask whether using the sum of the weights instead of $w^2_+$ would
lead to the same algorithm, that is, would induce the same improvements that
\SquarePImp\ does, for weighted $\{2,3\}$-intersection graphs.
However the claw from the example in Figure~\ref{fig:two_unmatched-case2} would
not be improving for the sum of the weights.
From that example, one can construct a larger example, with a weight-1 vertex
adjacent to each of~$q$, $s$, and~$t$, and include these in the dashed red
independent set, to show that the variant using the sum of the weights does not
achieve a ratio better than~$7/5$. 

The ideas used here to achieve a better approximation for wMIS on weighted
$\{2,3\}$-intersection graphs might lead to improvements for wMIS on $d$-claw
free graphs, or particularly to the weighted 3D-matching problem.
Also, our strategy applied to Neuwohner's algorithm~\cite{Neuwohner2021} 
might lead to a ratio better than $7/5$ for weighted $\{2,3\}$-intersection graphs,
which would imply an improvement for \maxleaves\ on rooted dags.

In the Maximum Leaf Spanning Tree problem, one is given a connected undirected
graph~$G$ and wants to find a spanning tree of~$G$ with the maximum number of
leaves, where a leaf is a vertex of degree~1.
The best known approximation algorithm for this problem has ratio~2 and it was
proposed by Solis-Oba more than~20 years
ago~\cite{SolisOba1998,SolisObaBL2017}.
It would be nice also if some of the ideas explored in this paper were helpful
to obtain a better approximation for the maximum leaf spanning tree, or for
\maxleaves\ for general rooted digraphs.  
For both of these, however, the idea of using wMIS seems harder to be applied.

\bibliographystyle{plain}
\bibliography{maxleaves}

\newpage

\appendix

\section{Proof of Theorem~\ref{thm:weighted_approx}}
\label{proofsDAM}

In a previous work, we presented a theorem for \maxleaves\ involving
approximations for the weighted
3D-matching~\cite[Theorem~4.3]{FernandesL2021}.
That theorem can be adapted to address approximations for the wMIS on weighted
$\{2,3\}$-intersection graphs, resulting in Theorem~\ref{thm:weighted_approx}. 

In this appendix, we present the proof for Theorem~\ref{thm:weighted_approx},
whose proof relies on the adaptation of two lemmas from~\cite[Lemmas~4.1
and~4.2]{FernandesL2021}.
Though the statement of these three results are different from their
corresponding versions in~\cite{FernandesL2021}, their proofs depend on
certain variables defined using the algorithm addressed.
Once we define these variables using \Call{MaxLeaves-12MIS}{}, the proofs are
(essentially) the same.
Yet, for completeness, we include them here. 

We start by presenting the two adapted lemmas.
For that, we establish the notation. 
Let us denote by $\calA$-\Call{MaxLeaves-12MIS}{} a version of
\Call{MaxLeaves-12MIS}{} that uses an algorithm~$\calA$ for the wMIS on
weighted $\{2,3\}$-intersection graphs in Line~\ref{alg:call_squareimp} instead
of \SquareImp.

Let~$D$ be a rooted \Dag\ and consider a call
$\calA$-\Call{MaxLeaves-12MIS}{$D$}.
Let~$F_1$ and~$T$ be the branchings produced in Lines~\ref{line:F1}
and~\ref{line:T} respectively. 
Let~$F_3$ be the state of the branching~$F_2$ just before Line~\ref{line:T}.
In what follows, let~$F_2$ denote the branching obtained from~$F_1$ if
Lines~\ref{line:forF2}-\ref{line:EndForF2} were executed only for vertices~$v$
with $w_v = 2$.  
For~$i=1,2,3$, let~$k_i$ be the number of non-trivial components of~$F_i$
and~$N_i$ be the number of vertices in such components.  

\newcommand{\FF}{F}
\newcommand{\NN}{N}
\newcommand{\kk}{k}

\begin{lemma}\label{lem:lower_w3dm}
  Let $T$ be the arborescence produced by $\calA$-\Call{MaxLeaves-12MIS}{$D$}.
  Then \[\ell(T) \ \geq \ \frac{\NN_1-\kk_1}{12} + \frac{\NN_2-\kk_2}{6} + \frac{\NN_3-\kk_3}{2} + 1 \; .\]
\end{lemma}
\begin{proof}
  Let $n$ be the number of vertices of $D$. 
  Let $T_1,\ldots,T_{\kk_1}$ be the non-trivial arborescences in $\FF_1$. 
  Note that $\ell(T_j) \geq \frac{1+3|V(T_j)|}4$ because all internal vertices
  of~$T_j$ have out-degree at least~4.
  Therefore, 
  \begin{align*}
    \ell(\FF_1) & \ = \ n - \NN_1 + \sum_{j=1}^{\kk_1} \ell(T_j)
                  \ \geq \  n - \NN_1 + \sum_{j=1}^{\kk_1} \frac{1+3|V(T_j)|}4 \\
                & \ = \ n - \NN_1 + \frac{3\NN_1}4 + \frac{\kk_1}4 
                  \ = \ n - \frac{\NN_1-\kk_1}4 \; .
  \end{align*}
  
  The number of components in $\FF_i$ is $n-\NN_i+\kk_i$ for $i=1,2,3$.
  Hence, the number of leaves lost from $\FF_1$ to $\FF_2$ is exactly 
  \[\frac{(n-\NN_1+\kk_1) - (n-\NN_2+\kk_2)}3 \ = \ \frac{\NN_2-\kk_2}3 - \frac{\NN_1-\kk_1}3 \; .\] 
  Similarly, the number of leaves lost from $\FF_2$ to $\FF_3$ is exactly
  \[\frac{(n-\NN_2+\kk_2) - (n-\NN_3+\kk_3)}2 \ = \ \frac{\NN_3-\kk_3}2 - \frac{\NN_2-\kk_2}2 \; .\] 
  Also, the number of leaves lost from $\FF_3$ to $T$ is exactly $n - \NN_3 +
  \kk_3 - 1 = n - (\NN_3-\kk_3) - 1$.
  Thus
  \begin{align*}
    \ell(T) & \ \geq \ n - \frac{\NN_1-\kk_1}4 - \left(\frac{\NN_2-\kk_2}3 - \frac{\NN_1-\kk_1}3\right) \\
            & \phantom{\ \geq \ n \ } - \left(\frac{\NN_3-\kk_3}2 - \frac{\NN_2-\kk_2}2\right) - (n-(\NN_3-\kk_3)-1) \\
            & \ = \ \frac{1}{12} (\NN_1-\kk_1) + \frac16 (\NN_2-\kk_2) + \frac12 (\NN_3-\kk_3) + 1 \; .
  \end{align*}
\vspace{-5mm} 
\end{proof}

Now we present an upper bound on $\opt(D)$ that relates to the lower bound
presented in Lemma~\ref{lem:lower_w3dm}.

\begin{lemma}\label{lemma:w3dm_upper_opt}
  If the algorithm $\calA$ used in $\calA$-\Call{MaxLeaves-12MIS}{$D$} is an
  $\alpha$-approximation algorithm for the wMIS on weighted
  $\{2,3\}$-intersection graphs, then 
  $$\opt(D) \ \leq \ \frac{3-2\alpha}{3}(\NN_1-\kk_1) + \frac{\alpha}{6}(\NN_2-\kk_2) + \frac{\alpha}{2}(\NN_3-\kk_3) + 1.$$ 
\end{lemma}
\begin{proof}
  Let $T^*$ be a spanning arborescence of $D$ with the maximum number of
  leaves.  
  Call $R$ the set of all roots of non-trivial components of $\FF_1$. 
  Call $L$ the set of leaves of $T^*$ that are isolated vertices of $\FF_1$. 
  Let $Z := L \cup R \setminus \{r\}$, where $r$ is the root of $D$. 
  The witness of a vertex $z \in Z$ is the closest proper predecessor $q(z)$ of
  $z$ in $T^*$ which is in a non-trivial component of $\FF_1$.
  Note that each witness is an internal vertex of $T^*$.
  
  We will prove that the number $\psi$ of distinct witnesses is
  \begin{align}
    \label{eq:witness_number_w3dm}
    \psi & \geq |Z| - 2\alpha\left(\frac{\NN_2-\kk_2}3 - \frac{\NN_1-\kk_1}3\right) - \alpha\left(\frac{\NN_3-\kk_3}2 - \frac{\NN_2-\kk_2}2\right) \\ 
         & = |Z| + 2\alpha\,\frac{\NN_1-\kk_1}3 - \alpha\,\frac{\NN_2-\kk_2}6 - \alpha\,\frac{\NN_3-\kk_3}2 \; \nonumber. 
  \end{align}
  Because $|Z| = \kk_1 - 1 + |L|$ and each witness lies in a non-trivial
  component of~$\FF_1$ and is internal in $T^*$, we deduce that
  \begin{align*}
    \opt(D) & \ \leq \ \NN_1 - \psi + |L| \\
            & \ \leq \ \NN_1 - |Z| - 2\alpha\,\frac{\NN_1-\kk_1}3 + \alpha\,\frac{\NN_2-\kk_2}6 + \alpha\,\frac{\NN_3-\kk_3}2 + |L| \\
            & \ = \ \NN_1 - \kk_1 - 2\alpha\,\frac{\NN_1-\kk_1}3 + \alpha\,\frac{\NN_2-\kk_2}6 + \alpha\,\frac{\NN_3-\kk_3}2 + 1 \\
            & \ = \ \frac{3-2\alpha}{3}(\NN_1-\kk_1) + \frac{\alpha}{6}(\NN_2-\kk_2) + \frac{\alpha}{2}(\NN_3-\kk_3) + 1 \; .
  \end{align*}
  It remains to prove~\eqref{eq:witness_number_w3dm}.
  
  For a witness $s$, let $Z_s := \{ z \in Z: q(z) = s\}$ and let $T^*_s$ be the
  subarborescence of~$T^*$ induced by the union of all paths in $T^*$ from $s$
  to each vertex in~$Z_s$.  
  The number of such arborescences $T^*_s$ is exactly $\psi$.  
  The only internal vertex of~$T^*_s$ that is in a non-trivial component
  of~$\FF_1$ is its root $s$, which is necessarily a leaf of~$\FF_1$. 
  So the maximum out-degree in~$T^*_s$ is at most three. 
  
  Again, no $z \in Z_s$ is a predecessor in $T^*_s$ of another $z' \in Z_s$. 
  Indeed, suppose by contradiction that $z$ is in the path from $s$ to $z'$.  
  Then $z$ is not a leaf of~$T^*$, and is in $R$, thus being in a non-trivial component of~$\FF_1$, 
  which is a contradiction, because~$z$, and not~$s$, would be the witness for~$z'$.
  Hence $T^*_s$ has exactly $|Z_s|$ leaves.  

  Let $(G,w)$ be the weighted $\{2,3\}$-intersection graph built in
  Lines~\ref{line9}-\ref{line11}, and let $I$ be the independent set in~$G$
  computed by $\calA$ in Line~\ref{alg:call_squareimp} of
  $\calA$-\Call{MaxLeaves-12MIS}{$D$}.
  For every~$v$ such that~$U_v = \{a,b,c\}$, the algorithm includes $v_a,v_b,v_c$ in $\Candidates$. 
  At most one in~$\{v, v_a, v_b, v_c\}$ is included in~$I$.
  Let~$B_i$ be the set of vertices~$v$ in $I$ such that $|U_v| = i$, for~$i = 2,3$. 
  Note that~$|B_3|$ is exactly the number of leaves lost from branching~$\FF_1$ to $\FF_2$, so 
  \begin{equation}
    \label{eq:w3dm_b3}
    |B_3| \ = \ \frac{\NN_2-\kk_2}3 - \frac{\NN_1-\kk_1}3 \; .
  \end{equation}
  Also, $|B_2|$ is exactly the number of leaves lost from branching~$\FF_2$ to $\FF_3$, so
  \begin{equation}
    \label{eq:w3dm_b2}
    |B_2| \ = \ \frac{\NN_3-\kk_3}2 - \frac{\NN_2-\kk_2}2 \; .
  \end{equation}
  Finally, $|I| = |B_3| + |B_2|$ and $w(I) = 2|B_3| + |B_2|$.
  
  Vertices with out-degree two and three in $T^*_s$ are all in the
  set~$\Candidates$.
  Indeed, let~$v$ be one such vertex.
  Either~$v$ is an isolated vertex or~$v$ is a leaf of a non-trivial component
  of~$\FF_1$.  
  So~$d_{\FF_1}^+(v) = 0$.  
  As the children of $v$ in $T^*_s$ have in-degree 0 in~$\FF_1$, they are all
  in~$U_v$. 
  Hence~$v \in \Candidates$. 
  
  For $i = 2,3$, let $C^i_s$ be the set of vertices of $\Candidates$ with
  out-degree~$i$ in~$T^*_s$, and let~$C = \cup_s C^i_s$.  
  The number of leaves in~$T^*_s$ is ${|Z_s| = 2|C^3_s|+|C^2_s|+1}$. 
  The set of internal vertices of~$T^*_s$ and of~$T^*_{s'}$ are disjoint for
  distinct witnesses~$s$ and~$s'$.  
  So the sets~$C^i_s$ and $C^i_{s'}$ are disjoint.  
  Note that $C$ is an independent set in $G$, thus~$w(C) = 2|C^3_s| + |C^2_s|
  \leq w(I^*) \leq \alpha\,w(I)$, where $I^*$ is a maximum weight independent
  set in $(G,w)$.
  Hence
  \begin{align*}
    |Z| \ &= \ \sum_s |Z_s| \ = \ \sum_s (2|C^3_s|+|C^2_s|+1) \ = \ w(C) + \psi \\ 
        &\leq \ \alpha\,w(I) + \psi \ = \ 2\alpha|B_3| + \alpha|B_2| + \psi \,.
  \end{align*}
  Therefore, 
  \begin{align*}
    \psi & \ \geq \ |Z| - 2\alpha|B_3| - \alpha|B_2| \\
         & \ = \ |Z| - 2\alpha\left(\frac{\NN_2-\kk_2}3 
           - \frac{\NN_1-\kk_1}3\right) - \alpha\left(\frac{\NN_3-\kk_3}2 - \frac{\NN_2-\kk_2}2\right)\,,
  \end{align*}
  which completes the proof of~\eqref{eq:witness_number_w3dm}. 
\end{proof}

Now we are ready to present the proof of Theorem~\ref{thm:weighted_approx}. 

\begin{proof}[of Theorem~\ref{thm:weighted_approx}]
  First, suppose $\alpha \geq \frac43$. 
  In this case, $\frac{3-2\alpha}{3} \leq \frac{\alpha}{12}$ and, 
  by Lemmas~\ref{lem:lower_w3dm} and~\ref{lemma:w3dm_upper_opt}, 
  \begin{align*}
    \opt(D) & \ \leq \ \frac{3-2\alpha}{3}(N_1-k_1) + \frac{\alpha}{6}(N_2-k_2) + \frac{\alpha}{2}(N_3-k_3) + 1 \\
            & \ \leq \ \frac{\alpha}{12} (N_1-k_1) + \frac{\alpha}6 (N_2-k_2) + \frac{\alpha}2 (N_3-k_3) + \alpha \\
            & \ \leq \ \alpha\,\ell(T)\; . 
  \end{align*}

  Now, suppose $\alpha < \frac43$, and let $\beta = \frac43 - \alpha$.
  By Lemma~\ref{lemma:w3dm_upper_opt},
  \begin{align}
    \opt(D) & \ \leq \ \frac{3-2\alpha}{3}(N_1-k_1) + \frac{\alpha}{6}(N_2-k_2) + \frac{\alpha}{2}(N_3-k_3) + 1 \nonumber \\
            & \ \leq \ \left(\frac19+\frac23\,\beta\right)(N_1-k_1) + \left(\frac29-\frac16\,\beta\right)(N_2-k_2) \nonumber \\
            & \phantom{\ \leq \ } + \left(\frac23-\frac12\,\beta\right)(N_3-k_3) + 1 \nonumber \\
            & \ = \ \frac19(N_1-k_1) + \frac29(N_2-k_2) + \frac23(N_3-k_3) + \frac43 \nonumber \\ 
            & \phantom{\ \leq \ } + \frac23\,\beta(N_1-k_1) - \frac16\,\beta(N_2-k_2) - \frac12\,\beta(N_3-k_3) - \frac13 \nonumber \\
            & \ \leq \ \frac43\,\ell(T) + \frac23\,\beta\left((N_1{-}k_1) - \frac14(N_2{-}k_2) - \frac34(N_3{-}k_3)\right) - \frac13\label{eq:lT}\\
            & \ \leq \ \frac43\,\ell(T)\;, \label{eq:negative} 
  \end{align}
  where~\eqref{eq:lT} holds by Lemma~\ref{lem:lower_w3dm}
  and~\eqref{eq:negative} holds because the number of components in~$F_1$,
  $F_2$, and $F_3$ is so that $n-N_1+k_1 \geq n-N_2+k_2 \geq n-N_3+k_3$, and
  this implies that $N_1-k_1 \leq N_2-k_2 \leq N_3-k_3$, and therefore $N_1-k_1
  \leq \frac14(N_2-k_2) + \frac34(N_3-k_3)$. 
\end{proof}

\end{document}